\documentclass[twocolumn,showpacs,showkeys,preprintnumbers,amsmath,amssymb,floatfix,aps,nofootinbib]{revtex4-1}
\usepackage[latin9]{inputenc}
\usepackage{color}
\usepackage{graphicx}

\usepackage[colorlinks,urlcolor=blue]{hyperref}
\usepackage{breakurl}

\makeatletter

\@ifundefined{textcolor}{}
{%
 \definecolor{BLACK}{gray}{0}
 \definecolor{WHITE}{gray}{1}
 \definecolor{RED}{rgb}{1,0,0}
 \definecolor{GREEN}{rgb}{0,1,0}
 \definecolor{BLUE}{rgb}{0,0,1}
 \definecolor{CYAN}{cmyk}{1,0,0,0}
 \definecolor{MAGENTA}{cmyk}{0,1,0,0}
 \definecolor{YELLOW}{cmyk}{0,0,1,0}
}

\newcounter{univ_counter}
\addtocounter{univ_counter} {1} 
\edef\ANL{$^{\arabic{univ_counter}}$ }

\addtocounter{univ_counter} {1} 
\edef\BRESCIA{$^{\arabic{univ_counter}}$ }

\addtocounter{univ_counter} {1} 
\edef\UCR{$^{\arabic{univ_counter}}$ }

\addtocounter{univ_counter} {1} 
\edef\CSUDH{$^{\arabic{univ_counter}}$ }

\addtocounter{univ_counter} {1} 
\edef\CANISIUS{$^{\arabic{univ_counter}}$ }

\addtocounter{univ_counter} {1} 
\edef\UCONN{$^{\arabic{univ_counter}}$ }

\addtocounter{univ_counter} {1} 
\edef\FU{$^{\arabic{univ_counter}}$ }

\addtocounter{univ_counter} {1}
\edef\FERRARAU{$^{\arabic{univ_counter}}$ }

\addtocounter{univ_counter} {1} 
\edef\FIU{$^{\arabic{univ_counter}}$ }

\addtocounter{univ_counter} {1} 
\edef\FSU{$^{\arabic{univ_counter}}$ }

\addtocounter{univ_counter} {1} 
\edef\GIESSEN{$^{\arabic{univ_counter}}$ }

\addtocounter{univ_counter} {1} 
\edef\GWUI{$^{\arabic{univ_counter}}$ }

\addtocounter{univ_counter} {1} 
\edef\GLASGOW{$^{\arabic{univ_counter}}$ }

\addtocounter{univ_counter} {1} 
\edef\HAMPTON{$^{\arabic{univ_counter}}$ }

\addtocounter{univ_counter} {1} 
\edef\INFNFE{$^{\arabic{univ_counter}}$ }

\addtocounter{univ_counter} {1} 
\edef\INFNFR{$^{\arabic{univ_counter}}$ }

\addtocounter{univ_counter} {1} 
\edef\INFNGE{$^{\arabic{univ_counter}}$ }

\addtocounter{univ_counter} {1} 
\edef\INFNPAV{$^{\arabic{univ_counter}}$ }

\addtocounter{univ_counter} {1} 
\edef\INFNRO{$^{\arabic{univ_counter}}$ }

\addtocounter{univ_counter} {1} 
\edef\INFNTUR{$^{\arabic{univ_counter}}$ }

\addtocounter{univ_counter} {1} 
\edef\ITEP{$^{\arabic{univ_counter}}$ }

\addtocounter{univ_counter} {1} 
\edef\JUELICH{$^{\arabic{univ_counter}}$ }

\addtocounter{univ_counter} {1} 
\edef\KNU{$^{\arabic{univ_counter}}$ }

\addtocounter{univ_counter} {1} 
\edef\LAMAR{$^{\arabic{univ_counter}}$ }

\addtocounter{univ_counter} {1} 
\edef\MIT{$^{\arabic{univ_counter}}$ }

\addtocounter{univ_counter} {1} 
\edef\MISS{$^{\arabic{univ_counter}}$ }

\addtocounter{univ_counter} {1} 
\edef\UNH{$^{\arabic{univ_counter}}$ }

\addtocounter{univ_counter} {1} 
\edef\NMSU{$^{\arabic{univ_counter}}$ }

\addtocounter{univ_counter} {1} 
\edef\NSU{$^{\arabic{univ_counter}}$ }

\addtocounter{univ_counter} {1} 
\edef\OHIOU{$^{\arabic{univ_counter}}$ }

\addtocounter{univ_counter} {1} 
\edef\ODU{$^{\arabic{univ_counter}}$ }

\addtocounter{univ_counter} {1} 
\edef\ORSAY{$^{\arabic{univ_counter}}$ }

\addtocounter{univ_counter} {1} 
\edef\URICH{$^{\arabic{univ_counter}}$ }

\addtocounter{univ_counter} {1} 
\edef\ROMAII{$^{\arabic{univ_counter}}$ }

\addtocounter{univ_counter} {1} 
\edef\SACLAY{$^{\arabic{univ_counter}}$ }

\addtocounter{univ_counter} {1} 
\edef\MSU{$^{\arabic{univ_counter}}$ }

\addtocounter{univ_counter} {1} 
\edef\SCAROLINA{$^{\arabic{univ_counter}}$ }

\addtocounter{univ_counter} {1} 
\edef\TEMPLE{$^{\arabic{univ_counter}}$ }

\addtocounter{univ_counter} {1} 
\edef\UTFSM{$^{\arabic{univ_counter}}$ }

\addtocounter{univ_counter} {1}
\edef\JLAB{$^{\arabic{univ_counter}}$ }

\addtocounter{univ_counter} {1} 
\edef\VIRGINIA{$^{\arabic{univ_counter}}$ }

\addtocounter{univ_counter} {1} 
\edef\WM{$^{\arabic{univ_counter}}$ }

\addtocounter{univ_counter} {1} 
\edef\YEREVAN{$^{\arabic{univ_counter}}$ }

\addtocounter{univ_counter} {1} 
\edef\YORK{$^{\arabic{univ_counter}}$ }

\begin{document}

\preprint{Phys. Rev. C}

\title{Beam-Recoil Transferred Polarization in $K^+Y$ Electroproduction in the Nucleon Resonance Region with CLAS12}

\author{D.S.~Carman$^\dag$,\JLAB\
A.~D'Angelo,\INFNRO$\!\!^,$\ROMAII\
L.~Lanza,\INFNRO\
V.I.~Mokeev,\JLAB\
K.P.~Adhikari,\HAMPTON\
M.J.~Amaryan,\ODU\
W.R.~Armstrong,\ANL\ 
H.~Atac,\TEMPLE\ 
H.~Avakian,\JLAB\
C. Ayerbe Gayoso,\MISS$\!\!^,$\WM\ 
N.A.~Baltzell,\JLAB
% ...
\\
(CLAS Collaboration)}

\affiliation{\ANL Argonne National Laboratory, Argonne, Illinois 60439}
\affiliation{\BRESCIA Universit\`{a} degli Studi di Brescia, 25123 Brescia, Italy}
\affiliation{\UCR University of California Riverside, 900 University Avenue, Riverside, California 92521}
\affiliation{\CSUDH California State University, Dominguez Hills, Carson, California 90747}
\affiliation{\CANISIUS Canisius College, Buffalo, New York 14208}
\affiliation{\UCONN University of Connecticut, Storrs, Connecticut 06269}
\affiliation{\FU Fairfield University, Fairfield, Connecticut 06824}
\affiliation{\FERRARAU Universit\`{a} di Ferrara, 44121 Ferrara, Italy}
\affiliation{\FIU Florida International University, Miami, Florida 33199}
\affiliation{\FSU Florida State University, Tallahassee, Florida 32306}
\affiliation{\GIESSEN II Physikalisches Institut der Universitaet Giessen, 35392 Giessen, Germany}
\affiliation{\GWUI The George Washington University, Washington, D.C. 20052}
\affiliation{\GLASGOW University of Glasgow, Glasgow G12 8QQ, United Kingdom}
\affiliation{\HAMPTON Hampton University, Hampton, VA 23669}
\affiliation{\INFNFE INFN, Sezione di Ferrara, 44100 Ferrara, Italy}
\affiliation{\INFNFR INFN, Laboratori Nazionali di Frascati, 00044 Frascati, Italy}
\affiliation{\INFNGE INFN, Sezione di Genova, 16146 Genova, Italy}
\affiliation{\INFNPAV INFN,  Sezione di Pavia, 27100 Pavia, Italy}
\affiliation{\INFNRO INFN, Sezione di Roma Tor Vergata, 00133 Rome, Italy}
\affiliation{\INFNTUR INFN, Sezione di Torino, 10125 Torino, Italy}
\affiliation{\ITEP National Research Center Kurchatov Institute, Institute of Theoretical and Experimental Physics, 117218 Moscow, Russia}
\affiliation{\JUELICH Institute f\"{u}r Kernphysik, Forschungszentrum J\"{u}lich, 52425 J\"{u}lich, Germany}
\affiliation{\KNU Kyungpook National University, Daegu 702-701, Republic of Korea}
\affiliation{\LAMAR Lamar University, 4400 MLK Blvd, Beaumont, Texas 77710}
\affiliation{\MIT Massachusetts Institute of Technology, Cambridge, Massachusetts 02139}
\affiliation{\MISS Mississippi State University, Mississippi State, Mississippi 39762}
\affiliation{\UNH University of New Hampshire, Durham, New Hampshire 03824}
\affiliation{\NMSU New Mexico State University, Las Cruces, New Mexico 88003}
\affiliation{\NSU Norfolk State University, Norfolk, Virginia 23504}
\affiliation{\OHIOU Ohio University, Athens, Ohio 45701}
\affiliation{\ODU Old Dominion University, Norfolk, Virginia 23529}
\affiliation{\ORSAY Universit\'{e} Paris-Saclay, CNRS/IN2P3, IJCLab, 91405 Orsay, France}
\affiliation{\URICH University of Richmond, Richmond, Virginia 23173}
\affiliation{\ROMAII Universit\`{a} di Roma Tor Vergata, 00133 Rome, Italy}
\affiliation{\SACLAY IRFU, CEA, Universit\'{e} Paris-Saclay, F-91191 Gif-sur-Yvette, France}
\affiliation{\MSU Skobeltsyn Nuclear Physics Institute and Physics Department at Lomonosov Moscow State University, 119899 Moscow, Russia}
\affiliation{\SCAROLINA University of South Carolina, Columbia, South Carolina 29208}
\affiliation{\TEMPLE Temple University, Philadelphia, Pennsylvania 19122}
\affiliation{\UTFSM Universidad T\'{e}cnica Federico Santa Mar\'{i}a, Casilla 110-V Valpara\'{i}so, Chile}
\affiliation{\JLAB Thomas Jefferson National Accelerator Facility, Newport News, Virginia 23606}
\affiliation{\VIRGINIA University of Virginia, Charlottesville, Virginia 22901}
\affiliation{\WM College of William and Mary, Williamsburg, Virginia 23187}
\affiliation{\YEREVAN Yerevan Physics Institute, 375036 Yerevan, Armenia}
\affiliation{\YORK University of York, York YO10 5DD, United Kingdom}

\date{\today}

\begin{abstract}
    Beam-recoil transferred polarizations for the exclusive electroproduction of $K^+\Lambda$ and $K^+\Sigma^0$ final states from an
    unpolarized proton target have been measured using the CLAS12 spectrometer at Jefferson Laboratory. The measurements at beam energies 
    of 6.535~GeV and 7.546~GeV span the range of four-momentum transfer $Q^2$ from 0.3 to 4.5~GeV$^2$ and invariant energy $W$ from 1.6 to 
    2.4~GeV, while covering the full center-of-mass angular range of the $K^+$. These new data extend the existing hyperon polarization 
    data from CLAS in a similar kinematic range but from a significantly larger dataset. They represent an important addition to the world 
    data, allowing for better exploration of the reaction mechanism in strangeness production processes, for further understanding of the 
    spectrum and structure of excited nucleon states, and for improved insight into the strong interaction in the regime of non-perturbative dynamics.
\end{abstract}


\maketitle

\section{Introduction}
\label{intro}

Over the past decade new precise data from exclusive meson photo- and electroproduction have resulted in significant progress in mapping
out the spectrum of excited nucleon states ($N^*$s) and understanding their structure. These detailed studies hold the key to gaining
insight into the nature of the strong interaction dynamics that govern these systems~\cite{capstick,azbu12,Burkert:2019bhp,fbs-carman}.

Based mainly on exclusive meson electroproduction data acquired with the CLAS detector in Hall~B at Jefferson Laboratory (JLab), the
nucleon resonance electroexcitation amplitudes, {\it i.e.} the $\gamma_v p N^*$ electrocouplings, have become available for most $N^*$
states in the mass range up to 1.8~GeV for photon virtualities $Q^2$ up to $\sim$5~GeV$^2$~\cite{azbu12,fbs-carman}. These data offer
unique information on the strong interaction in the regime of large QCD running coupling, the so-called strong QCD (sQCD) regime, which
is responsible for the generation of these $N^*$ states as bound systems of quarks and gluons, with different quantum numbers and
distinctively different structural features. See Refs.~\cite{Burkert:2019bhp,fbs-carman,qcd2019} for recent reviews of the field. The
resonant contributions to the inclusive proton $F_2$ and $F_L$ structure functions have recently been computed from the experimental
results on the $\gamma_vpN^*$ electrocouplings, paving a way for the exploration of the nucleon parton distribution functions in the 
resonance region along with quark-hadron duality~\cite{blin21}.

Mapping out the spectrum of $N^*$ excited states is necessary to explore approximate symmetries relevant for the sQCD regime. Both 
constituent quark models and lattice QCD approaches predict many more $N^*$ states than have been unraveled from analysis of the
experimental data, with a rich spectrum of states predicted in the mass range above 1.8~GeV. This is known as the ``missing'' resonance
problem. Assessing the experimental evidence of higher-mass excited states is also critical for models probing the transition from the
deconfined quark/gluon phase to the hadron phase in the early $\mu$s-old universe~\cite{Burkert:2020}.

The recent progress in understanding the structure of the nucleon excited states has mainly been provided by advanced analyses of the
CLAS data for exclusive electroproduction of the $\pi^+n$, $\pi^0 p$, $\eta p$, and $\pi^+ \pi^- p$ channels from a proton target. However, 
high-precision data from the CLAS Collaboration on exclusive photoproduction of $K^+Y$ ($Y=\Lambda,\Sigma^0$)
\cite{mcnabb,bradford06,bradford07,mccracken,dey,paterson} have been crucial in the exploration of the $N^*$ spectrum: nine new baryon 
states have recently been discovered within global multi-channel analyses of the exclusive photoproduction data with a decisive impact 
from the $K^+Y$ polarization observables~\cite{Burkert:2020,Bur17}. Table~\ref{nstar-evol} shows a comparison of the current Particle 
Data Group~\cite{pdg} listings to that from just a decade ago for twelve $N^*$ and $\Delta^*$ states in the mass range up to 2.2~GeV. For 
many of these states the addition of the $KY$ channels proved important~\cite{ronchen}. Note that although the two ground-state 
hyperons have the same $uds$ valence quark content, they have different isospins ($I$=0 for $\Lambda$ and $I$=1 for $\Sigma^0$), so that 
$N^*$ states of $I=1/2$ can decay to $K^+\Lambda$, but $\Delta^*$ states cannot. Since both $N^*$ and $\Delta^*$ resonances can couple to 
the $K^+\Sigma^0$ final state, the hyperon final state selection is equivalent to an isospin filter.
  
\begin{table}[htbp]
\centering
\begin{tabular}{c|c||c|c|c|c|c} \hline
State               & PDG  & PDG  & $\pi N$ & $K\Lambda$ & $K \Sigma$ & $\gamma N$ \\
$N(mass)J^P$        & 2010 & 2020 &         &            &            &            \\ \hline
$N(1710)1/2^+$      & ***  & **** & **** & ** & *  & **** \\ \hline
$N(1875)3/2^-$      &      & ***  & **   & *  & *  & **   \\ \hline
$N(1880)1/2^+$      &      & ***  & *    & ** & ** & **   \\ \hline
$N(1895)1/2^-$      &      & **** & *    & ** & ** & **** \\ \hline
$N(1900)3/2^+$      & **   & **** & **   & ** & ** & **** \\ \hline
$N(2000)5/2^+$      & *    & **   & *    &    &    & **   \\ \hline
$N(2100)1/2^+$      & *    & ***  & ***  & *  &    & **   \\ \hline
$N(2120)3/2^-$      &      & ***  & **   & ** & *  & ***  \\ \hline
$N(2060)5/2^-$      &      & ***  & **   & *  & *  & ***  \\ \hline
$\Delta(1600)3/2^+$ & ***  & **** & ***  &    &    & **** \\ \hline
$\Delta(1900)1/2^-$ & **   & ***  & ***  &    & ** & ***  \\ \hline
$\Delta(2200)7/2^-$ & *    & ***  & **   &    & ** & ***  \\ \hline
\end{tabular}
\caption{Evolution of our understanding of the excited $N^*$ and $\Delta^*$ spectra over the past decade and the available evidence from 
  different initial/final states based on the PDG * ratings in the listings from a decade ago and the current listings~\cite{fbs-carman}.
  The $KY$ channels represent a crucial inclusion in this expansion of our understanding.}
\label{nstar-evol}
\end{table}

The CLAS $\gamma p \to K^+ Y$ ($Y\!=\!\Lambda,\Sigma^0$) data based on high-statistics experiments have allowed for precision measurements 
with fine binning in the relevant $(W,\cos \theta_K^{c.m.})$ kinematic phase space ({\it n.b.}~\!data in these channels from MAMI, SAPHIR,
and GRAAL are available as well - see the available review papers~\cite{crede13,klempt10} for details). In addition, CLAS has also provided 
most of the available world data results on cross sections~\cite{5st,carman13} and polarization observables
\cite{carman03,rauecarman,sltp,carman09,ipol} for $K^+Y$ electroproduction in the nucleon resonance region. These measurements span $Q^2$ 
from 0.3 to 4.5~GeV$^2$, invariant mass $W$ from 1.6 to 3.0~GeV, and cover the full center-of-mass (c.m.) angular range of the $K^+$. $KY$ 
exclusive production is sensitive to coupling to higher-lying $N^*$ states for $W > 1.6$~GeV, which is precisely the mass range where the
understanding of the $N^*$ spectrum is most limited. See Refs.~\cite{carman16,carman18} for recent reviews on the CLAS electroproduction 
datasets.

The available $K^+Y$ electroproduction data from CLAS have comparable bin widths and statistical uncertainties as for the available CLAS 
$\pi^+\pi^-p$ electroproduction data and can be used to confirm the $\gamma_vpN^*$ electrocouplings for the resonances in the mass
range $>$1.6~GeV that have been obtained from $\pi^+\pi^-p$ electroproduction~\cite{Mokeev:2018zxt,Mok20}. 

Recently a new $N'(1720)3/2^+$ baryon state has been discovered from the combined studies of $\pi^+\pi^-p$ photo- and electroproduction
data from protons~\cite{Mok201}. Similarly, signals of new baryon states observed in photoproduction data can be investigated in a
complementary fashion using electroproduction data by ensuring that, at fixed $Q^2$, the determined states have the same masses and
total decay widths from analyses of both the $\pi^+\pi^-p$ and $K^+Y$ electroproduction channels. However, to be most beneficial in this
regard, it is critical to further develop the existing $K^+Y$ reaction models ({\it e.g.} Refs.~\cite{maxwell,rpr,skoupil,doring,mart21})
to determine the $\gamma_vpN^*$ electrocouplings and to make stronger claims on the $N^* \to KY$ couplings. Improving the statistical
precision and extending the kinematic range of the electroproduction data on the $K^+Y$ differential cross sections and polarization
observables will be critical to foster these efforts. One of the goals of measuring $K^+Y$ electroproduction with the new CLAS12
spectrometer in Hall~B at JLab is to provide electroproduction data in the $Q^2$ range up to 2-3~GeV$^2$ at the same level of accuracy as 
the available photoproduction data, while ultimately extending the available data up to $Q^2$ of 10-12~GeV$^2$. This present measurement
is meant to move in that direction.

The beam-recoil transferred polarization observable has been reported in two previous CLAS electroproduction publications. In 
Ref.~\cite{carman03}, results from a CLAS dataset taken with an electron beam energy of 2.567~GeV were made available for the 
$K^+\Lambda$ final state spanning $Q^2$ from 0.3 to 1.5~GeV$^2$ and $W$ from 1.6 to 2.15~GeV. These data provided the first-ever 
measurement for the $K^+\Lambda$ transferred polarization in electroproduction. In a follow-up paper, additional data from the same
experiment and from a larger dataset taken at beam energies of 4.261~GeV and 5.754~GeV~\cite{carman09}, were reported for the
transferred polarization of the $K^+\Lambda$ final state in the range of $Q^2$ from 0.7 to 5.4~GeV$^2$ and $W$ from 1.6 to 2.6~GeV.
In addition, the first-ever measurement for the $K^+\Sigma^0$ final state in electroproduction was provided in these same kinematics,
although with precision barely sufficient to determine the sign of the polarization.

In this work, measurement of the beam-recoil transferred polarization for the $K^+\Lambda$ and $K^+\Sigma^0$ final states is provided
over a kinematic range of $Q^2$ from 0.3 to 4.5~GeV$^2$ and $W$ from 1.6 to 2.4~GeV with a dataset from CLAS12 that is five times larger 
than any electroproduction dataset available from CLAS for these channels. These data significantly reduce the uncertainties on the 
available $K^+\Lambda$ beam-recoil transferred polarization measurements, while providing the first statistically meaningful measurements 
for the $K^+\Sigma^0$ final state.

The organization for the remainder of this paper is as follows. In Section~\ref{formalism} the definition of the transferred polarization
in terms of the underlying response functions is presented along with the coordinate systems in which the polarization components are
expressed and Section~\ref{approach} provides details on the approach used to extract the polarization components from the data.
Section~\ref{details} provides an overview of the CLAS12 detector and the datasets employed for this work, followed in
Section~\ref{analysis} with details regarding the analysis cuts and corrections, as well as the yield extraction procedure. A discussion
of the sources of systematic uncertainty is provided in Section~\ref{syserr}. Section~\ref{results} presents the measured beam-recoil
transferred polarizations from the CLAS12 data compared with several model predictions that are available at this time. Finally, a summary
of this work and our conclusions are given in Section~\ref{summary}.


\section{Formalism}
\label{formalism}

\begin{figure}[htbp]
\centering
\includegraphics[width=1.0\columnwidth]{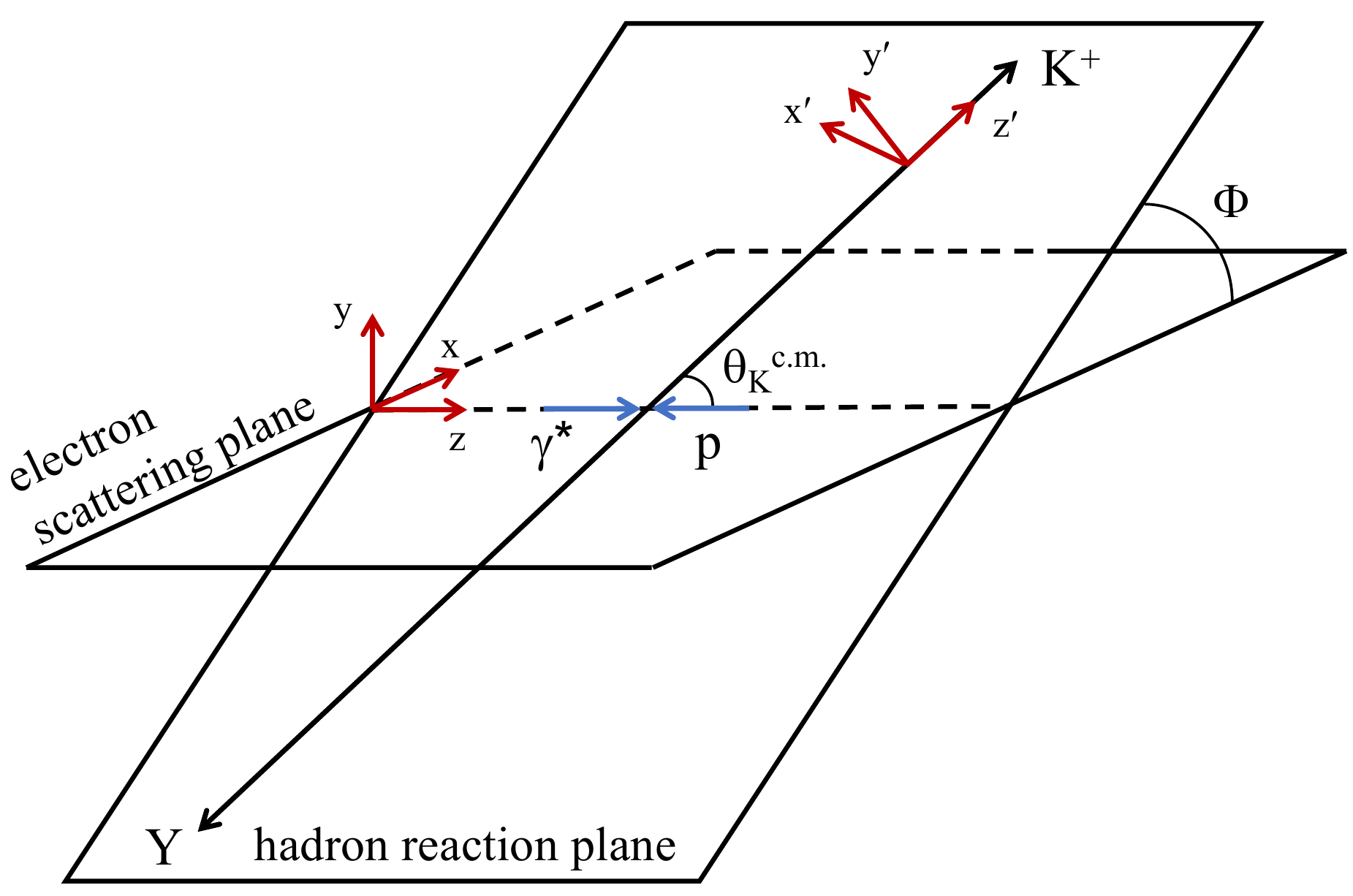}
\vspace{-4mm}
\caption{Kinematics for $K^+Y$ electroproduction defining the c.m.~angles and coordinate systems used to express the formalism and to
  present the polarization components extracted in the analysis.}
\label{coor4}
\end{figure}

Following the notation of Ref.~\cite{knochlein}, the most general form for the $K^+Y$ virtual photo-absorption cross section from a
proton target, allowing for a polarized electron beam, target proton, and recoiling hyperon, is given by:
\begin{widetext}
\begin{eqnarray}
\label{csec1}
\frac{d\sigma_v}{d\Omega_K^{c.m.}} = K_f \sum_{\alpha,\beta} P_\alpha P_\beta \Bigl[ R_T^{\beta\alpha} &+&
  \epsilon R_L^{\beta\alpha} \sqrt{\epsilon (1+\epsilon)} ( ^c\!R_{LT}^{\beta\alpha} \cos{\Phi} +
  \!^s\!R_{LT}^{\beta\alpha} \sin{\Phi}) + \epsilon(^c\!R_{TT}^{\beta\alpha} \cos{2\Phi} +
  \!^s\!R_{TT}^{\beta\alpha} \sin{2\Phi}) \nonumber \\
  &+& h\sqrt{\epsilon (1-\epsilon)} (^c\!R_{LT'}^{\beta\alpha} \cos{\Phi} + \!^s\!R_{LT'}^{\beta\alpha}
  \sin{\Phi}) + h \sqrt{1-\epsilon^2} R_{TT'}^{\beta\alpha} \Bigr].
\end{eqnarray}
\end{widetext}
\noindent
In this expression, the terms $R^{\beta \alpha}$ represent the response functions that account for the full complexity of the reaction 
dynamics expressed in terms of bilinear combinations of the hadronic current. The components of the hadronic current are related to
the reaction amplitudes. The superscripts $\alpha$ and $\beta$ refer to coordinate systems in which the target and hyperon polarizations
are expressed, respectively. The leading $c$ and $s$ superscripts on the response functions indicate whether they multiply a cosine or
sine dependence of the term on the angle $\Phi$ between the electron scattering and hadron reaction planes (see Fig.~\ref{coor4}). Here 
$h$ is the helicity of the beam electron and $K_f$ is a kinematic factor given by the ratio of the c.m. momenta of the outgoing kaon and 
the virtual photon, and $\epsilon$ is the virtual photon transverse polarization parameter:
\begin{equation}
\epsilon=\left(1+2\frac{\nu^2}{Q^2}\tan^2{\frac{\theta_{e'}}{2}} \right)^{-1}.
\end{equation}
\noindent
Here $\nu=E_e - E_{e'}$ is the energy transfer to the target proton and $\theta_{e'}$ is the electron scattering angle in the laboratory
frame.

It is important to point out that the coefficients of the response function terms can be expressed differently in the formalism
presented in different sources. Some authors use a pre-factor for the $\sigma_L$ ($\sigma_{LT}$) term of $\epsilon_L$
($\sqrt{2\epsilon_L(\epsilon+1)}$) instead, where $\epsilon_L$ parameterizes the longitudinal polarization of the virtual photon. Some
also take a $\sin\theta_K^{c.m.}$ ($\sin^2\theta_K^{c.m.}$) term out of the definition of $\sigma_{LT}$ ($\sigma_{TT}$). Eq.(\ref{csec1}) 
avoids the use of $\epsilon_L$ and includes the $\theta_K^{c.m.}$-dependent terms within the response functions themselves.

In Eq.(\ref{csec1}) the target polarization is expressed in the coordinate system $(x,y,z)$ with the $z$-axis along the virtual
photon direction and the $y$-axis normal to the electron scattering plane. The hyperon polarization is expressed in the coordinate
system $(x',y',z')$ with the $z'$-axis along the outgoing $K^+$ direction and the $y'$ axis normal to the hadron production plane
(see Fig.~\ref{coor4}).

The terms $P_\alpha$ and $P_\beta$ in Eq.(\ref{csec1}) are polarization projection operators and are written as $P_\alpha=(1,\vec{P})$ and
$P_\beta=(1,\vec{P}')$. The zero components $P_0$ give rise to cross section contributions present in the polarized as well as the
unpolarized case. In an experiment without beam (target) polarization $\alpha$ ($\beta$) = 0. 

In an experiment in which the beam, target, and recoil particles are unpolarized, Eq.(\ref{csec1}) can be written as:
\begin{multline}
\label{csec2}
\sigma_0 \equiv \left( \frac{d\sigma_v}{d\Omega_K^{c.m.}} \right)^{\!\!00} = K_f \Bigl[ R_T^{00} +   \epsilon R_L^{00} + \\
  \sqrt{\epsilon (1+\epsilon)} R_{LT}^{00} \cos{\Phi} + \epsilon R_{TT}^{00} \cos{2\Phi} \Bigr].
\end{multline}
Of direct interest for this work is the extraction of the hyperon polarization. Each of the hyperon polarization components, $P_{x'}$, 
$P_{y'}$, $P_{z'}$, can be split into a beam helicity independent part, called the {\em induced} polarization, and a beam helicity dependent 
part, called the {\em transferred} polarization. The three beam-recoil transferred polarization components are written in the $(x',y',z')$ 
system as:
\begin{eqnarray}
\label{ptran1}
P_{x'}' &=& \frac{K_f}{\sigma_0} \left( \sqrt{\epsilon (1-\epsilon)} R_{LT'}^{x'0} \cos{\Phi} +
\sqrt{1-\epsilon^2} R_{TT'}^{x'0} \right) \nonumber \\
P_{y'}' &=& \frac{K_f}{\sigma_0} \sqrt{\epsilon (1-\epsilon)} R_{LT'}^{y'0} \sin{\Phi} \nonumber \\
P_{z'}' &=& \frac{K_f}{\sigma_0} \left( \sqrt{\epsilon (1-\epsilon)} R_{LT'}^{z'0} \cos{\Phi} +
\sqrt{1-\epsilon^2} R_{TT'}^{z'0} \right).
\end{eqnarray}
To accommodate finite bin sizes in the relevant kinematic variables $Q^2$, $W$, and the polar angle of the $K^+$ in the c.m.~(actually 
$\cos \theta_K^{c.m.}$ is employed here) and to improve statistics, this analysis presents the transferred polarization components summed
over all angles $\Phi$. These $\Phi$-integrated polarization transfer components in the $(x',y',z')$ system are given by:
\begin{eqnarray}
\label{xpypzp}
{\cal P}_{x'}' &=& K_I \sqrt{1 - \epsilon^2} R_{TT'}^{x'0} \nonumber \\
{\cal P}_{y'}' &=& 0 \nonumber \\
{\cal P}_{z'}' &=& K_I \sqrt{1 - \epsilon^2} R_{TT'}^{z'0},
\end{eqnarray}
\noindent
where $K_I = 1/(R_T^{00} + \epsilon R_L^{00})$. Note that the $\Phi$-integrated transferred polarization components are now written using
the notation ${\cal P}'$.

The transferred polarization components can also be expressed in the $(x,y,z)$ system. To express these terms, the components defined for 
the $(x',y',z')$ system in Eq.(\ref{ptran1}) must undergo a transformation that performs a rotation of $\theta_K^{c.m.}$ about $\hat{y}'$ 
followed by a rotation of $\Phi$ about $\hat{z}'$. With this transformation the $(x,y,z)$ polarization components integrated over $\Phi$ 
can be expressed as:
\begin{eqnarray}
\label{xyz}
{\cal P}_{x}' &=&  \sqrt{\epsilon (1-\epsilon)} \frac{K_I}{2}\!\left(\!R_{LT'}^{x'0}\!\cos{\theta_K^{c.m.}}\!-\!
R_{LT'}^{y'0}\!+\!R_{LT'}^{z'0}\!\sin{\theta_K^{c.m.}}\!\right) \nonumber \\
{\cal P}_{y}' &=& 0 \nonumber \\
{\cal P}_{z}' &=& \sqrt{1-\epsilon^2} K_I\!\left( -R_{TT'}^{x'0} \sin{\theta_K^{c.m.}} + R_{TT'}^{z' 0} \cos{\theta_K^{c.m.}} \right).
\end{eqnarray}
\noindent
As in the primed system, the $y$ component of the polarization transfer in the unprimed system ${\cal{P}}'_y$ is constrained to be zero.

The transferred polarization components are presented in both the primed and unprimed systems shown in Fig.~\ref{coor4}. In 
the primed system, the $\Phi$-integrated transferred polarization components are sensitive to the response functions $R_{TT'}^{x'0}$ and 
$R_{TT'}^{z'0}$. However, in the unprimed system the components are also sensitive to $R_{LT'}^{x'0}$, $R_{LT'}^{y'0}$, and $R_{LT'}^{z'0}$. 
Note that $R_{LT'}^{y'0}$ is equivalent to $-R_{LT'}^{0y}$~\cite{knochlein}, which is accessible in an experiment with an unpolarized beam 
and polarized target. The structure functions $R_T^{00}$ and $R_L^{00}$ available from the measurements with unpolarized beam and target are 
required for the computation of the term $K_I$ in Eq.(\ref{xpypzp}).


\section{Hyperon Polarization Extraction Approach}
\label{approach}

\subsection{Decay Angular Distributions}

The $\Lambda$ hyperon decays weakly into a pion and a nucleon with a branching ratio of 64\% into $p\pi^-$ and 36\% into $n\pi^0$. In
these decays, the nucleon has an asymmetric angular distribution with respect to its spin direction. This asymmetry is the result of an
interference between parity non-conserving ($s$-wave) and parity-conserving ($p$-wave) amplitudes in the weak decay. In the hyperon 
rest frame, the angular distribution of the $\Lambda$ decay nucleon for each spin quantization axis can be written as~\cite{bonner}:
\begin{equation}
\label{ang-dist}
\frac{dN}{d \cos \theta_N^{RF}} = N_0 \left( 1 + \alpha P_\Lambda \cos \theta_N^{RF} \right),
\end{equation}
\noindent
where $N_0$ is the yield integral, $P_\Lambda$ is the $\Lambda$ polarization component, and $\theta_N^{RF}$ is the angle between the 
polarization vector and the decay-nucleon momentum in the $\Lambda$ rest frame. In this work we focus solely on the $\Lambda \to p \pi^-$ 
decay branch and explicitly replace $\theta_N^{RF}$ with $\theta_p^{RF}$. The $\Lambda$ weak decay asymmetry parameter $\alpha$ is given
in the PDG as 0.732$\pm$0.014~\cite{pdg}, and is based on the average determination from measurements of BESIII~\cite{bes-alpha} and 
CLAS~\cite{ireland}. 

The hyperon polarization in Eq.(\ref{ang-dist}) is the sum of the induced and transferred polarization:
\begin{equation}
\label{pol1}
{\vec P}_\Lambda = {\vec P}_\Lambda^0 \pm h {\vec P}_\Lambda'.
\end{equation}
\noindent
However, as the electron beam was not 100\% polarized, the helicity term $h$ in the hyperon polarization must be replaced by the
longitudinal electron beam polarization $P_b$. Combining Eq.(\ref{ang-dist}) and Eq.(\ref{pol1}), the $\Phi$-integrated decay proton
angular distribution to determine the transferred polarization is given by:
\begin{equation}
\label{ang_distl}
\frac{dN}{d \cos \theta_p^{RF}} = N_0 [1 + \alpha P_b {\cal P}'_\Lambda \cos \theta_p^{RF}].
\end{equation}
The $\Sigma^0$ decays into $\gamma \Lambda$ (branching ratio 100\%). A $\Sigma^0$ with polarization $P_\Sigma$ will yield a decay $\Lambda$ 
that retains some of the polarization of its parent. As shown in Ref.~\cite{gatto}, we can expect that on average for the decay $\Lambda$
in its rest frame, $P_\Lambda = -\frac{1}{3} P_\Sigma$. For the case of a final state $\Sigma^0$, the $\Lambda$ rest frame can be calculated
only if in addition to the detection of the electron, kaon, and decay proton, either the decay pion of the $\Lambda$ or the decay $\gamma$
from the $\Sigma^0$ is detected. Due to the small acceptance of CLAS12 for such a final state this is not practical. In
Ref.~\cite{bradford07} it was shown that the polarization of the daughter $\Lambda$ from the $\Sigma^0$ decay can be measured without
boosting the detected proton to the reference frame of the $\Lambda$. This gives rise to a dilution factor of the weak decay asymmetry
parameter for the $\Sigma^0$ that is reduced from $-0.333 \alpha$ to $-0.256 \alpha$.

One method to access the hyperon polarization components is by forming the beam spin asymmetry of the decay proton angular distribution.
Writing this to be generally applicable to extract the transferred polarization for either the $\Lambda$ or the $\Sigma^0$ hyperon gives:
\begin{equation}
\label{asm-pol}
A = \frac{N^+ - N^-}{N^+ + N^-} = \nu_Y \alpha P_b {\cal{P}}'_Y \cos \theta_p^{RF},
\end{equation}
\noindent
where $\nu_Y = 1.0$ for the $\Lambda$ measurement and $\nu_Y = -0.256$ for the $\Sigma^0$ measurement. From Eq.(\ref{asm-pol}) it is apparent 
that the slope of the measured asymmetry of the decay proton as a function of $\cos \theta_p^{RF}$ is directly proportional to the 
$\Phi$-integrated hyperon transferred polarization for a given coordinate system axis choice.

\begin{figure}[htbp]
\centering
\includegraphics[width=0.85\columnwidth,height=7cm]{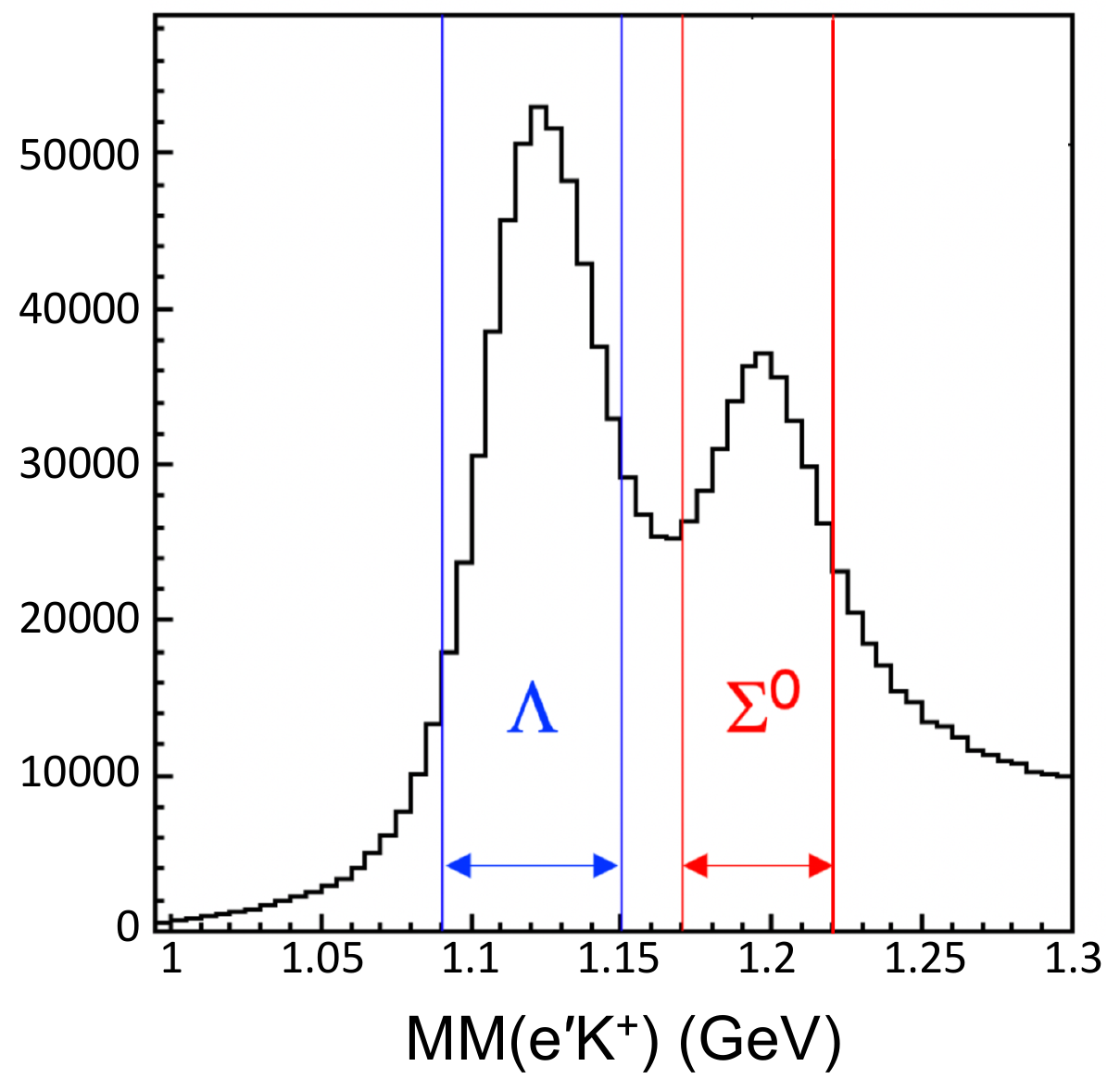}
\vspace{-2mm}
\caption{The $e'K^+$ missing mass distribution after all particle identification and exclusivity cuts described in Section~\ref{analysis}
  for the 6.535~GeV dataset summed over all kinematics. The vertical lines about the $\Lambda$ and $\Sigma^0$ hyperon peaks identify the
  analysis ranges used to select the event samples.}
\label{mm-regions}
\end{figure}

Practically, the hyperon transferred polarization is extracted by analyzing data binned in the relevant kinematic variables $Q^2$, $W$,
and $\cos \theta_K^{c.m.}$. For the $\Lambda$ and $\Sigma^0$ analyses the reactions are selected in the $e'K^+$ missing mass distributions in
mass regions about the individual hyperon peaks. As shown in Fig.~\ref{mm-regions}, the nominal $\Lambda$ mass region was chosen in the
range from 1.09-1.15~GeV and the nominal $\Sigma^0$ mass region was chosen in the range from 1.17-1.22~GeV. The exact choices are somewhat
arbitrary but were selected to maximize the event yields for the hyperons of interest, while minimizing the contamination of the
contributing backgrounds. See Section~\ref{syserr} for details on the systematic uncertainty regarding the hyperon mass regions chosen.

As shown in Fig.~\ref{mm-regions} the hyperon signals in each mass region are not pure. Underlying both the $\Lambda$ and $\Sigma^0$ peaks
is a background arising from the multi-pion events dominated by the exclusive reaction channel $ep \to e'\pi^+ \pi^- p$, where the $\pi^+$
is misidentified by CLAS12 as a $K^+$ due to the finite timing resolution of the CLAS12 time-of-flight measurements. Additionally, in the 
$\Lambda$ mass region the tail of the resolution-smeared $\Sigma^0$ peak contaminates the $\Lambda$ events. Within the $\Sigma^0$ mass
region, there is a more sizable contamination from $\Lambda$ radiative tail events. The cross contamination of the hyperons into the
neighboring mass regions must be accounted for as the hyperons typically have sizable polarizations. The yield extraction procedure is
described in Section~\ref{spec_fits}.

\boldmath
\subsection{$\Lambda$ Transferred Polarization}
\label{lform}
\unboldmath

The measured raw helicity-gated yield asymmetry, including all sources of background, can be written in a general way as:
\begin{equation}
\label{asymL1}
A_{raw} = \frac{(N_\Lambda^+ + N_\Sigma^+ + N_B^+)\: -\: (N_\Lambda^- + N_\Sigma^- + N_B^-)}{N_\Lambda + N_\Sigma + N_B},
\end{equation}
\noindent
where $N_\Lambda^\pm$, $N_\Sigma^\pm$, and $N_B^\pm$ refer to the $\Lambda$, $\Sigma^0$, and non-hyperon background yields, respectively, for 
the two beam helicity states and $N_\Lambda = N_\Lambda^+ + N_\Lambda^-$, $N_\Sigma = N_\Sigma^+ + N_\Sigma^-$, and $N_B = N_B^+ + N_B^-$ 
are the total yields for each of the three different contributions. These yields for the $\Lambda$ polarization analysis were determined 
within a mass window around the $\Lambda$ peak in the $MM(e'K^+)$ distribution as shown in Fig.~\ref{mm-regions}, binning in the appropriate 
kinematic variables ($Q^2$, $W$, $\cos \theta_K^{c.m.}$) of interest.

Rewriting the raw asymmetry in Eq.(\ref{asymL1}) we have:
\begin{equation}
\label{asymL2}
A_{raw} = \frac{A_\Lambda + A_\Sigma \cdot F_\Sigma + A_B \cdot F_B}{1 + F_\Sigma + F_B},
\end{equation}
\noindent
where the asymmetries for the individual contributions within the $\Lambda$ mass window are $A_\Lambda = (N_\Lambda^+ - N_\Lambda^-)/N_\Lambda$, 
$A_\Sigma = (N_\Sigma^+ - N_\Sigma^-)/N_\Sigma$, and $A_B = (N_B^+ - N_B^-)/N_B$. We have also adopted the notation 
$F_\Sigma = N_\Sigma/N_\Lambda$ and $F_B = N_B/N_\Lambda$ to represent the ratio of the $\Sigma^0$ contamination relative to the $\Lambda$ yield
and the ratio of the multi-pion background yield relative to the $\Lambda$ yield in the $\Lambda$ mass window, respectively. In this
analysis the form of Eq.(\ref{asymL2}) further simplifies given that the asymmetry $A_B$ associated with the underlying multi-pion
background contribution is consistent with zero (see Section~\ref{bck_pol} for details).

The link between the hyperon helicity asymmetries and the hyperon polarization is given in Eq.(\ref{asm-pol}). We can also generically
write for the measured raw helicity asymmetry without any background subtraction:
\begin{equation}
\label{asymL5}
A_{raw} = \alpha P_b [{\cal P}_{raw}'] \cos \theta_p^{RF}.
\end{equation}
\noindent
Expanding the asymmetry of Eq.(\ref{asymL2}) using the asymmetry contributions from Eq.(\ref{asm-pol}), we can write:
\begin{eqnarray}
\label{asymL6}
A_{raw} &=& \frac{\alpha P_b {\cal P}_\Lambda' \cos \theta_p^{RF} + \nu_\Sigma \alpha P_b {\cal P}_\Sigma' \cos \theta_p^{RF}
    \cdot F_\Sigma}{1 + F_\Sigma + F_B} \nonumber \\ [2ex]
\label{asymL7}
  &=& \alpha P_b \left[ \frac{{\cal P}_\Lambda' + \nu_\Sigma {\cal P}_\Sigma' F_\Sigma} {1 + F_\Sigma + F_B} \right] \cos \theta_p^{RF}.
\end{eqnarray}
\noindent
Comparing the form of Eq.(\ref{asymL7}) to Eq.(\ref{asymL5}) we can define the raw polarization for all events in the $\Lambda$ mass
window without any background subtraction as:
\begin{equation}
\label{pmeasL1}
  {\cal P}_{raw}' = \frac{{\cal P}_\Lambda' + \nu_\Sigma {\cal P}_\Sigma' F_\Sigma}{1 + F_\Sigma + F_B}.
\end{equation}
\noindent
Rearranging the terms in Eq.(\ref{pmeasL1}), we can solve for ${\cal P}_\Lambda'$:
\begin{equation}
  \label{pmeasL2}
  {\cal P}_\Lambda' = {\cal P}_{raw}' \left( 1 + F_\Sigma + F_B \right) - \nu_\Sigma {\cal P}_\Sigma' F_\Sigma.
\end{equation}
\noindent
In this expression the $\Lambda$ transferred polarization is determined from the measured raw polarization accounting for the polarization 
contamination from the $\Sigma^0$ tail beneath the $\Lambda$ peak. Note also that even though the asymmetry of the multi-pion background 
contribution is zero, this background still contributes to a dilution of the polarization of the $\Lambda$ events.

Based on Eq.(\ref{pmeasL2}), the statistical uncertainty of ${\cal P}_\Lambda'$ (neglecting the small correlation terms) is given by:
\begin{multline}
  \delta {\cal P}_\Lambda' = \Bigl( (1 + F_\Sigma + F_B)^2 (\delta {\cal P}_{raw}')^2 + \\
  \left( {\cal P}_{raw}' - \nu_\Sigma {\cal P}_{\Sigma}' \right)^2 (\delta F_\Sigma)^2 + ({\cal P}_{raw}')^2(\delta F_B)^2 +\\
  \left( \nu_\Sigma F_\Sigma \right)^2 (\delta {\cal P}_\Sigma')^2 \Bigr)^{1/2},
\end{multline}
\noindent
where $\delta {\cal P}_{raw}'$, $\delta F_\Sigma$, $\delta F_B$, and $\delta {\cal P}_\Sigma'$ represent the statistical uncertainties in 
the measured raw polarization, the $\Sigma^0$ to $\Lambda$ yield ratio, the multi-pion background to $\Lambda$ yield ratio, and the
measured $\Sigma^0$ polarization, respectively.

The measured $\Lambda$ polarization needs the measured $\Sigma^0$ polarization as an input. Using an iterative process,
the measured $\Lambda$ polarization, which only has a small contamination from the $\Sigma^0$, is used to determine the $\Sigma^0$
polarization (see Section~\ref{sform}). This $\Sigma^0$ polarization is then used to recompute the $\Lambda$ polarization. After several
iterations through the computation, the calculation converges for the computation of the polarization of both hyperons.

\boldmath
\subsection{$\Sigma^0$ Transferred Polarization}
\label{sform}
\unboldmath

The approach to measure the $\Sigma^0$ polarization using events within the $\Sigma^0$ mass window follows in the same way as outlined
for the $\Lambda$ polarization measurement in Section~\ref{lform}, again accounting for the background contributions beneath the
$\Sigma^0$ mass peak that arise from the radiative tail of the $\Lambda$ events and the multi-pion background contribution. Beginning
with the measured raw asymmetry defined in the $\Sigma^0$ mass window given by $A_{raw}$ in Eq.(\ref{asymL1}), the $\Sigma^0$ polarization
can be expressed as:
\begin{equation}
\label{pmeasS2}
  {\cal P}_\Sigma' = {\cal P}_{raw}' \left( 1 + F_\Lambda + F_B \right) - \frac{1}{\nu_\Sigma} {\cal P}_\Lambda' F_\Lambda,
\end{equation}
\noindent
where $F_\Lambda = N_\Lambda/N_\Sigma$ and $F_B = N_B/N_\Sigma$ are the yield ratios within the $\Sigma^0$ mass window, again using the fact 
that the asymmetry associated with the multi-pion contribution is consistent with $A_B = 0$. The corresponding statistical uncertainty
on ${\cal P}'_\Sigma$ (again, neglecting the small correlation terms) is given by:
\begin{multline}
  \delta {\cal P}_\Sigma' = \Bigl( (1 + F_\Lambda + F_B)^2 (\delta {\cal P}_{raw}')^2 + \\
  \left( {\cal P}_{raw}' - \frac{1}{\nu_\Sigma}{\cal P}_{\Lambda}' \right)^2 (\delta F_\Lambda)^2 + ({\cal P}_{raw}')^2(\delta F_B)^2 + \\
  \left( \frac{1}{\nu_\Sigma}F_\Lambda \right)^2 (\delta {\cal P}_\Lambda')^2 \Bigr)^{1/2}.
\end{multline}


\section{Experimental Details}
\label{details}

The study of both the spectrum and structure of nucleon excited states represents one of the founding experimental physics programs at 
JLab. Beginning in 1997 until it was decommissioned in 2012, the CEBAF Large Acceptance Spectrometer (CLAS)~\cite{mecking} located in
Hall~B was used for studies of inclusive, semi-inclusive, and exclusive reactions from a fixed target with beams of electrons and photons
at energies up to 6~GeV. Measurements with CLAS allowed for the study of exclusive reactions in the range of $Q^2$ up to 5~GeV$^2$ and $W$
up to 3~GeV, spanning nearly the full c.m.~angular range of the final state particles. The CLAS detector has provided the majority of the
available world data on the $\pi N$, $\eta p$, $K^+ \Lambda$, $K^+ \Sigma^0$, and $\pi^+\pi^- p$ electroproduction channels in the nucleon
resonance region. 

The CLAS detector was replaced with the large acceptance CLAS12 spectrometer~\cite{clas12-nim} as part of the JLab 12~GeV upgrade project
in the period from 2012-2017 with beam operations for physics beginning in 2018. The approved CLAS12 measurement program includes several 
experiments as part of the continuing effort to study the spectrum and structure of $N^*$ states with electron beams of energy up to
11~GeV. The data will span an unprecedented kinematic range of $Q^2$ from 0.05 to 12~GeV$^2$ in the nucleon resonance region, covering the
full c.m. angular range for the final state particles.

\begin{figure}[htbp]
\centering
\includegraphics[width=1.0\columnwidth]{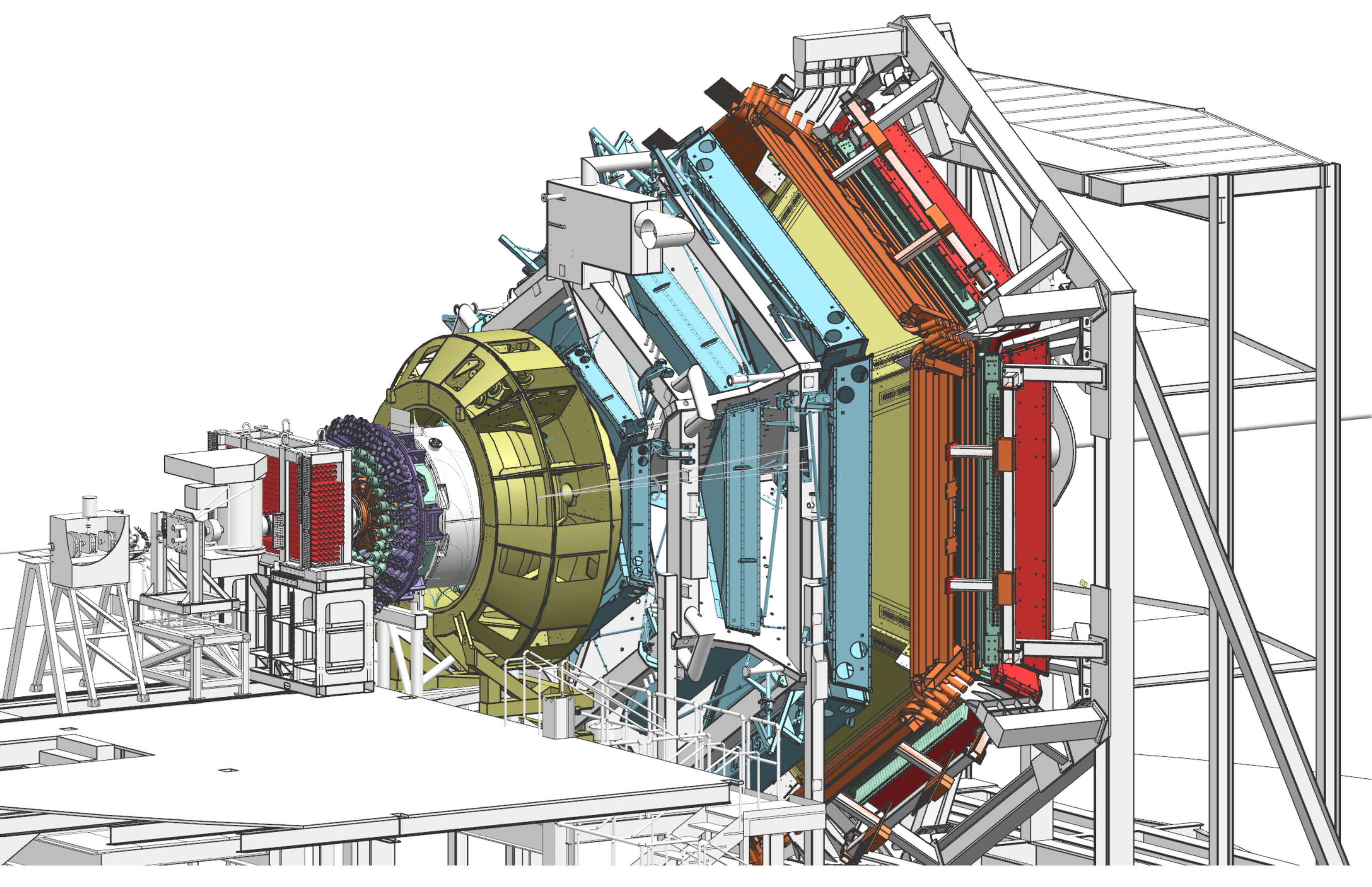}
\vspace{-8mm}
\caption{Model of the CLAS12 spectrometer in Hall~B at Jefferson Laboratory. The electron beam is incident from the left side of this
  figure. The CLAS12 detector is roughly 20~m in scale along the beam axis.}
\label{clas12-model}
\end{figure}

The CLAS12 spectrometer is comprised of a Forward Detector system built around a 6 coil superconducting torus magnet that divides the 
azimuthal acceptance into six 60$^\circ$-wide sectors and a Central Detector built around a superconducting solenoid magnet. 
Figure~\ref{clas12-model} shows a model representation of CLAS12. The Forward Detector covers polar angles from 5$^\circ$ to 35$^\circ$ and 
the Central Detector covers polar angles from 35$^\circ$ to 125$^\circ$. CLAS12 has been optimized for the reconstruction of exclusive 
reactions. In the forward direction, CLAS12 consists of 3 sets of multi-layer drift chambers~\cite{dc-nim} for charged particle tracking
that are placed before, within, and after the torus field. Downstream of the chambers, CLAS12 consists of multiple layers of a large-area 
scintillator hodoscope for precise timing measurements for charged particles~\cite{ftof-nim} and a sampling electromagnetic calorimeter
for electron and neutral identification~\cite{ecal-nim}. The Forward Detector also consists of different types of Cherenkov detectors. Of
relevance in this work is a CO$_2$-filled high threshold Cherenkov detector that spans the full azimuthal range, which is used as
part of the trigger selection for electrons~\cite{htcc-nim}. The Central Detector consists of a multi-layer vertex tracker
\cite{svt-nim,cvt-nim} surrounded by a barrel of scintillation counters for charged particle identification~\cite{ctof-nim} via precision
flight time measurements. Each of the active elements of these detectors resides within the 5-T solenoid field. This field is used both for 
momentum analysis of charged tracks in the Central Detector volume and as the confining field for the intense M{\o}ller background produced 
as the electron beam passes through the target. This low-energy radiation is directed along the beamline into a tungsten absorber to shield 
the CLAS12 detectors.

The data contained in this work was collected as part of the Run Group K (RG-K) set of experiments that took data in December 2018 as part
of a short 3-week test run. The experiment collected data with a longitudinally polarized electron beam on a 5-cm-long liquid-hydrogen target. 
Data have been acquired at beam energies of 6.535~GeV and 7.546~GeV. The 6.535~GeV (7.546~GeV) dataset was collected at an average beam-target 
luminosity of 1$\times$10$^{35}$~cm$^{-2}$s$^{-1}$ (5$\times$10$^{34}$~cm$^{-2}$s$^{-1}$) and amounted to 18.2~mC (10.7~mC) of accumulated 
electron charge. The torus magnet was set to its maximum field strength to optimize the reconstructed momentum resolution for charged particles 
and its polarity was set to bend negatively charged particles outward, away from the beamline. The electron beam polarization was measured 
periodically during the data run using the Hall~B M{\o}ller polarimeter~\cite{beamline-nim} and its value was found to be 86\% on average. 
The polarization of the beam was flipped at a rate of 30~Hz. To minimize any systematic effects associated with the helicity signal in Hall~B,
the signal itself was received by the CLAS12 data acquisition system in patterns delayed by 8 helicity windows, with the helicity of the
first window of each pattern determined by a pseudo-random generator in the JLab accelerator controls. The beam helicity charge asymmetry
was monitored throughout the run period and was at the level of $\pm$0.1\%.

For this experiment the event readout was triggered by a coincidence between a track candidate in the drift chamber, a signal in the
electron-sensitive Cherenkov detector, and a cluster in the forward electromagnetic calorimeter with a cut on the minimum number of 
photoelectrons in the Cherenkov detector. The sophisticated trigger system~\cite{trigger-nim} required a reconstructed charged track
candidate consistent with a negatively charged particle in the drift chambers that matched the calorimeter hit cluster, as well as a
matched hit in the forward timing hodoscope. These trigger requirements were designed to reduce the backgrounds and improve the trigger
purity. The CLAS12 data acquisition system (DAQ)~\cite{daq-nim} recorded data at rates up to 20~kHz based on multiple CLAS12 trigger
streams with a live-time greater than 90\%.


\section{Data Analysis}
\label{analysis}

\begin{figure}[htbp]
\centering
\includegraphics[width=1.0\columnwidth]{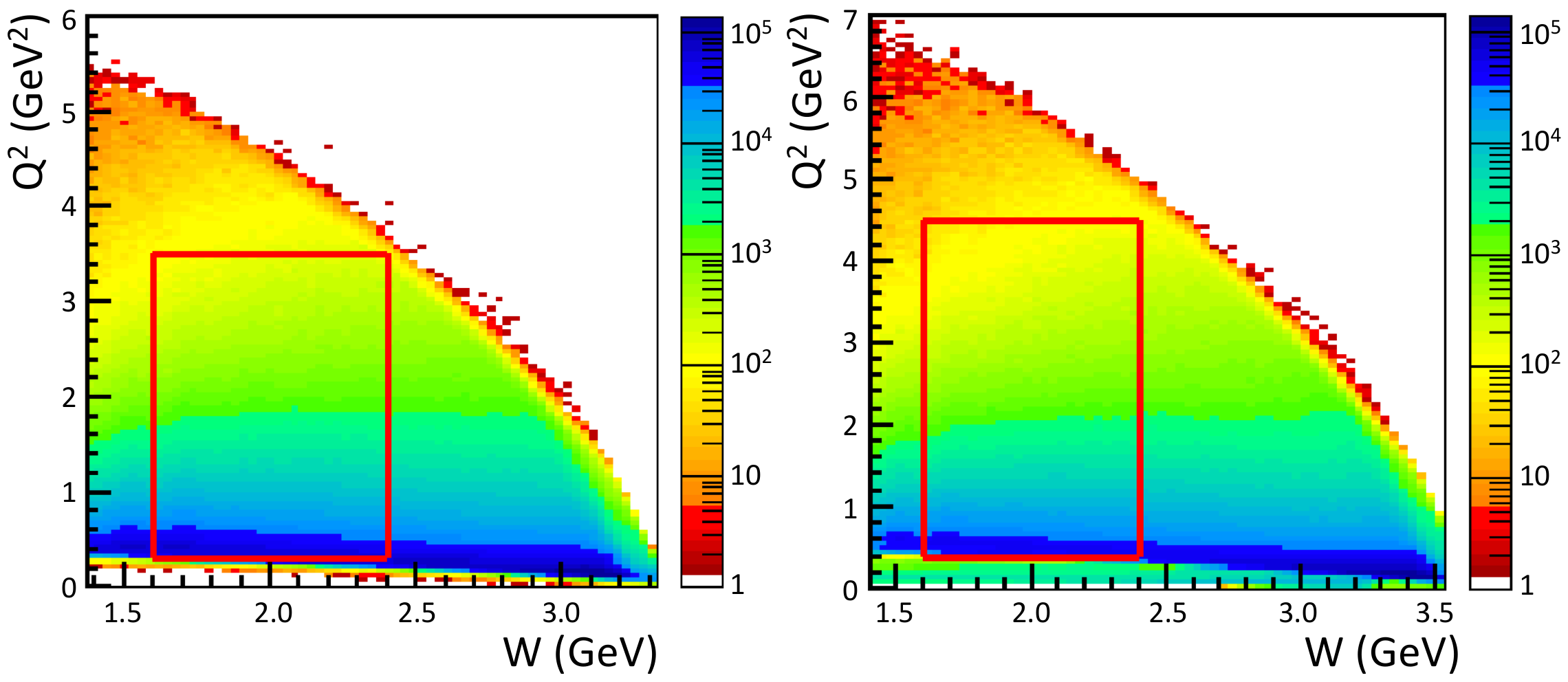}
\vspace{-6mm}
\caption{Kinematic coverage of the electron from the 6.535~GeV and 7.546~GeV datasets in terms of $Q^2$ vs.~$W$ (units GeV$^2$/GeV).
  The overlaid rectangular boxes highlight the analysis region in this work.}
\label{q2w}
\end{figure}

\begin{figure}[htbp]
\centering
\includegraphics[width=1.0\columnwidth]{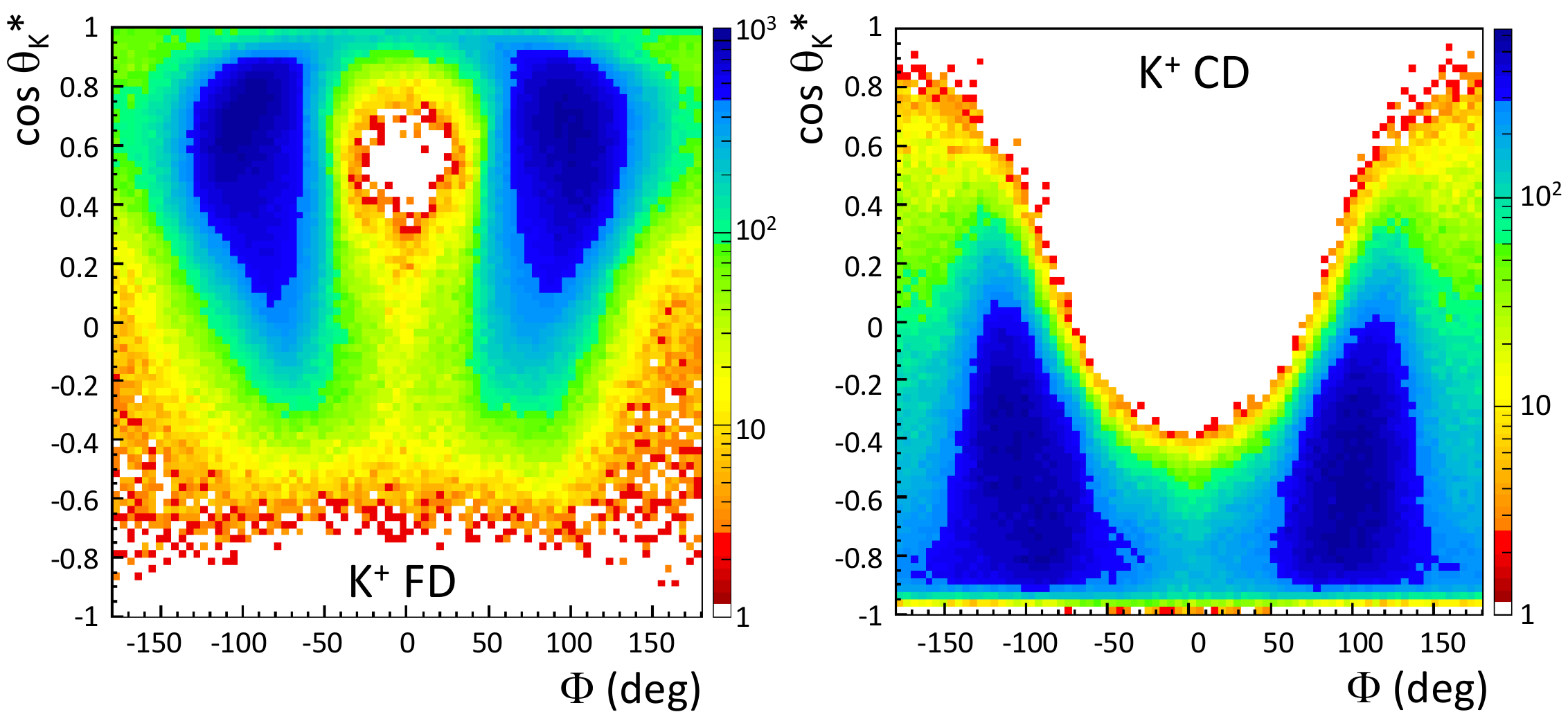}
\vspace{-6mm}
\caption{Kinematic coverage at 6.535~GeV of the electroproduced $K^+$ in terms of $\cos \theta_K^{c.m.}$ vs.~$\Phi$ (deg), where $\Phi$ is
  the angle between the lepton scattering plane and the hadronic reaction plane. The left plot is for the $K^+$ detected in the Forward 
  Detector (FD) and the right plot is for the $K^+$ detected in the Central Detector (CD).}
\label{thphi}
\end{figure}

Hyperon identification relies on missing-mass reconstruction of the reaction $ep\!\to\!e'K^+X$. In addition, for the polarization 
measurement, the reconstruction of the proton from the hyperon decay is required. The acceptance for this three-body $e'K^+p$ final state 
is on the order of 5\% to 20\% depending on $Q^2$, $W$, $\cos \theta_K^{c.m.}$, and $\cos \theta_p^{RF}$. The analysis results shown here 
span $Q^2$ from 0.3 to 4.5~GeV$^2$ and $W$ within the nucleon resonance region from 1.6 to 2.4~GeV. Figure~\ref{q2w} shows the electron 
acceptance of the datasets in terms of $Q^2$ vs.~$W$. Figure~\ref{thphi} shows the kinematic phase space for the electroproduced $K^+$ 
from the 6.535~GeV data, which is separated into the coverage for the Forward Detector and Central Detector of CLAS12.

In the kinematic region of interest, the 6.535~GeV (7.546~GeV) dataset contains 636k (260k) $K^+\Lambda$ events and 323k (122k) 
$K^+\Sigma^0$ events in the $e'K^+p$ topology. This data sample is roughly 5 times larger than that for the polarization analyses of
the available CLAS electroproduction datasets~\cite{carman03,carman09}. The data presented in this work represents only 10\% of the full
dataset ultimately planned for collection as part of this experiment over the next several years. In this section, details are provided
on our procedures for particle identification, on the cuts used to isolate the $K^+\Lambda$ and $K^+\Sigma^0$ final states, on the
hyperon spectrum fitting procedure, and on other cuts and corrections that are part of the data analysis.

\subsection{Particle Identification}

Event reconstruction began by selecting events with a viable electron candidate in the CLAS12 Forward Detector. The initial identification 
of electrons was performed by the CLAS12 Event Builder~\cite{recon-nim}. This required a negatively charged particle -- identified by its 
track curvature in the torus magnetic field -- that was matched with hits in the high-threshold Cherenkov detector, forward time-of-flight 
system, and calorimeter. The detected deposited energy in the sampling-type calorimeter was required to be consistent with the parameterized 
sampling fraction distribution vs.~deposited energy. This definition was already sufficient to remove the dominant pion contamination, 
however, the analysis applied further cuts to further purify the electron sample. Cuts were placed on the electron momentum as reconstructed 
in the drift chamber system, the particle flight time from the event vertex to the forward time-of-flight system, and the reconstructed event 
vertex distribution to be sure the track originated from the hydrogen target cell (the trace-back resolution at the target location is about 
1~cm). Finally, a shower profile cut was applied to further reduce the pion contamination as the CO$_2$ radiator of the Cherenkov detector 
gives signals for pions starting at around 4.5~GeV.

\begin{figure}[htbp]
\centering
\includegraphics[width=1.0\columnwidth]{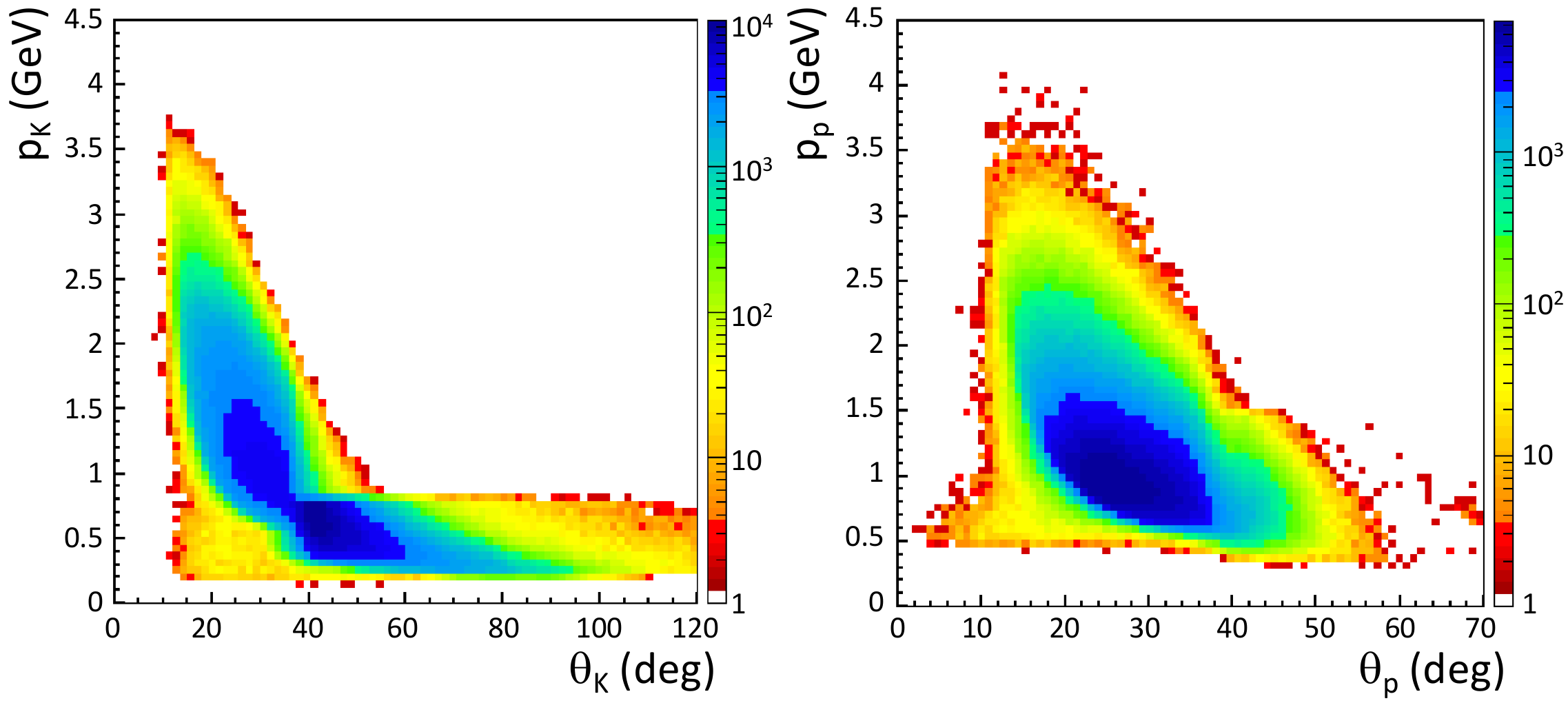}
\vspace{-6mm}
\caption{The kinematic phase space in terms of momentum vs.~lab polar angle $\theta$ of the reconstructed $K^+$ (left) and $p$ (right) in 
  CLAS12 for the 6.535~GeV dataset, combining events reconstructed in the Forward Detector and the Central Detector. The acceptance gap 
  between the two CLAS12 detector systems occurs at about 35$^\circ$.}
\label{pthh}
\end{figure}

After a viable electron candidate was identified in a given event, the hadron identification process searched within the selected event 
sample for events with one (and only one) reconstructed $K^+$ and $p$ candidate in CLAS12. The Event Builder algorithm for charged
hadrons compared the measured flight time for each track from the event vertex to the time-of-flight system, to the computed time for a
given hadron species, starting from its measured momentum and the assumed mass. The hypothesis that minimized the time difference was
assigned as the particle type. Additional cuts were applied to improve the hadron identification purity on the minimum particle momentum 
(0.4~GeV in the Forward Detector and 0.2~GeV in the Central Detector), the particle flight time to the time-of-flight systems, and the 
reconstructed $\beta = v/c$ (0.4-1.1 in the Forward Detector and 0.2-1.1 in the Central Detector) for the track. Figure~\ref{pthh} shows
the kinematic phase space for the reconstructed $K^+$ and $p$ from the 6.535~GeV dataset in terms of momentum vs.~laboratory polar angle. 
The $K^+$ sample also included a cut on the reconstructed event vertex to ensure the track originated from the hydrogen target cell.

\subsection{Additional Cuts and Corrections}
\label{add-cuts}

It is important to optimize the accuracy of the momentum reconstruction of the final state $e'$ and $K^+$ to maximize the hyperon signal to
background ratio in the $MM(e'K^+)$ spectra and to enable optimal separation of the $K^+\Lambda$ and $K^+\Sigma^0$ final states. It is also 
important to optimize the accuracy of momentum reconstruction of the final state particles in order to minimize the systematic uncertainties 
of the measured proton angular distribution used to determine the hyperon polarization.

The measured charged particle momenta in CLAS12 have inaccuracies due to unaccounted for geometrical misalignments of the tracking
detectors, calibration systematic biases, charged particle energy loss in the passive detector materials, and inaccuracies in the
magnetic field maps for the torus and solenoid used in the charged particle tracking. However, the systematics of the measured momenta
from CLAS12 were minimized using momentum corrections for the different final state particles based on exclusive event reconstruction 
kinematic constraints.

In each of the six sectors of the CLAS12 Forward Detector, the reconstructed electron momentum was scaled in order to properly position
the elastic proton peak in the invariant mass $W$ spectrum. These corrections were all below 0.5\%. In these CLAS12 kinematics, the missing
mass resolution is dominated by the reconstructed electron as it has the largest momentum.

The momenta of the $K^+$ and $p$ were corrected for energy loss in the CLAS12 detector passive material layers between the reaction vertex
and the time-of-flight systems. This correction was based on Monte Carlo event reconstruction relying on the accurate accounting of the
materials in the simulation. These corrections were less than 15-20~MeV over the full momentum range of the data. Then, in each of the 
six sectors of the Forward Detector and each of the three sectors of the tracker in the Central Detector, the $K^+$ momentum was scaled to 
position the $\Lambda$ peak in the $MM(e'K^+)$ spectrum at the correct mass. In the Forward Detector the corrections were less than 0.5\%
and in the Central Detector average $\sim$4\%. Similarly, the proton momentum was corrected in the different CLAS12 sectors selecting
$\Lambda$ events and scaling the proton momenta to position the $\pi^-$ peak in the $MM(e'K^+p)$ spectrum at its correct mass. The
corrections are $\sim$2\% and $\sim$7\% in the Forward and Central Detectors, respectively.

The accuracy of the momentum reconstruction was such that the residual distortions of the $MM(e'K^+)$ and $MM(e'K^+p)$ spectra were at a
level below $\pm$5~MeV over the full kinematic phase space of the data. The remaining residual distortions of the reconstructed momenta
were shown to have a minimal effect on the assigned systematic uncertainties of the extracted hyperon polarizations.

The reconstructed momentum of charged particles in the CLAS12 Forward Detector suffers from systematic inaccuracies at the boundaries of 
the azimuthal acceptance in each sector close to the torus coils. To remove these events, geometrical fiducial cuts were employed to
exclude tracks detected in these regions. For the electrons, a selection on the calorimeter fiducial volume was also applied to ensure
containment of the electromagnetic shower, such that the sampling fraction cuts allow for high purity of the electron candidate sample.

In the extraction of the hyperon polarization components no radiative corrections were applied to the data. The need for such corrections
is minimized by employing relatively strict hyperon selection cuts on the $MM(e'K^+)$ mass distributions to remove the radiative tail 
events. This is expected to be a reasonable approach as the radiative effects are independent of the beam helicity and thus should 
effectively cancel out of the asymmetry calculation. With our relatively tight hyperon mass cuts, the maximum radiated photon energy is 
only about 50~MeV, which has a negligible impact on our computed $\cos \theta_p^{RF}$ values with respect to each quantization axis.

\subsection{Final State Identification}
\label{hypid}

\begin{figure*}[htbp]
\centering
\includegraphics[width=1.00\textwidth]{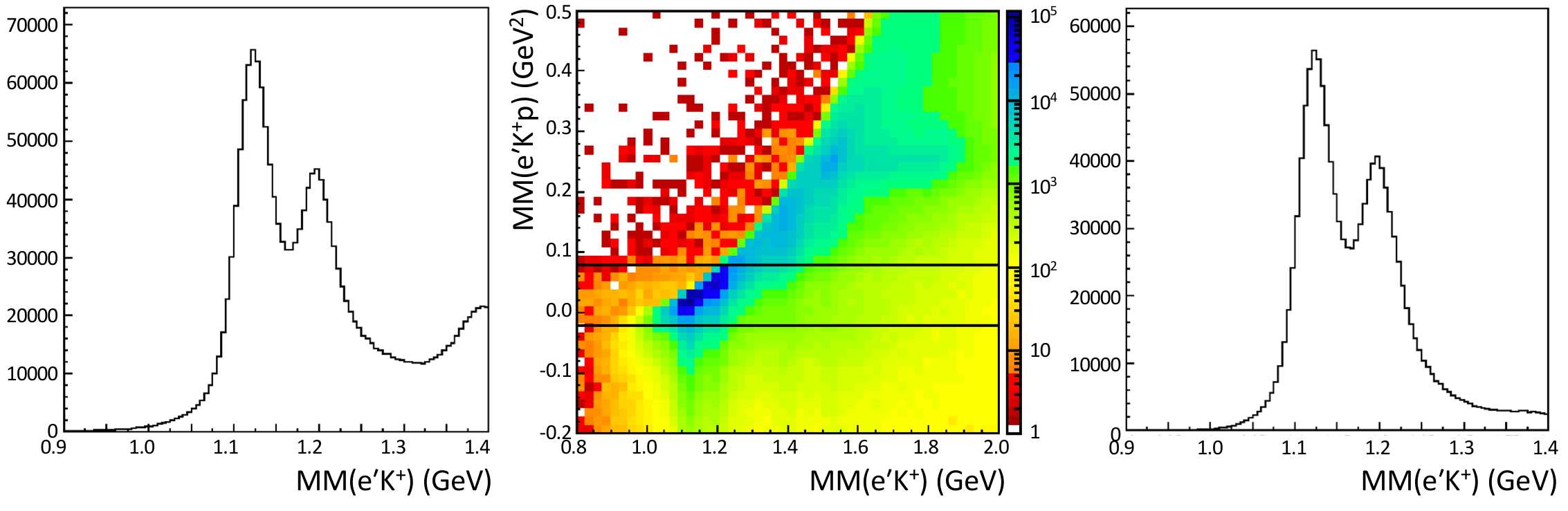}
\vspace{-8mm}
\caption{(Left) $MM(e'K^+)$ distribution requiring detection of a proton in the final state. (Middle) $MM^2(e'K^+p)$ vs.~$MM(e'K^+)$ phase 
space showing the cut employed on the $MM(e'K^+p)$ distribution to improve selection of the ground state hyperons. (Right) $MM(e'K^+)$ 
distribution shown in the left plot but with the additional cut on $MM^2(e'K^+p)$. Data are shown from the 6.535~GeV dataset.}
\label{hypcut}
\end{figure*}

The $K^+\Lambda$ and $K^+\Sigma^0$ final states were identified by selecting mass regions within the $MM(e'K^+)$ distribution as discussed 
in Section~\ref{approach}. The backgrounds in these spectra can be reduced using additional restrictions based on the reconstruction of the 
$e'K^+p$ final state. For $K^+\Lambda$ the $MM(e'K^+p)$ distribution should be consistent with a missing $\pi^-$ and for $K^+\Sigma^0$ it 
should be consistent with a missing $\pi^-$ and a low-momentum $\gamma$. Cuts were applied on $MM^2(e'K^+p)$ from $-0.02$ to 0.08~GeV$^2$ 
to select the ground state hyperon region. Figure~\ref{hypcut} shows the $MM^2(e'K^+p)$ vs.~$MM(e'K^+)$ distribution phase space from the 
6.535~GeV dataset with the cut applied, as well as the $MM(e'K^+)$ distribution before and after this additional cut. The $MM(e'K^+)$ 
spectrum before the cut shows an additional peak at about 1.4~GeV that arises due to the contributions of the $\Sigma^0(1385)$ and 
$\Lambda(1405)$ hyperon excited states. The cut also serves to significantly reduce the background beneath the hyperon peaks that arises 
primary from the multi-pion channels with the $\pi^+$ misidentified as a $K^+$.

For this analysis, three different hadronic event topologies were combined together. The dominant topologies with roughly equal statistics
are $e'K^+_Fp_F$ and $e'K^+_Cp_F$, where the hadron subscripts $F$ and $C$ refer to whether the hadron was detected in the CLAS12 Forward
Detector or Central Detector, respectively. The $e'K^+_Fp_C$ topology contains only about 10\% of the event yields. The $e'K^+_Cp_C$ topology
is kinematically disfavored due to energy/momentum conservation with the electron detected in the forward direction.

With the current status of the reconstruction of CLAS12 and the detector alignment (which at the current time is still not fully optimized
for the central tracking system), tracks reconstructed in the Forward Detector have significantly better momentum resolution than tracks
in the Central Detector - $\Delta p_F/p_F {\sim}1\%$ and $\Delta p_C/p_C {\sim}10\%$. The hyperon resolution in the $MM(e'K^+_F)$ topologies 
is $\sim$16-18~MeV and worsens to $\sim$18-20~MeV in the $MM(e'K^+_C)$ topology. The $MM(e'K^+)$ resolution of CLAS12 is relatively 
independent of $W$ and $\cos \theta_K^{c.m.}$ for the different $e'K^+$ topologies. However, the resolution degrades slowly vs.~$Q^2$ from 
16~MeV at $Q^2=0.3$~GeV$^2$ to 22~MeV at $Q^2=4.5$~GeV$^2$.

\subsection{Spectrum Fits for Yield Extraction}
\label{spec_fits}

As mentioned in Section~\ref{approach}, there are three contributions to the $MM(e'K^+)$ spectrum in the analysis range of interest for
the polarization measurement. These include the contribution from the $K^+\Lambda$ channel, the $K^+\Sigma^0$ channel, and the underlying 
multi-pion background that is present due to the finite timing resolution in the CLAS12 time-of-flight systems. At momenta above
$\sim$2.5~GeV in the Forward Detector and $\sim$0.8~GeV in the Central Detector, the misidentification of $\pi^+$ tracks as $K^+$ allows
the multi-pion topology to pollute the $K^+Y$ sample.

The approach to determine the three contributions to the $MM(e'K^+)$ spectrum relied on input from both Monte Carlo and data sources. The
hyperon contributions were accounted for by hyperon lineshape templates based on the realistic GEANT4 simulation of the CLAS12 detector
\cite{sim-nim} and the genKYandOnePion event generator~\cite{genky} that was developed by fitting the available $K^+Y$ four-fold
differential cross sections from CLAS. The event generator includes physically motivated extrapolations that span the entire kinematic
range and well reproduces the event distributions vs.~$Q^2$, $W$, and $\cos \theta_K^{c.m.}$. The $K^+Y$ simulations were generated with
radiative effects turned on in order to account for the radiative tails on the high-mass side of the hyperon peaks. For both the 
$K^+\Lambda$ and $K^+\Sigma^0$ final states 200M events were generated at each beam energy.

As the momentum resolution of the reconstructed Monte Carlo for charged tracks was better than that of the data, the Monte Carlo $K^+Y$
template spectra were Gaussian smeared bin-by-bin in the mass spectra to minimize the fit $\chi^2$ in the template fits. The Gaussian
smearing was optimized individually for each bin in $Q^2$, $W$, and $\cos \theta_K^{c.m.}$ and for each hadron topology.

\begin{figure}[htbp]
\centering
\raisebox{0mm}{\includegraphics[width=1.00\columnwidth]{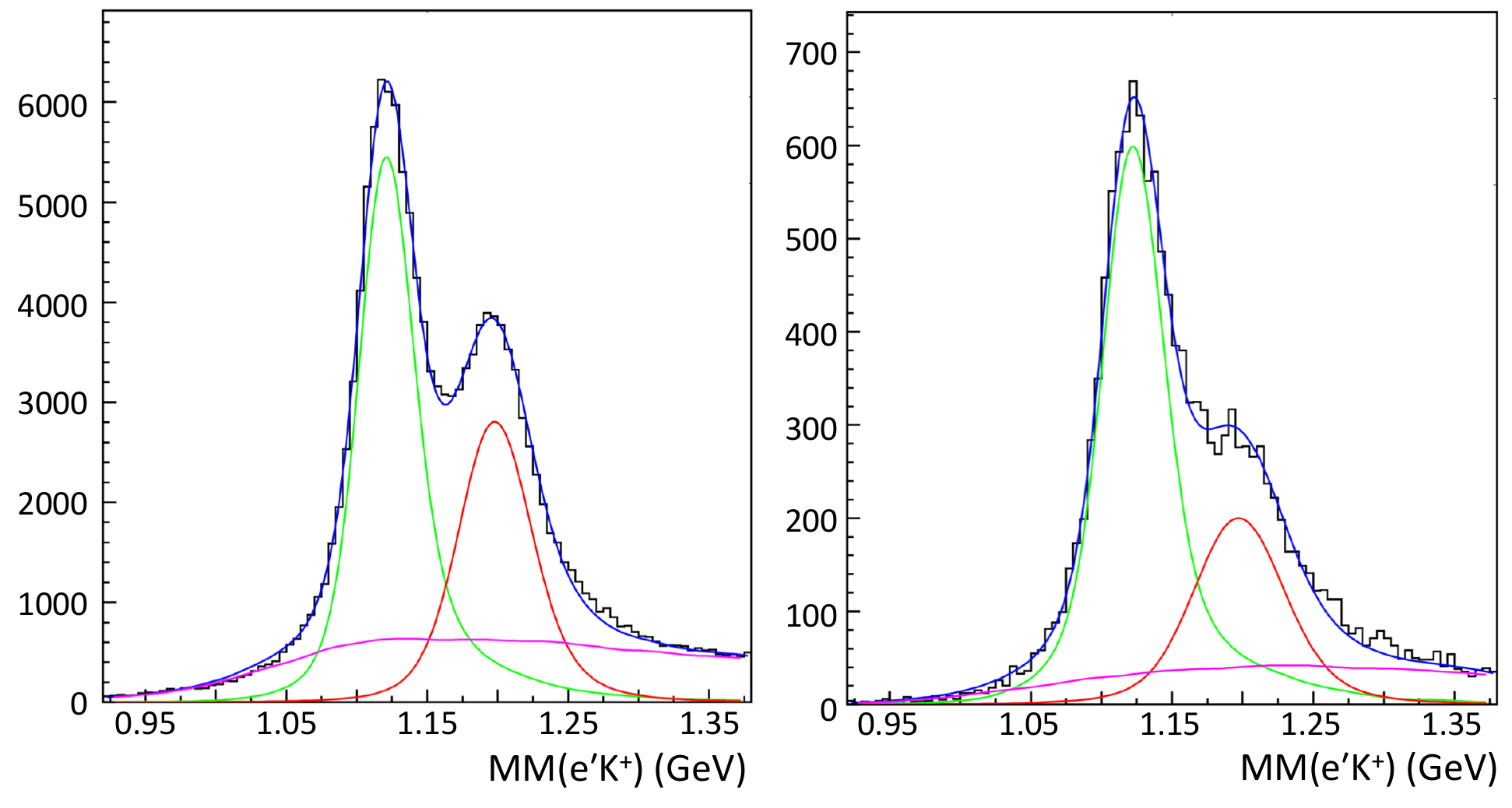}}
\vspace{-7mm}
\caption{Representative $MM(e'K^+)$ fit results using hyperon templates derived from Monte Carlo ($\Lambda$: green curve, $\Sigma^0$: red 
  curve) and a background template based on beam data (magenta curve). The blue curve shows the full fit result. The fits shown are from
  the 1D analysis binned in $Q^2$ with the left plot for $Q^2$ from 0.6-0.7~GeV$^2$ and the right plot for $Q^2$ from 2.8-3.1~GeV$^2$ from 
  the 6.535~GeV dataset.}
\label{fit_examples}
\end{figure}

For the multi-pion background, $ep \to e'\pi^+ p X$ events from data were used with the $\pi^+$ re-assigned the $K^+$ mass.
The same analysis code used for the $K^+Y$ events was used to sort the $MM(e'\pi^+)$ distributions. The $MM(e'K^+)$ spectrum in each
analysis bin was then fit with a function of the form:
\begin{equation}
MM = A \cdot \Lambda_{tmpl} + B \cdot \Sigma_{tmpl} + C \cdot BCK_{tmpl},
\end{equation}
\noindent
where $\Lambda_{tmpl}$ and $\Sigma_{tmpl}$ are the simulated hyperon distributions with weighting factors $A$ and $B$, respectively, 
and $BCK_{tmpl}$ is the template for the multi-pion background with a weighting factor of $C$. Figure~\ref{fit_examples} shows 
representative spectrum fits to determine the hyperon yields and yield ratios within the $\Lambda$ and $\Sigma^0$ mass regions as defined
in Section~\ref{approach}. The statistical uncertainties on the different contributions were determined using the MINUIT~\cite{minuit}
fit uncertainties on the template scale factors.

\begin{figure}[htbp]
\centering
\includegraphics[width=1.0\columnwidth]{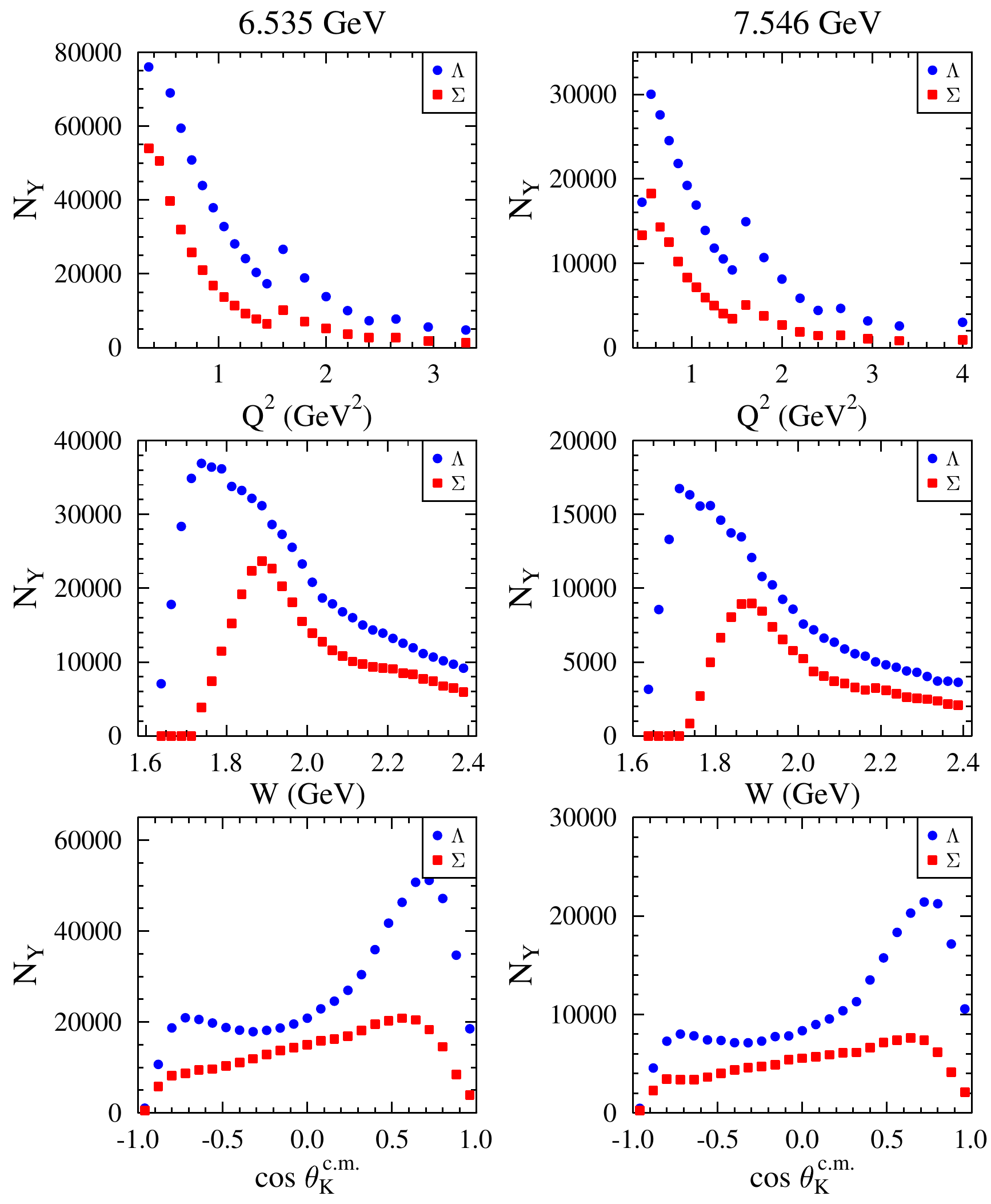}
\vspace{-7mm}
\caption{Hyperon yields from the 6.535~GeV (left) and 7.546~GeV (right) datasets vs.~$Q^2$, $W$, and $\cos \theta_K^{c.m.}$ summed over the 
  other two variables. The blue (red) data points are for the $K^+\Lambda$ ($K^+\Sigma^0$) events in the $\Lambda$ ($\Sigma^0$) mass region. 
  Note that the abrupt shift at $Q^2$=1.5~GeV$^2$ occurs due to the change in $Q^2$ bin size at this point.}
\label{totals}
\end{figure}

Figure~\ref{totals} shows the extracted $\Lambda$ and $\Sigma^0$ yields for both beam energies. These distributions are for the 1D polarization
analysis (detailed in Section~\ref{binning}) sorting the polarization vs.~$Q^2$, $W$, and $\cos \theta_K^{c.m.}$, integrated over the other two
variables (and the angle $\Phi$ between the electron scattering and hadron reaction planes). The yields decrease rapidly with increasing $Q^2$
due to the roughly monopole fall-off of the kaon form factor. To compensate for this the bin sizes were chosen to increase with $Q^2$, with
larger bins starting at $Q^2$=1.5~GeV$^2$. The yields vs.~$W$ rise rapidly for the $K^+\Lambda$ and $K^+\Sigma^0$ channels within the first
100~MeV of their respective reaction thresholds, peaking at $\sim$1.7~GeV for $K^+\Lambda$ and at $\sim$1.9~GeV for $K^+\Sigma^0$. The yields
then gradually fall off with increasing $W$. The yields for both hyperon channels show a strong forward peaking in $\cos \theta_K^{c.m.}$ due 
to the importance of $t$-channel kaon exchange contributions. The very rapid fall-off just as $\cos \theta_K^{c.m.} \to 1$ is due to the 
forward acceptance hole of CLAS12 below $\theta \sim 5^\circ$.

\begin{figure}[htbp]
\centering
\includegraphics[width=1.0\columnwidth]{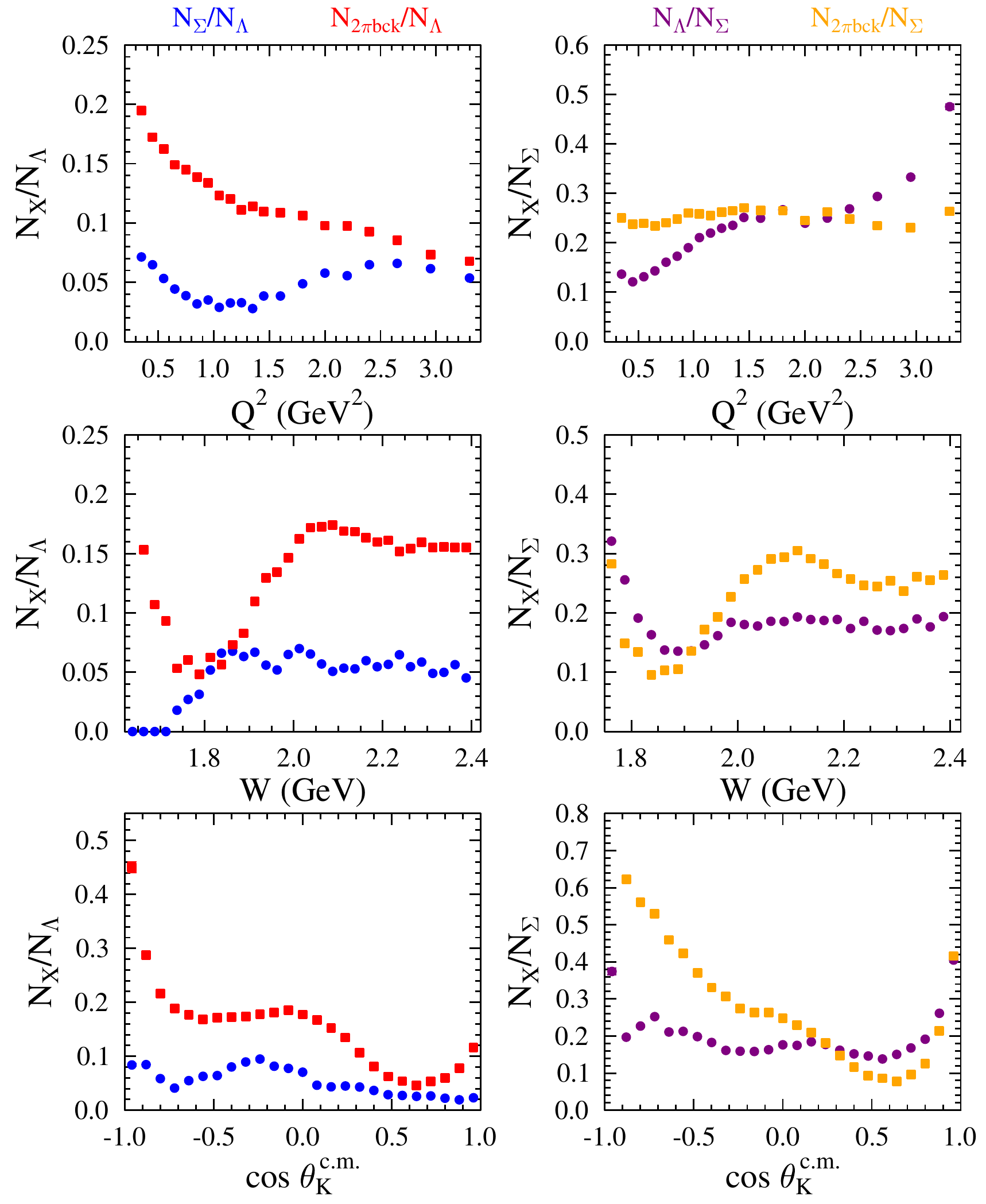}
\vspace{-7mm}
\caption{Yield ratios for the 6.535~GeV dataset showing the $N_\Sigma/N_\Lambda$ and $N_B/N_\Lambda$ ratios in the $\Lambda$ mass region 
(left) and the $N_\Lambda/N_\Sigma$ and $N_B/N_\Sigma$ ratios in the $\Sigma^0$ mass region (right).}
\label{ratios}
\end{figure}

As discussed in Section~\ref{approach}, the ratios of the yields $N_\Sigma/N_\Lambda$ and $N_B/N_\Lambda$ in the $\Lambda$ mass region and
$N_\Lambda/N_\Sigma$ and $N_B/N_\Sigma$ in the $\Sigma^0$ mass region are the relevant quantities for the polarization determination (see
Eq.(\ref{pmeasL2}) and Eq.(\ref{pmeasS2})). These yield ratios for the 6.535~GeV data are shown in Fig.~\ref{ratios}. In the $\Lambda$
mass region the average $\Sigma^0$ tail contamination is $\sim$5-10\% and the multi-pion contamination is $\sim$5-15\% depending on the 
kinematics. In the $\Sigma^0$ mass region, the $\Lambda$ radiative tail accounts for up to 40\% of the yield and the multi-pion contribution 
is on the order of 5-30\% depending on the kinematics.

\subsection{Data Binning}
\label{binning}

The results shown in this work are limited to the nucleon resonance region, spanning invariant mass $W$ from the $K^+Y$ threshold to
2.4~GeV. The $\Phi$-integrated beam-recoil transferred polarization components for the $K^+\Lambda$ and $K^+\Sigma^0$ final states are
presented in a 1D binning scenario vs.~$Q^2$, $W$, and $\cos \theta_K^{c.m.}$, integrated over the other two variables. The observables are
also presented in a 3D binning scenario divided into 2 bins in $Q^2$ of different extents to allow for comparable statistics in each
bin and 4 equal bins of $\cos \theta_K^{c.m.}$. In this multi-dimensional binning, the polarization observables are shown as a function of
$W$. Table~\ref{1d-bin} and Table~\ref{3d-bin} present the 1D and 3D binning choices, respectively. The multi-dimensional analysis is
not included here for the 7.546~GeV dataset, but is included along with all of the extracted observables from this analysis in the CLAS
physics database~\cite{physicsdb}.

\begin{table}[htbp]
\centering
\begin{tabular}{c|c|c} \hline
Dependence       & Range                   & Bin Size \\ \hline
$Q^2$            & [$Q^2_{min}$:1.5] GeV$^2$ & 0.1~GeV$^2$ \\
                 & [1.5:2.5] GeV$^2$       & 0.2~GeV$^2$ \\
                 & [2.5:3.1] GeV$^2$       & 0.3~GeV$^2$ \\
                 & [3.1:3.5] GeV$^2$       & 0.4~GeV$^2$ \\
                 & [3.5:4.5] GeV$^2$       & 1.0~GeV$^2$ \\ \hline
$W$              & [$W_{min}$:2.4] GeV    & 25~MeV \\ \hline
$\cos \theta_K^{c.m.}$ & [$-1$:1]               & 0.08 \\ \hline
\end{tabular}
\caption{Bin sizes for the 1D polarization analysis vs.~$Q^2$, $W$, and $\cos \theta_K^{c.m.}$. The analysis for all variables is limited to
  the kinematic phase space from $Q^2_{min}$ to $Q^2_{max}$ where $Q^2_{min}$/$Q^2_{max}$=0.3~GeV$^2$/3.5~GeV$^2$ for the 6.535~GeV dataset and
  0.4~GeV$^2$/4.5~GeV$^2$ for the 7.546~GeV dataset, and from $W_{min}$ to 2.4~GeV where $W_{min}$=1.625~GeV (1.725~GeV) for the $K^+\Lambda$
  ($K^+\Sigma^0$) final state.}
\label{1d-bin}
\end{table}

\begin{table}[htbp]
\centering
\begin{tabular}{c|c|c} \hline
  Variable  & \multicolumn{2} {c} {Bin Choices} \\ \hline
  $E_b$   & 6.535~GeV & 7.546~GeV \\ \hline
  $Q^2$  & [0.3:0.9] GeV$^2$ & [0.4:1.0] GeV$^2$ \\
         & [0.9:3.5] GeV$^2$ & [1.0:4.5] GeV$^2$ \\ \hline
  $W$    & \multicolumn{2} {c} {[$W_{min}$:2.4] GeV in 80~MeV bins} \\ \hline
  $\cos \theta_K^{c.m.}$ & \multicolumn{2} {c} {[$-1$:1] in 0.5 bins} \\ \hline
\end{tabular}
\caption{Bin sizes for the 3D polarization analysis in $Q^2$, $W$, and $\cos \theta_K^{c.m.}$ for the $E_b$=6.535~GeV and 7.546~GeV datasets, 
where $W_{min}$=1.625~GeV (1.725~GeV) for the $K^+\Lambda$ ($K^+\Sigma^0$) final state.}
\label{3d-bin}
\end{table}

The bin sizes are kept uniform in $W$ and $\cos \theta_K^{c.m.}$ in the 1D and 3D sorts. However, the $Q^2$ bin sizes increase with 
increasing $Q^2$ to compensate for the fall-off of the cross section. The results for all polarization components are reported at the 
geometric center of the kinematic bins.

\subsection{Multi-Pion Background Polarization Studies}
\label{bck_pol}

In Section~\ref{approach} the formalism to connect the measured raw yield helicity asymmetries to the $\Lambda$ and $\Sigma^0$
polarization was developed accounting for the different background contributions in the $\Lambda$ and $\Sigma^0$ mass regions of the
$MM(e'K^+)$ distribution as defined in Fig.~\ref{mm-regions}. The forms of Eq.(\ref{pmeasL2}) for ${\cal P}'_\Lambda$ and
Eq.(\ref{pmeasS2}) for ${\cal P}'_\Sigma$ were written assuming $A_B=0$, {\it i.e.} the asymmetry for the multi-pion background
contribution that underlies the hyperon peaks is zero.

\begin{figure}[htbp]
\centering
\includegraphics[width=0.95\columnwidth]{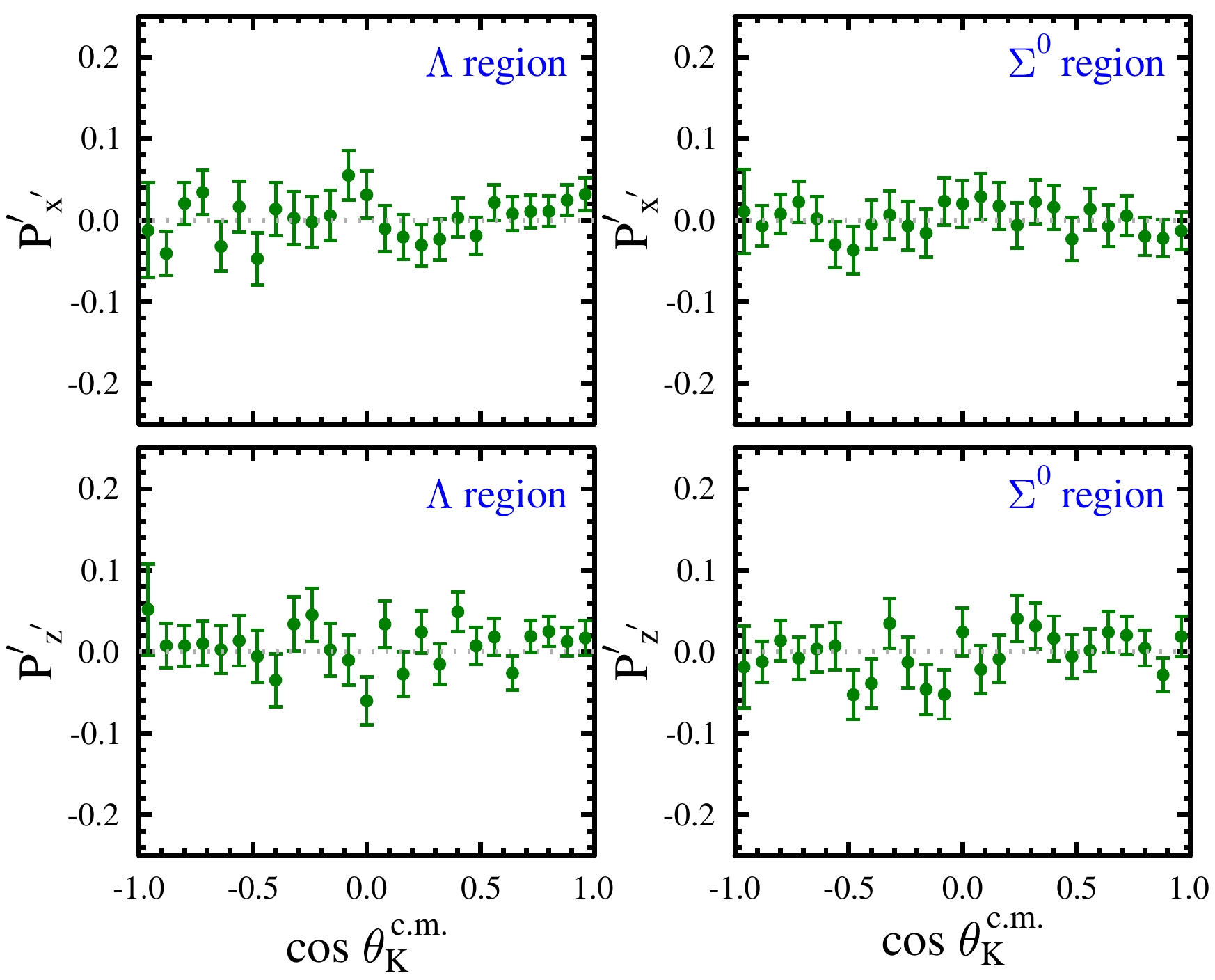}
\vspace{-3mm}
\caption{Measured polarizations determined for the multi-pion background in the 6.535~GeV dataset that underlies the hyperon peaks in the 
  $MM(e'K^+)$ distributions for the 1D analysis vs.~$\cos \theta_K^{c.m.}$ summing over $Q^2$ and $W$ for both the $\Lambda$ (left) and
  $\Sigma^0$ (right) mass regions defined in Section~\ref{approach} for the primed system. The error bars include the statistical 
  uncertainties only.}
\label{pol_bck}
\end{figure}

The assumption that $A_B=0$ can be directly tested sorting the helicity asymmetries for the $ep \to e'\pi^+pX$ final state, reassigning
the reconstructed $\pi^+$ with the $K^+$ mass. This was done using the same analysis code with the same binning, cuts, and conditions
as for the $K^+Y$ analysis. The asymmetries were measured for this channel and were found to be consistent with zero to within the
statistical uncertainties. Representative results for the measured background polarization are shown for the 1D sort 
vs.~$\cos \theta_K^{c.m.}$ in Fig.~\ref{pol_bck} in the $\Lambda$ and $\Sigma^0$ mass regions for the primed system defined in 
Fig.~\ref{coor4}.

\subsection{Polarization for Combined Hadron Topologies}
\label{3top}

As detailed in Section~\ref{hypid}, the analysis was based on combining together the three hadron event topologies $K^+_Fp_F$, $K^+_Fp_C$,
and $K^+_Cp_F$ (F = Forward Detector, C = Central Detector). To determine the hyperon polarization ${\cal P}'$ in each kinematic bin for
the 1D and 3D binning scenarios, it is not strictly appropriate to combine the different hadronic topologies based only on their
statistical uncertainties. The proper manner to determine the final polarization is to weight the results for the different topologies
accounting for their individual cross sections and detector acceptance functions. The ${\cal P}'$ value in each kinematic bin has been 
determined using:
\begin{equation}
  \label{pavg}
  {\cal P}'_{avg} = \frac{\sum\limits_{i=1}^3 \sigma_i \cdot ACC_i \cdot {\cal P}'_i}
  {\sum\limits_{i=1}^3 \sigma_i \cdot ACC_i},
\end{equation}
\noindent
where the sum is over the results from the three hadronic topologies in a given kinematic bin, $\sigma_i = d\sigma/d\Omega_i$ and $ACC_i$ 
are the differential cross section and acceptance for topology $i$ averaged over the bin, and ${\cal P}'$ is the hyperon polarization for 
the bin determined for topology $i$. This approach actually gives results fully consistent with combining the event yields for the three
hadron topologies using a statistical weight as the two dominant hadron topologies $K^+_Fp_F$ and $K^+_Cp_F$ cover essentially complementary
ranges in $\cos \theta_K^{c.m.}$. In the computation of Eq.(\ref{pavg}) the cross sections were determined using the CLAS data-based event
generator genKYandOnePion~\cite{genky} that was developed from fits to the available $KY$ cross section data from CLAS.


\section{Systematic Uncertainties}
\label{syserr}

In this section we define and quantify the sources of systematic uncertainty that affect the measured hyperon polarization observables for 
the 6.535~GeV and 7.546~GeV datasets. The contributions to the total systematic uncertainty belong to one of four general categories:

\begin{itemize}
\item Polarization Extraction
\item Beam-Related Factors
\item Acceptance Function
\item Background Contributions
\end{itemize}

\noindent
Each of these different sources is discussed in the subsections that follow. 

The procedure used to quantify the systematic uncertainty associated with each source was to compare the measured polarization ${\cal P}'$ 
for all kinematic bins with the nominal analysis cuts and procedures ($nom$) to that with modified cuts or procedures ($mod$). The average
difference of $\Delta {\cal P}' = {\cal P}'_{nom} - {\cal P}'_{mod}$ over all data points was used as a measure of the systematic uncertainty
for a given source, where we have used the weighted root-mean-square (RMS) of $\Delta {\cal P}'$ for all points given by:
\begin{equation}
\delta {\cal P}'_{sys} = \sqrt{ \frac{\sum_{i=1}^{N} (\Delta {\cal P}'_i)^2/(\delta {\cal P}'_i)^2}
{\sum_{i=1}^{N} 1/(\delta {\cal P}'_i)^2}}.
\end{equation}
\noindent 
Here the sums are over all $N$ data points and $\delta {\cal P}'_i$ is the statistical uncertainty of the $i^{th}$ data point. In each of 
the systematic uncertainty studies performed, the widths of the $\Delta {\cal P}'$ distributions were much larger than the measured centroids, 
which were all consistent with zero. In general, the systematic uncertainties are comparable to the statistical uncertainties for the 1D 
analysis binning and dominated by the statistical uncertainties for the 3D analysis binning for the 6.535~GeV dataset. For the 7.546~GeV 
dataset, the statistical uncertainties are dominant for both the 1D and 3D analysis binning. Our final systematic uncertainty accounting 
for the $\Lambda$ and $\Sigma^0$ polarization measurements for the 6.535~GeV and 7.546~GeV datasets is included in Table~\ref{systab} 
listing all sources. The final value in the table adds all the individual contributions in quadrature.

\begin{table*}[htbp]
\begin{center}
\begin{tabular} {|c|c|c|} \hline
Category                 & Contribution          & Systematic Uncertainty \\ \hline
Polarization Extraction  & Functional Form       & 0.005 \\
                         & Bin Size              & 0.004 \\ 
                         & Asymmetry Parameter   & 0.019 \\
                         & Model Dependence      & 0.010 ($\Lambda$), 0.030 ($\Sigma^0$) \\ \hline
Beam-Related Factors     & Beam Polarization     & 0.035 \\ \hline
Acceptance Function      & Fiducial Cut Form     & 0.007 \\ \hline
Background Contributions & Analysis Region       & 0.011 ($\Lambda$), 0.066 ($\Sigma^0$) 1D bins \\
                         &                       & 0.017 ($\Lambda$), 0.099 ($\Sigma^0$) 3D bins \\ \hline
\multicolumn{3} {|r|} {{\bf $\langle$Total Systematic Uncertainty$\rangle$} 0.044 ($\Lambda$), 0.078 ($\Sigma^0$) 1D bins}\\
\multicolumn{3} {|r|} {0.045 ($\Lambda$), 0.108 ($\Sigma^0$) 3D bins} \\ \hline
\end{tabular}
\caption{Summary table of the individual systematic uncertainty sources and the average total systematic uncertainty. Separate
  systematics were determined for the different hyperons and for the 1D and 3D binning sorts.} 
\label{systab}
\end{center}
\end{table*}

\subsection{Polarization Extraction}
\label{pol_sys}

The extracted polarization components have been compared using two different analysis approaches. The nominal technique is the asymmetry 
approach described in Section~\ref{approach}, which relates the hyperon polarization to the asymmetry of the difference divided by the
sum of the helicity-gated hyperon yields. An alternative approach is to extract the polarization from the ratio of the helicity-gated
yields via:
\begin{equation}
R = \frac{N^+}{N^-} = \frac{1 + \nu_Y \alpha P_b  {\cal P}_\Lambda' \cos \theta_p^{RF}}{1 - \nu_Y \alpha P_b  
{\cal P}_\Lambda' \cos \theta_p^{RF}}.
\end{equation}
\noindent
The difference between these two techniques resulted in a weighted RMS of $\delta {\cal P}'_{sys} = 0.005$, which is assigned as the 
systematic uncertainty.

A systematic uncertainty contribution arises due to binning choices made during the data sorting. The nominal analysis approach sorted 
the helicity-gated yields in $\cos \theta_K^{c.m.}$ into 6 bins. A comparison of the polarization components with the extraction from a 
sort with 8 and 10 bins in $\cos \theta_p^{RF}$ resulted in a weighted RMS of $\delta {\cal P}'_{sys} = 0.004$. The difference in the 
observables arises due to the fitting algorithm employed in which the centroids of the $\cos \theta_p^{RF}$ bins are assigned to the 
center of the bin. When the number of bins is reduced, the fit results are more sensitive to the bin content.

Another systematic uncertainty is due to the uncertainty in the weak decay asymmetry parameter $\alpha$. This uncertainty gives rise to
a scale-type uncertainty on the polarization components that is the same for both the $\Lambda$ and $\Sigma^0$ hyperons and is given by:
\begin{equation}
  \delta {\cal P}'_{sys} = \vert {\cal P}'_Y \vert \frac{\delta \alpha} {\alpha} = 0.019 \vert {\cal P}'_Y \vert.
\end{equation}
The final systematic contribution in this section arises due to the weighting factors used to combine the measured ${\cal P}'$ values
for the three different hadronic topologies in the detector ($K^+_Fp_F$, $K^+_Fp_C$, $K^+_Cp_F$) as discussed in Section~\ref{3top}. The
weighting factors (cross section$\times$acceptance) for the nominal analysis were determined from the CLAS data-based event generator
genKYandOnePion~\cite{genky}. The ${\cal P}'$ values were compared to the results deriving the weight factors using an alternative event
generator based on the Ghent RPR model~\cite{egky}. The assigned systematic for the model dependence was 0.010 for the $\Lambda$ analysis
and 0.030 for the $\Sigma^0$ analysis.

\subsection{Beam-Related Factors}
\label{beam_sys}

Two contributions were considered related to the systematic uncertainty of beam-related factors. The first was associated with the
beam polarization measurement from the M{\o}ller polarimeter system. This arises from the uncertainty in the M{\o}ller target foil
polarization, the statistical uncertainty in the measurements, as well as from variations of the polarization measurements over time.
These contributions have been estimated to be 3\%. This scale-type uncertainty results in an uncertainty in the hyperon polarization of:
\begin{equation}
\delta {\cal P}'_{sys} = \vert {\cal P}'_Y \vert \frac{\delta P_b}{P_b} = 0.035 \vert {\cal P}'_Y \vert.
\end{equation}
The second beam-related effect that contributes is the beam charge asymmetry that results from a systematic difference in the electron
beam intensity for the two different beam helicity states recorded by the data acquisition during production data taking. The helicity
asymmetry was measured throughout the data taking and its effect was shown to have a negligible effect on the polarization results.

\subsection{Acceptance Function}

The nominal analysis method does not apply acceptance corrections to the helicity-gated yields as the helicity asymmetries for the 
beam-recoil transferred polarization were shown to be insensitive to acceptance corrections. Studies correcting the helicity-gated
yields using a realistic acceptance function based on our nominal event generator (discussed in Section~\ref{spec_fits}) were found
to have a smaller effect on the extracted polarization components than varying the fiducial region in which the particles were
reconstructed. These studies were carried out by applying both loose and tight cuts on the $\theta/\phi$ range of the accepted
particles in the forward direction. The difference of 0.007 was assigned as the systematic uncertainty for all analysis bins.

\subsection{Background Contributions}
\label{bck_sys}

The approach to separate the $\Lambda$, $\Sigma^0$, and particle misidentification background within the $\Lambda$ and $\Sigma^0$ 
mass regions was detailed in Section~\ref{approach}. To check the stability of the yield extraction, the $\Lambda$ and $\Sigma^0$ analysis
regions were made both looser and tighter than the nominal ranges. The RMS width of the difference distribution for the extracted
polarizations was assigned as the associated systematic uncertainty for the yield stability. The RMS difference for the $\Lambda$ is
0.011 and for the $\Sigma^0$ is 0.066 for the 1D data sort. It should be expected that the $\Sigma^0$ result is more sensitive to the
definition of the analysis region due to the very strong (and highly polarized) $\Lambda$ contribution that gives rise to a larger
systematic effect. However, assigning a single systematic uncertainty to all analysis bins was found to be insufficient. The size of
the systematic was found to be correlated with the signal impurity within the analysis region, {\it i.e.} with
${\cal I}_\Lambda = 1 - N_\Lambda/(N_\Lambda + N_\Sigma + N_{bck})$ within the $\Lambda$ mass window and with
${\cal I}_\Sigma = 1 - N_\Sigma/(N_\Lambda + N_\Sigma + N_{bck})$ within the $\Sigma^0$ mass region. The assigned systematics for the $\Lambda$
and $\Sigma^0$ 1D analyses scaled ${\cal I}_\Lambda$ and ${\cal I}_\Sigma$ by multiplicative factors to reproduce the average RMS values
for the $\Lambda$ and $\Sigma^0$ analyses from varying the hyperon analysis regions. For the 3D analyses the assignment of a
corresponding systematic uncertainty using the same approach is dominated by statistical effects due to the smaller samples due to the
increased binning. A conservative choice was to multiply the associated factors by 1.5 relative to the 1D sorts.

\begin{figure}[htbp]
\centering
\includegraphics[width=0.95\columnwidth]{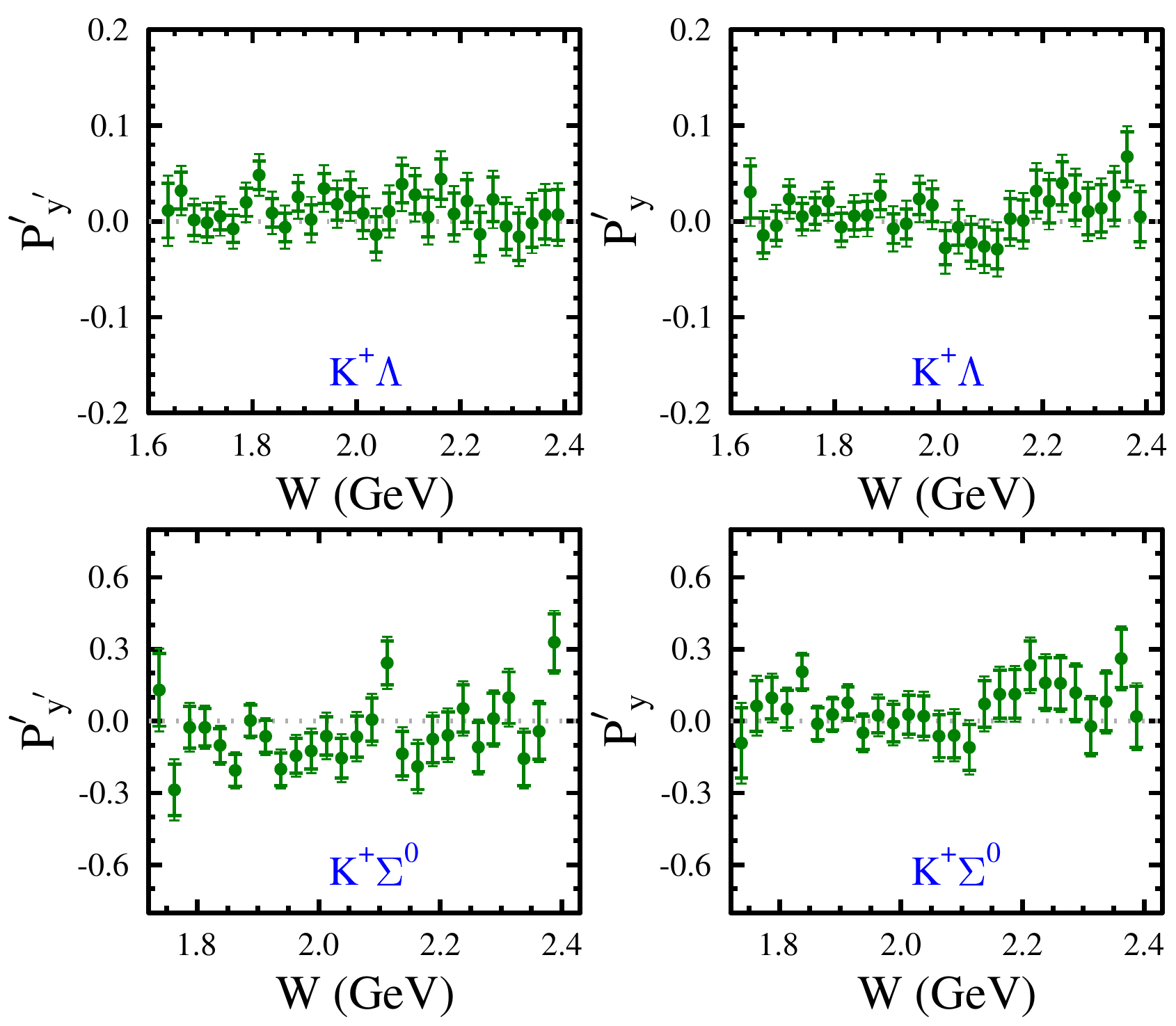}
\vspace{-3mm}
\caption{Distributions of the transferred $\Lambda$ (top) and $\Sigma^0$ (bottom) polarization components in relative to the $y'$-axis 
(left) and $y$-axis (right) vs.~$W$ from the 1D data analysis. The data shown are from the higher statistics 6.535~GeV dataset. The 
inner error bars on each data point represent the statistical uncertainties and the outer error bars represent the total uncertainties.}
\label{norm}
\end{figure}

\subsection{Other Checks}

After the investigation of the different systematic sources, a technique to verify the accuracy of the final systematic uncertainty 
assignment is to look at the deviations of the normal components of the extracted $\Lambda$ and $\Sigma^0$ polarizations ({\it i.e.}
along the $y'$ and $y$ axes). By definition as discussed in Section~\ref{formalism}, these components should be equal to zero. In
this analysis the weighted means of the ${\cal P}_{y'}'$ and ${\cal P}_y$ components for the $\Lambda$ and $\Sigma^0$ components were
consistent with zero with an RMS width consistent with the total uncertainty (statistical + systematic) assignments, which provides
confidence in the assignments. The extracted normal components for one of our data sorts for the $\Lambda$ and $\Sigma^0$ hyperons
are shown in Fig.~\ref{norm}.

Finally, another check of the analysis results included in this work is that the polarization components were extracted independently
by two different approaches. The nominal analysis approach to determine the hyperon polarization components was detailed in
Section~\ref{approach}. In the independent analysis the hyperon yields and backgrounds were fit in bins of $Q^2$, $W$,
$\cos \theta_K^{c.m.}$, $\cos \theta_p^{RF}$, and helicity $h$ using an analytic functional for the hyperons (Gaussian on the low-mass
side of the peaks, Landau on the high mass side) with a second-order polynomial for the underlying background. The comparison of the
results showed good agreement over the full kinematic phase space. This second analysis further served to verify that the systematic
uncertainty assignments were justified and served to cross check all data analysis selections and analysis routines.


\section{Data Results}
\label{results}

\boldmath
\subsection{$\Lambda$ Polarization Transfer}
\label{lambda}
\unboldmath

\begin{figure*}[htbp]
\centering
\includegraphics[width=0.45\textwidth]{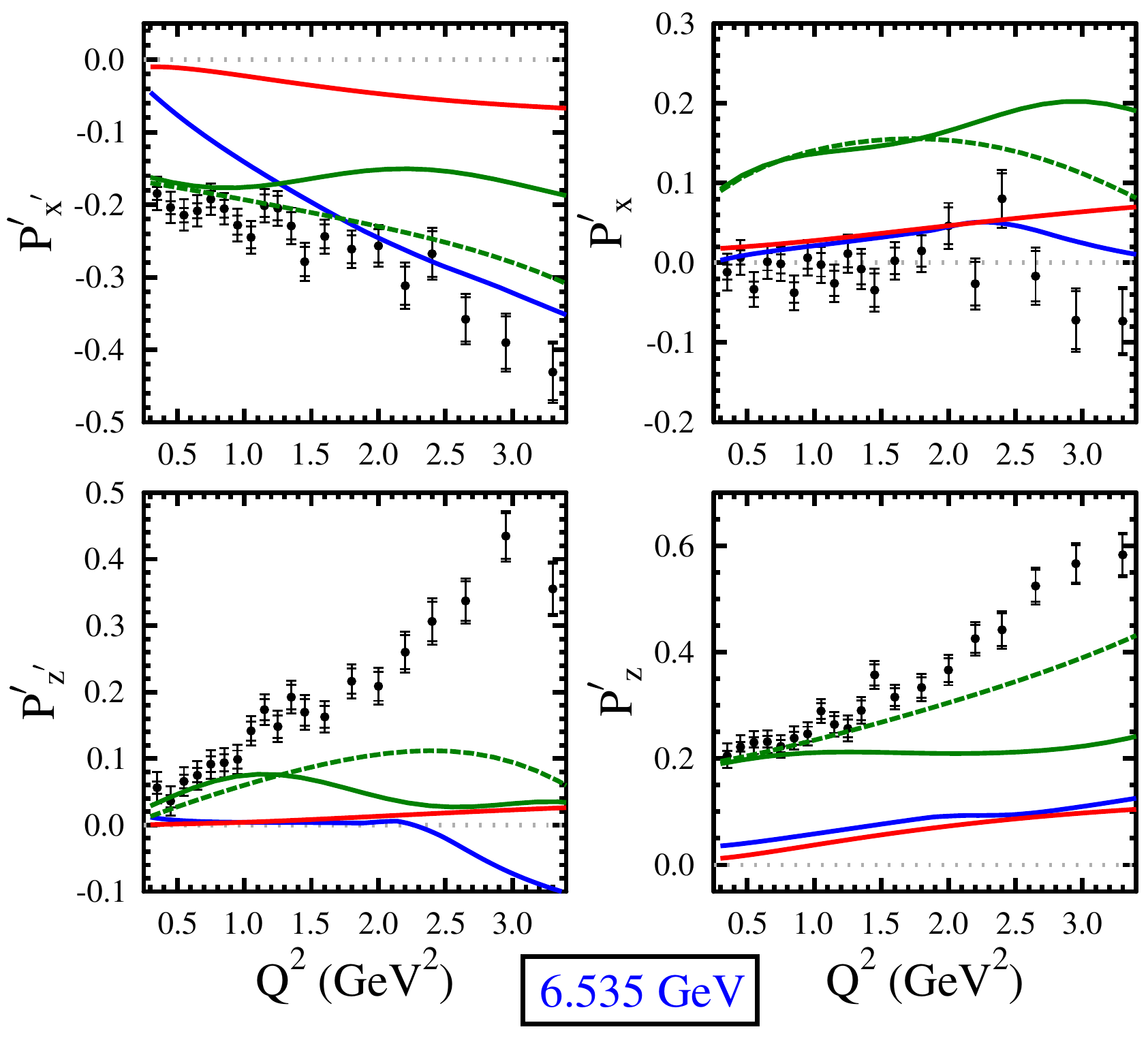}
\raisebox{0.5mm}{\includegraphics[width=0.45\textwidth]{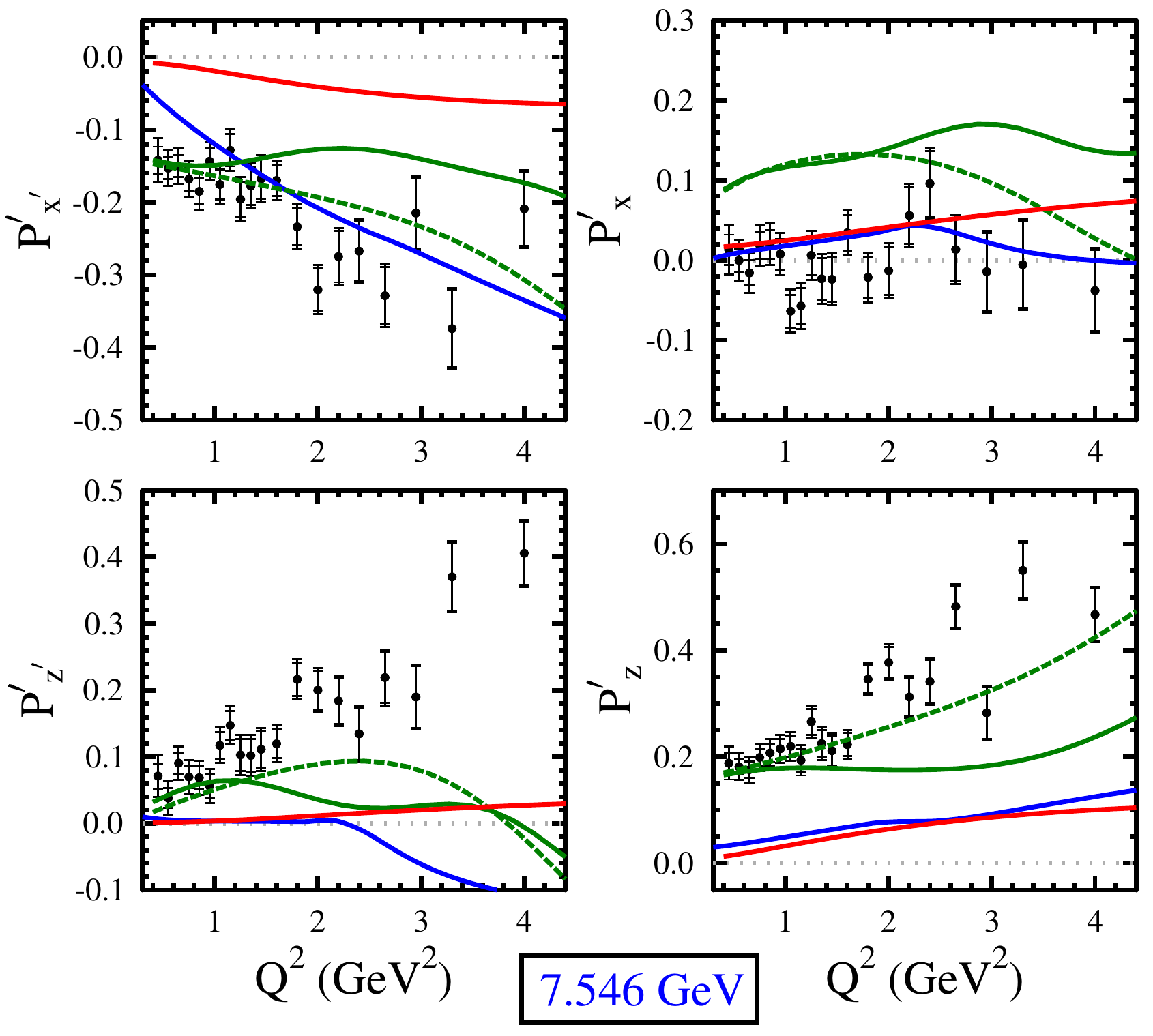}}
\vspace{-4mm}
\caption{Transferred $\Lambda$ polarization components ${\cal P}'$ with respect to the $(x',z')$ and $(x,z)$ axes vs.~$Q^2$ for a
  beam energy of 6.535~GeV (left) and 7.546~GeV (right). The data are limited to $Q^2$ from 0.3 to 3.5~GeV$^2$ (6.535~GeV) and from
  0.4 to 4.5~GeV$^2$ (7.546~GeV), and $W$ from 1.625 to 2.4~GeV. In the text this is referred to as the 1D sort (see Section~\ref{binning}
  for details). The inner error bars on the data points represent the statistical uncertainties and the outer error bars represent the
  total uncertainties. The curves are calculations from RPR~\cite{rpr} (solid green - full RPR-2011 model, dashed green - RPR-2011 model 
  with resonance terms off), BS3~\cite{skoupil} (solid red), and Kaon-MAID~\cite{kaon-maid1,kaon-maid2,kaon-maid3} (solid blue).}
\label{ltpol1a}
\end{figure*}

\begin{figure*}[htbp]
\centering
\includegraphics[width=0.45\textwidth]{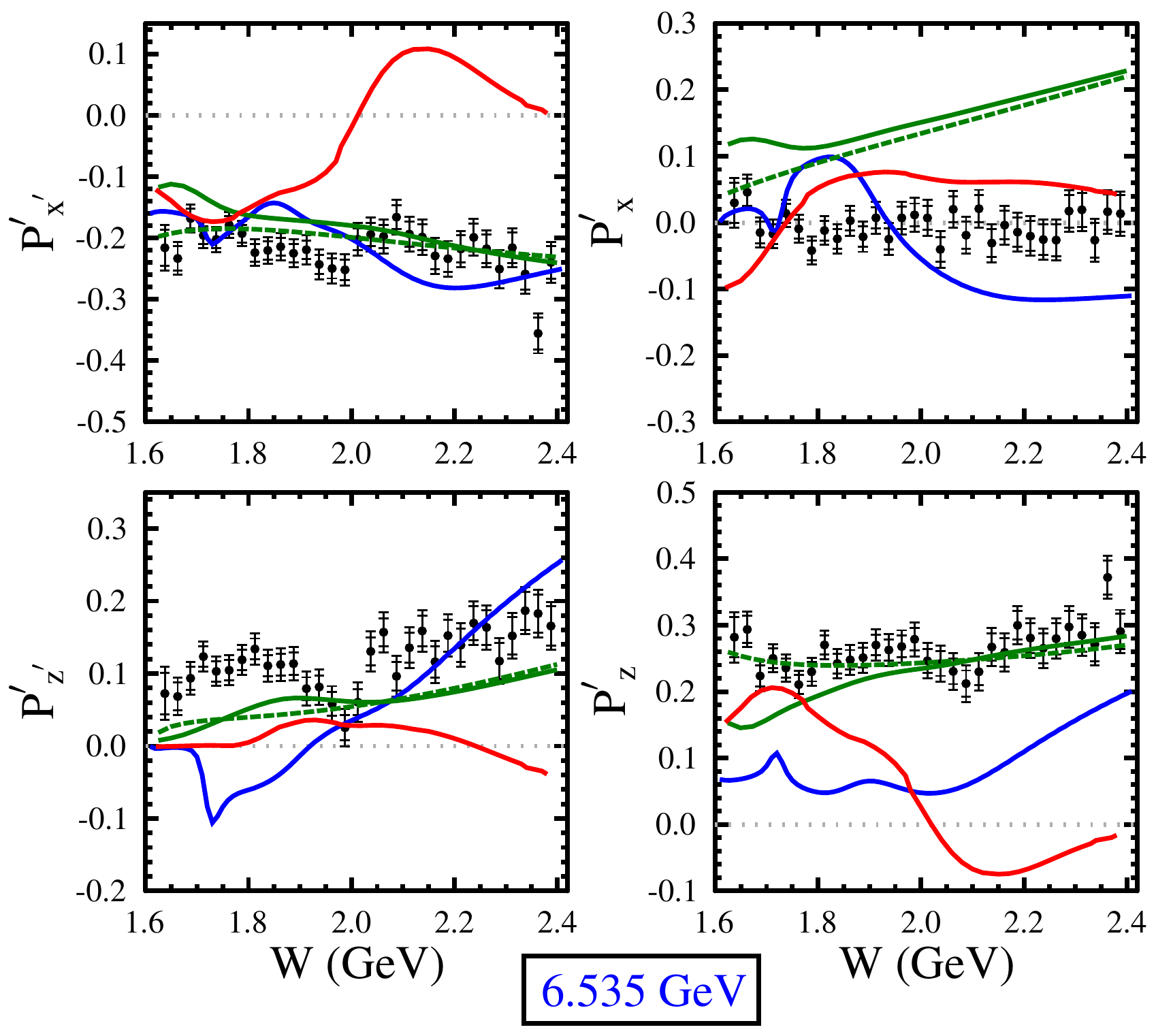}
\includegraphics[width=0.45\textwidth]{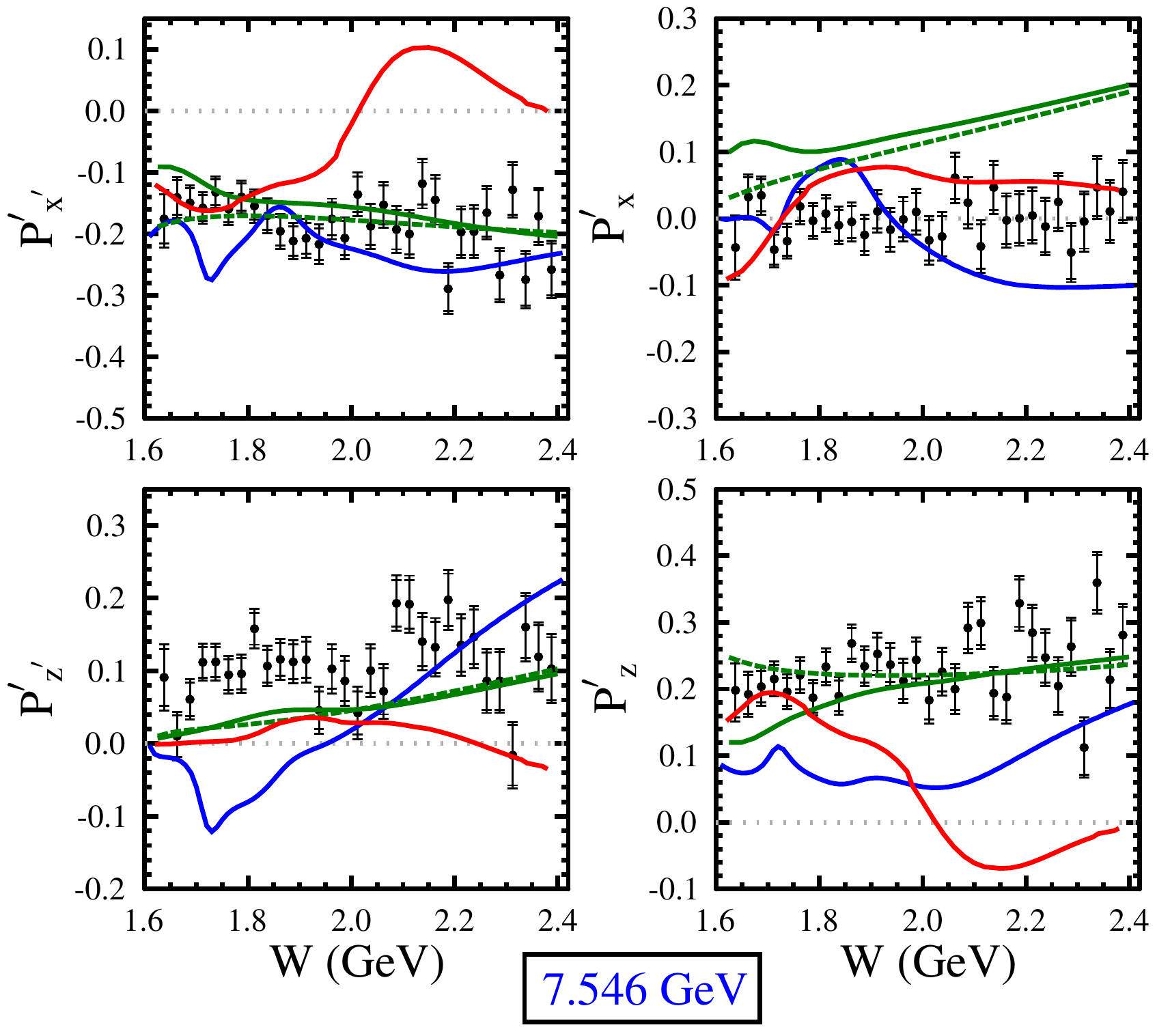}
\vspace{-4mm}
\caption{Transferred $\Lambda$ polarization components ${\cal P}'$ with respect to the $(x',z')$ and $(x,z)$ axes vs.~$W$ for a beam
  energy of 6.535~GeV (left) and 7.546~GeV (right). See the Fig.~\ref{ltpol1a} caption for details.}
\label{ltpol1b}
\end{figure*}

\begin{figure*}[htbp]
\centering
\includegraphics[width=0.45\textwidth]{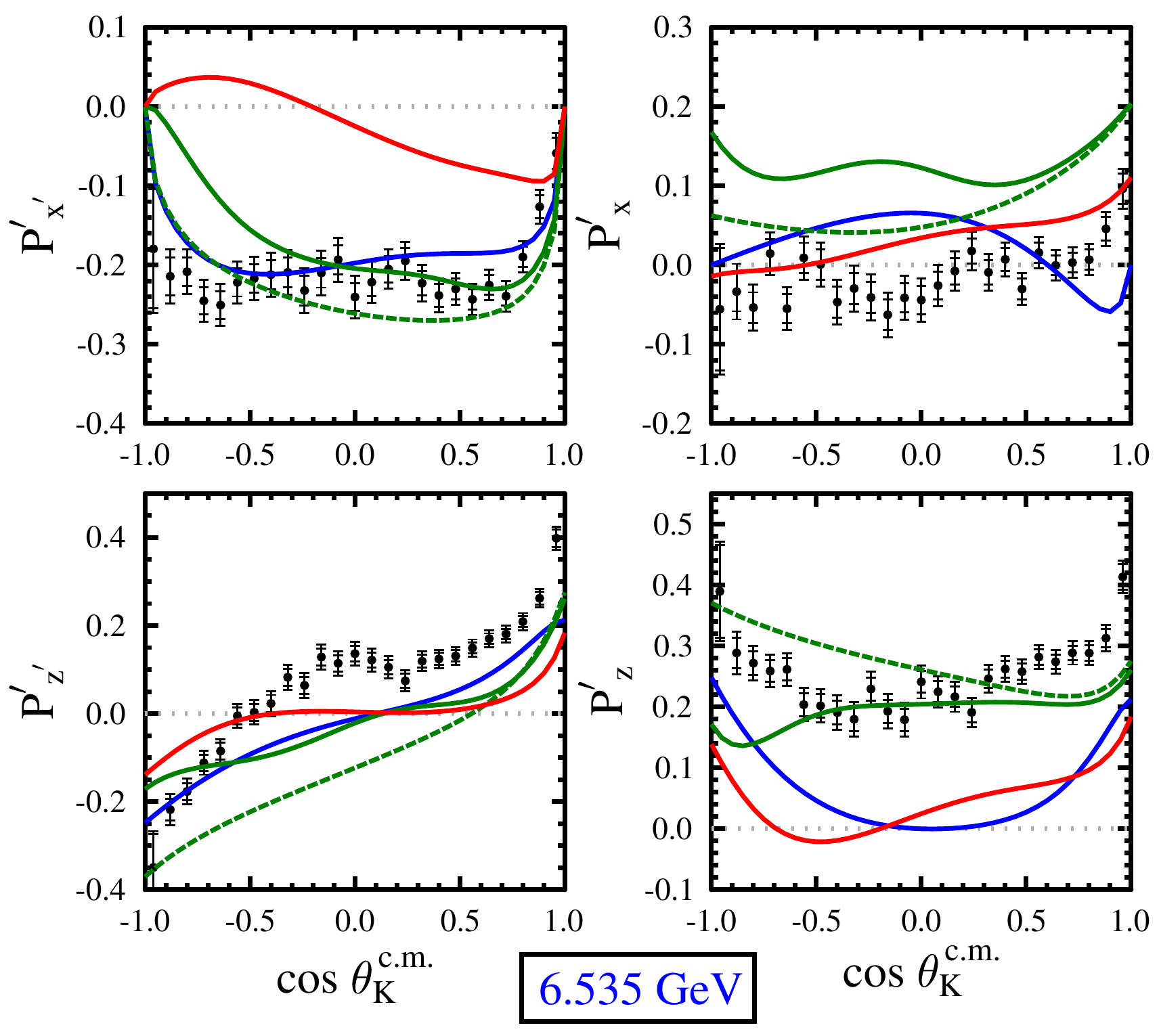}
\raisebox{0.5mm}{\includegraphics[width=0.45\textwidth]{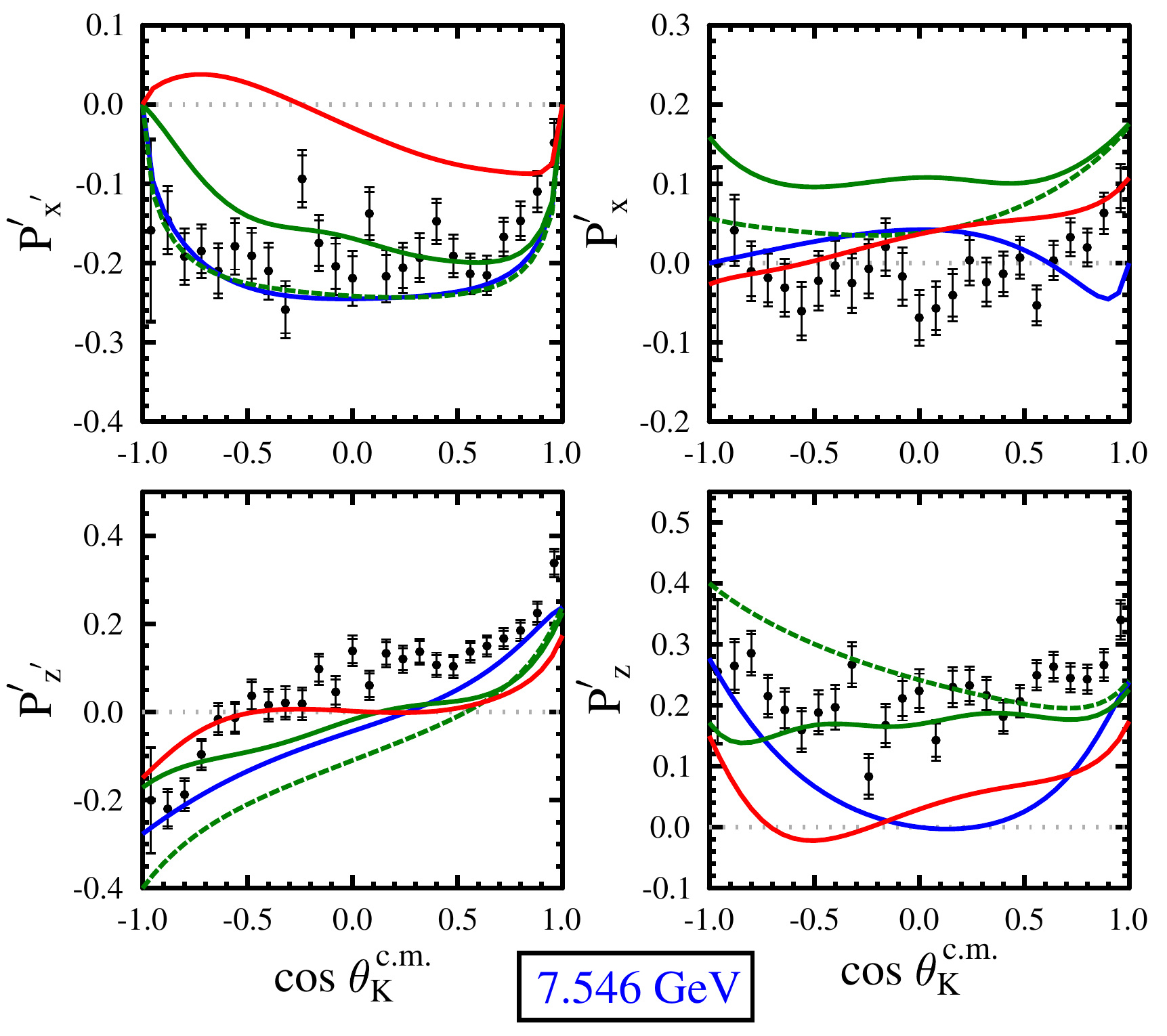}}
\vspace{-4mm}
\caption{Transferred $\Lambda$ polarization components ${\cal P}'$ with respect to the $(x',z')$ and $(x,z)$ axes vs.~$\cos \theta_K^{c.m.}$
  for a beam energy of 6.535~GeV (left) and 7.546~GeV (right). See the Fig.~\ref{ltpol1a} caption for details.}
\label{ltpol1c}
\end{figure*}

\begin{figure*}[htbp]
\centering
\includegraphics[width=0.45\textwidth]{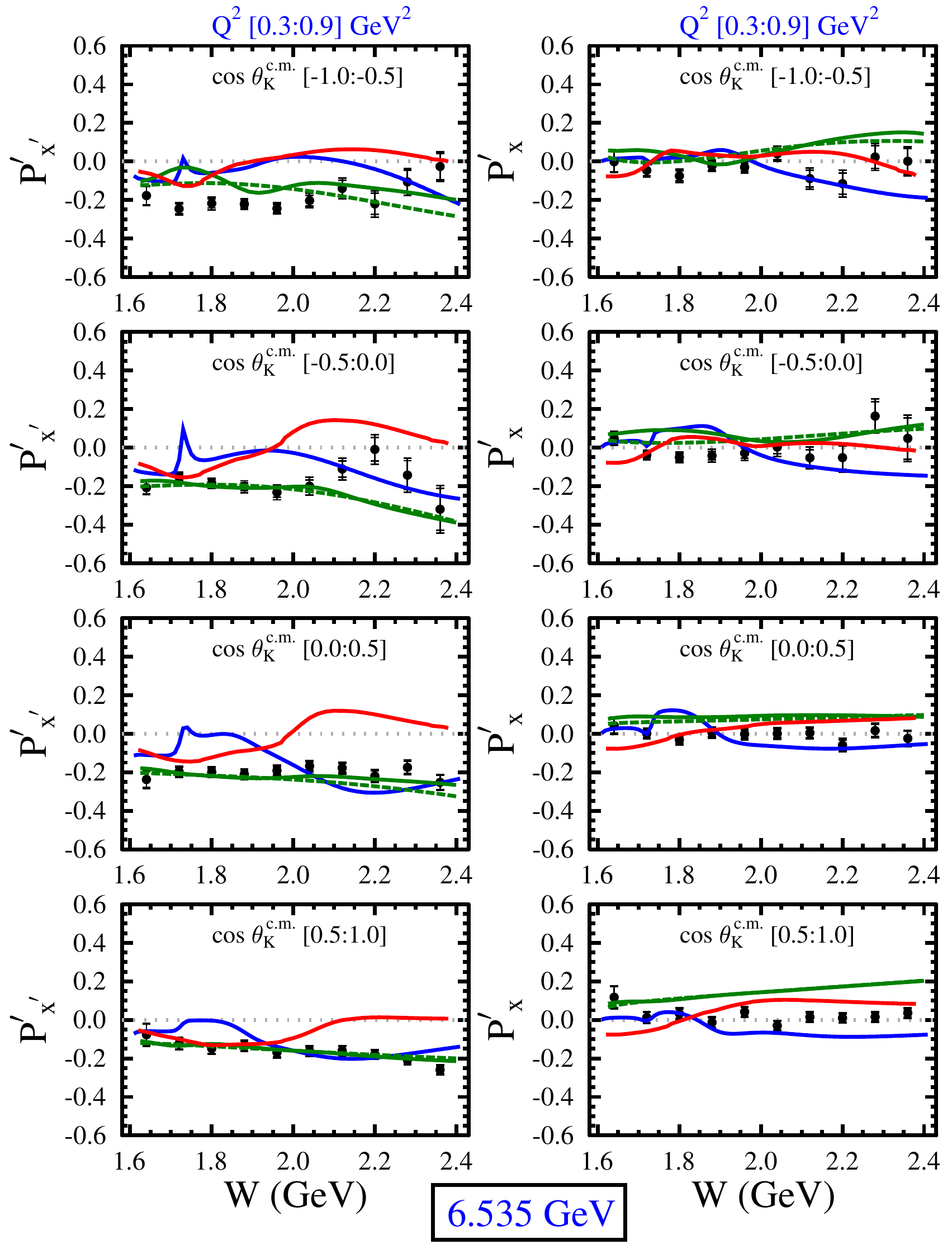}\hspace{5mm}
\includegraphics[width=0.45\textwidth]{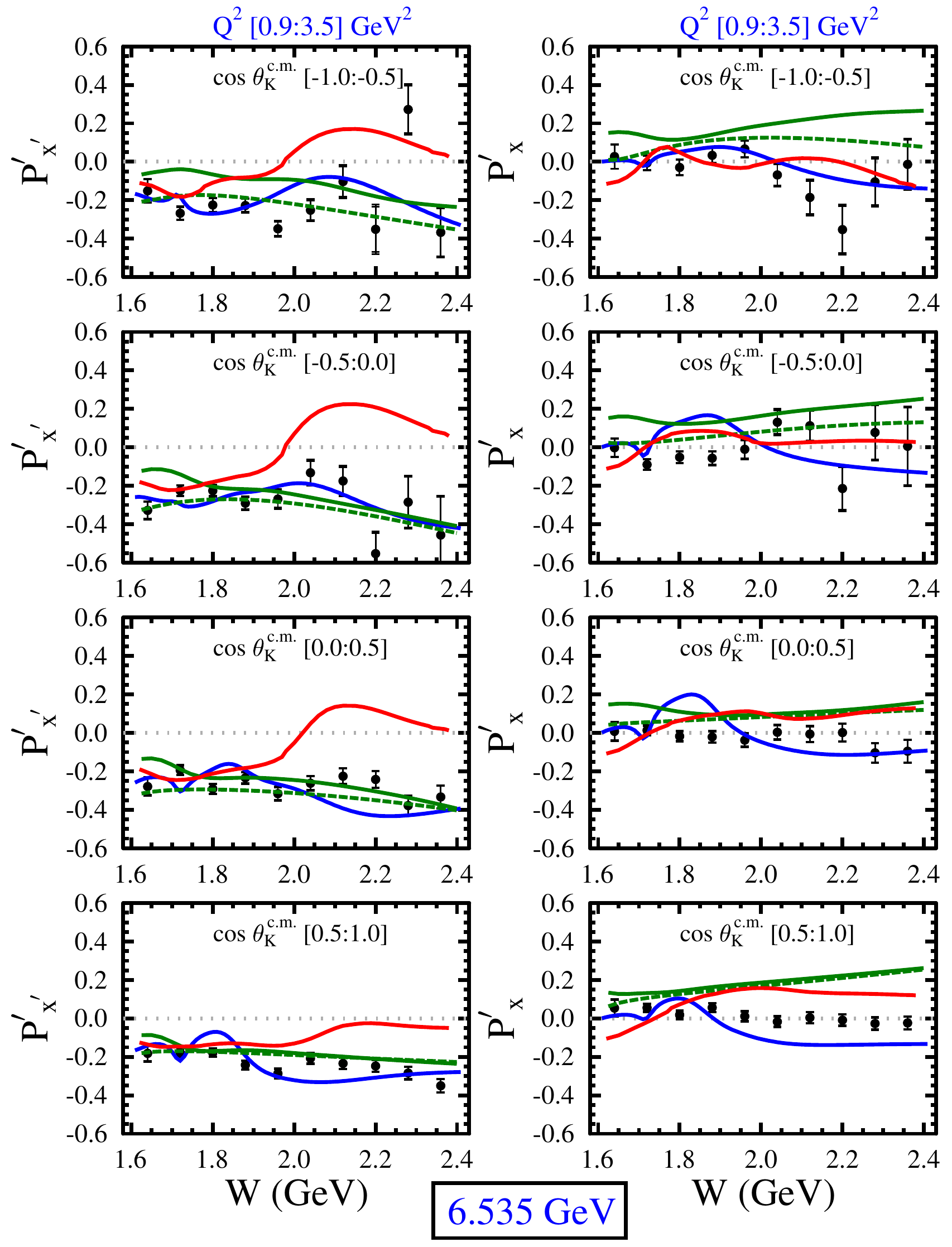}
\vspace{-4mm}
\caption{Transferred $\Lambda$ polarization components ${\cal P}'_{x'}$ and ${\cal P}'_x$ vs.~$W$ for a beam energy of 6.535~GeV. The
  data are binned in $Q^2$ from 0.3 to 0.9~GeV$^2$ (left 2$\times$4 plots) and $Q^2$ from 0.9 to 3.5~GeV$^2$ (right 2$\times$4 plots)
  for four different bins in $\cos \theta_K^{c.m.}$. In the text this is referred to as the 3D sort. The inner error bars on the data 
  points represent the statistical uncertainties and the outer error bars represent the total uncertainties. See the Fig.~\ref{ltpol1a} 
  caption for a description of the model curves.}
\label{ltpol3}
\end{figure*}

\begin{figure*}[htbp]
\centering
\includegraphics[width=0.455\textwidth]{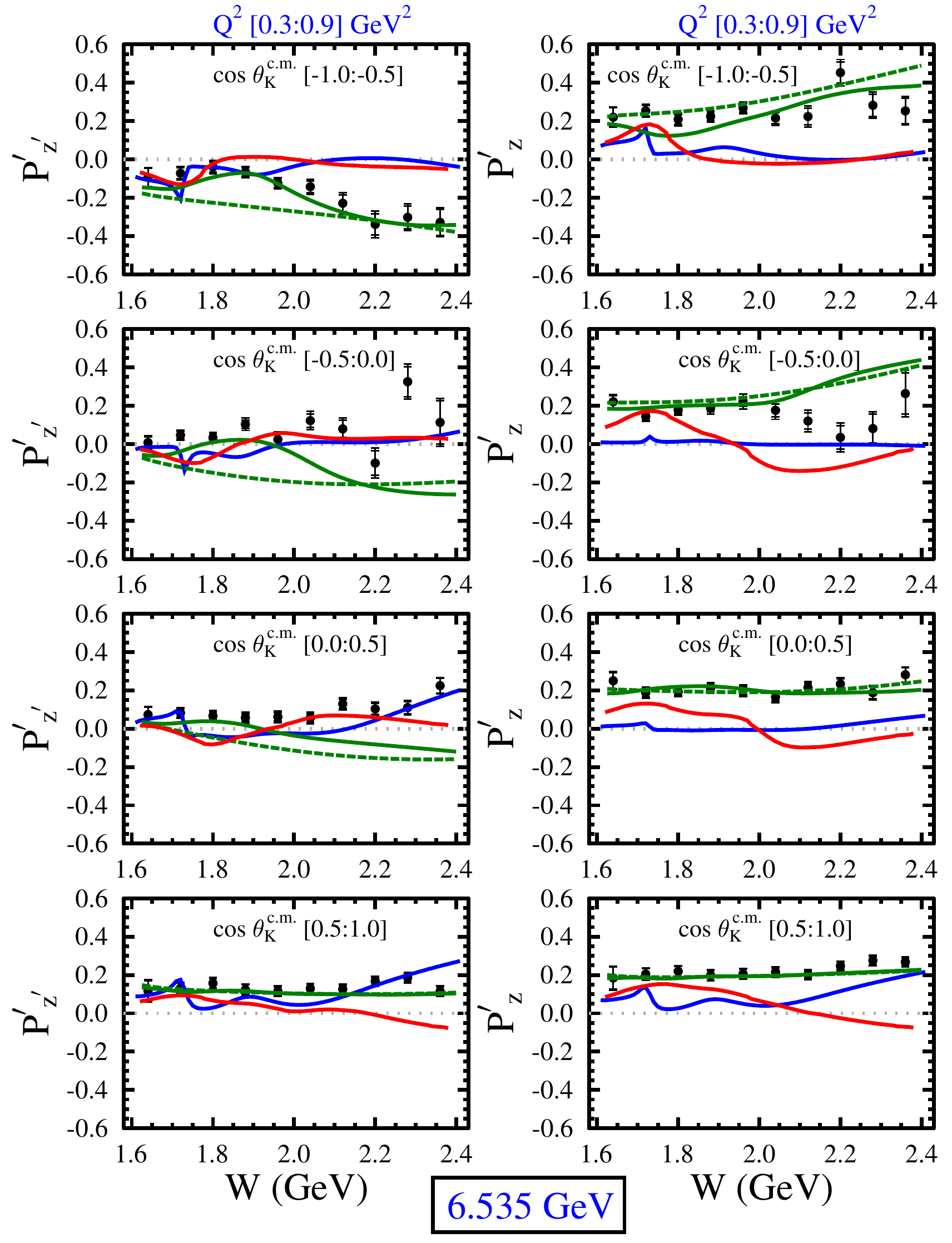}\hspace{5mm}
\includegraphics[width=0.45\textwidth]{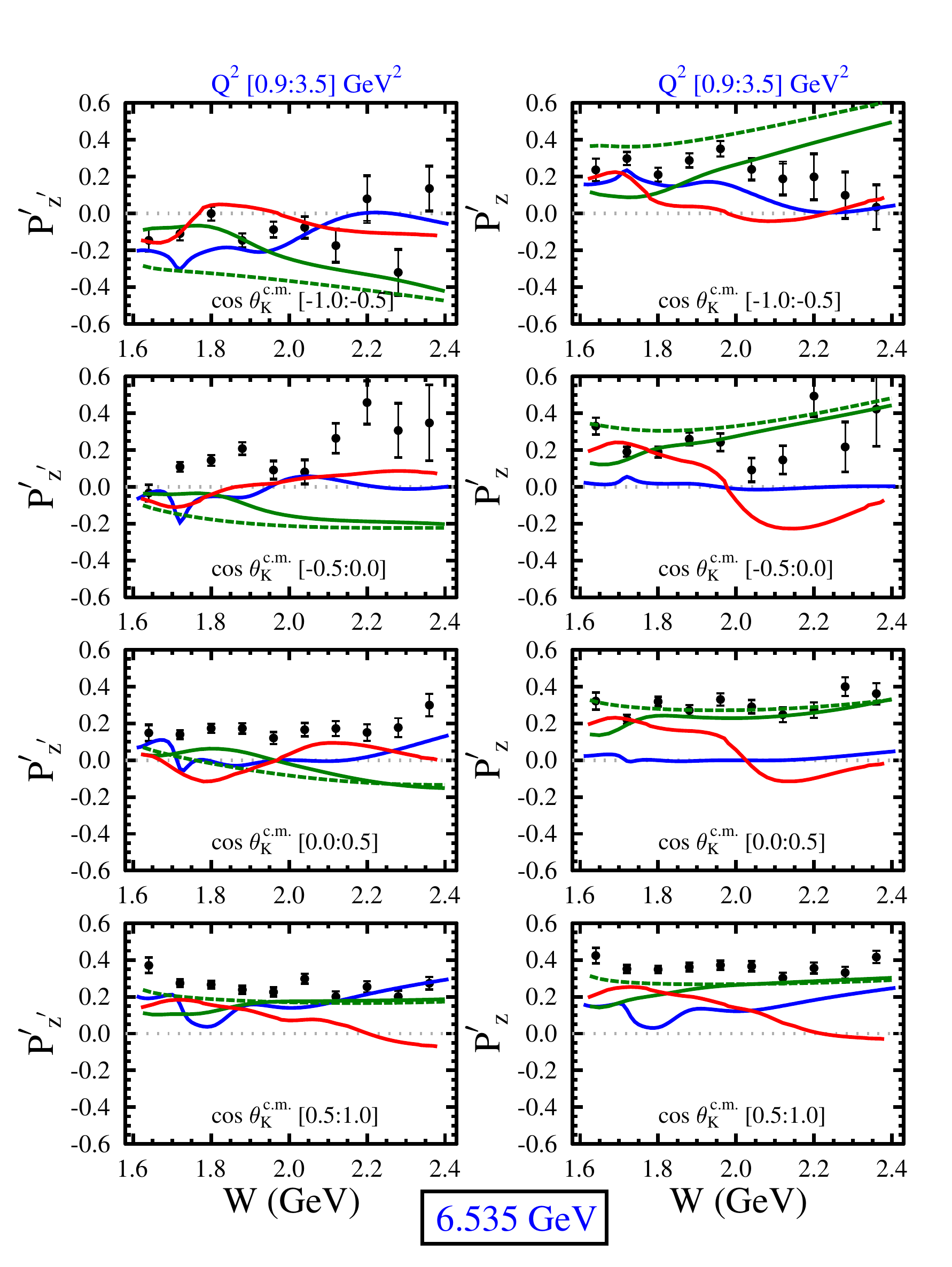}
\vspace{-4mm}
\caption{Transferred $\Lambda$ polarization components ${\cal P}'_{z'}$ and ${\cal P}'_z$ vs.~$W$ for a beam energy of 6.535~GeV. See
  the Fig.~\ref{ltpol1a} caption for a description of the model curves and the Fig.~\ref{ltpol3} caption for details on the data.}
\label{ltpol4}
\end{figure*}

The results for the beam-recoil transferred polarization to the $\Lambda$ hyperon in the $K^+\Lambda$ final state in the primed and
unprimed coordinate systems (see Fig.~\ref{coor4}) are shown for the datasets at electron beam energies of 6.535~GeV and 7.546~GeV in 
Figs.~\ref{ltpol1a} through \ref{ltpol4} compared to several model calculations. The error bars in these figures include statistical
and total uncertainties (statistical added in quadrature with the point-to-point systematics). The scale type uncertainties (due to
the asymmetry parameter and beam polarization) are not included and amount to an absolute scale uncertainty of 0.04 on the polarization. 
The full set of results is contained in the CLAS physics database~\cite{physicsdb}.

Generally speaking, in the 1D analyses shown in Figs.~\ref{ltpol1a} to ~\ref{ltpol1c} for the 6.535~GeV and 7.546~GeV datasets,
the transferred polarization to the $\Lambda$ vs.~the different kinematic variables is either relatively flat or 
smoothly/monotonically changing in magnitude. The ${\cal P}'_x$ components are consistent with zero over the full kinematic phase
space investigated and ${\cal P}'_{x'} \sim -0.2$ (flat) vs.~$W$ and $\cos \theta_K^{c.m.}$, however, it increases slowly in magnitude
vs.~$Q^2$. The components ${\cal P}'_{z'}$ and ${\cal P}'_z$ are generally positive in the range from 0 to 0.6, monotonically increasing
vs.~$Q^2$, but with a richer, more involved dependence vs.~$W$ and $\cos \theta_K^{c.m.}$, displaying a pronounced dip in ${\cal P}'_{z'}$
at $W \sim 1.9-2.0$~GeV. Both ${\cal P}'_{z'}$ and ${\cal P}'_z$ show a strong dependence on $\cos \theta_K^{c.m.}$. Within the
uncertainties the polarization components from the 6.535~GeV and 7.546~GeV datasets agree, showing a weak dependence on beam energy.

The kinematic trends in these observables are reasonably consistent with the CLAS analyses of these same observables acquired at beam 
energies of 2.567~GeV, 4.261~GeV, and 5.754~GeV in Refs.~\cite{carman03,carman09}. However, the present data have reduced statistical 
uncertainties and much improved coverage for $\cos \theta_K^{c.m.} < 0$, a region where the relative strength of $s$-channel contributions 
grows relative to the $t$-channel contributions that dominate at more forward $\theta_K^{c.m.}$ angles, and where effects from $u$-channel 
processes may emerge.

The present dataset from CLAS12 is valuable as it has sufficient statistics to enable a meaningful multi-dimensional analysis
for the first time for this observable. This is referred to in this work as the 3D analysis with binning as defined in
Section~\ref{binning}. Figures~\ref{ltpol3} and \ref{ltpol4} show the results of the 3D analysis of the 6.535~GeV dataset for the
beam-recoil $\Lambda$ polarization vs.~$W$ for two $Q^2$ bins and 4 equal-size $\cos \theta_K^{c.m.}$ bins from $-1 \to 1$.
Figure~\ref{ltpol3} shows that for the ${\cal P}'_{x'}$ and ${\cal P}'_x$ components, the general trends seen in the 1D analysis are
followed here with no strong $\cos \theta_K^{c.m.}$ dependence. Figure~\ref{ltpol4} shows that ${\cal P}'_{z'}$ has a strong dependence
vs.~$\cos \theta_K^{c.m.}$ with ${\cal P}'_{z'}$ negative at backward angles and positive at forward angles. However, ${\cal P}'_z$ is
flat vs.~$W$ and relatively independent vs.~$\cos \theta_K^{c.m.}$. 

The further increase in statistics foreseen from the full CLAS12 RG-K $K^+Y$ dataset will allow us to decrease the bin sizes over $Q^2$,
$W$, and $\cos \theta_K^{c.m.}$. This is necessary for the eventual extraction of the nucleon resonance electroexcitation amplitudes from
analysis of the data binned in 3D space.

\boldmath
\subsection{$\Sigma^0$ Polarization Transfer}
\label{sigma}
\unboldmath

\begin{figure*}[htbp]
\centering
\includegraphics[width=0.453\textwidth]{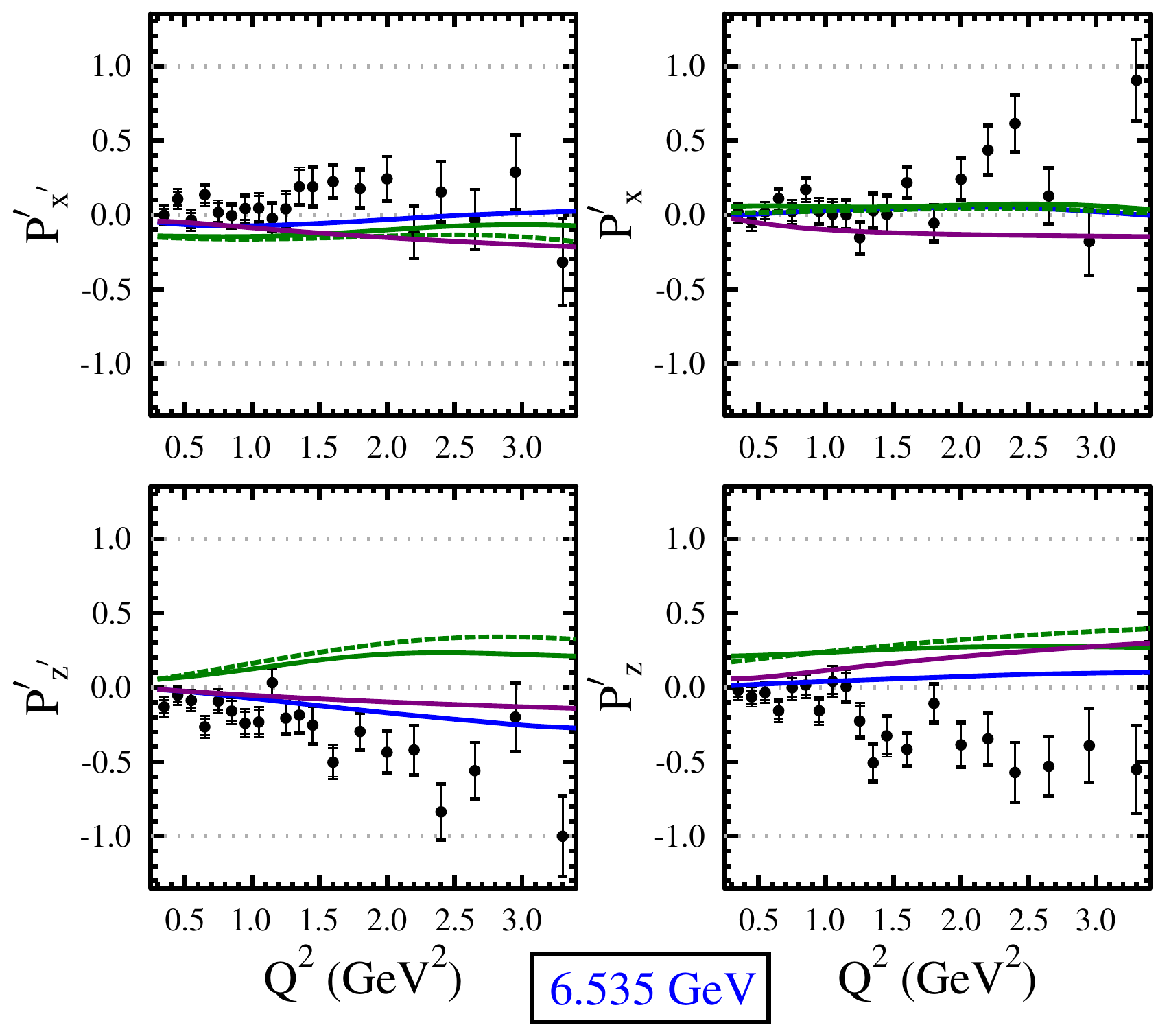}
\includegraphics[width=0.45\textwidth]{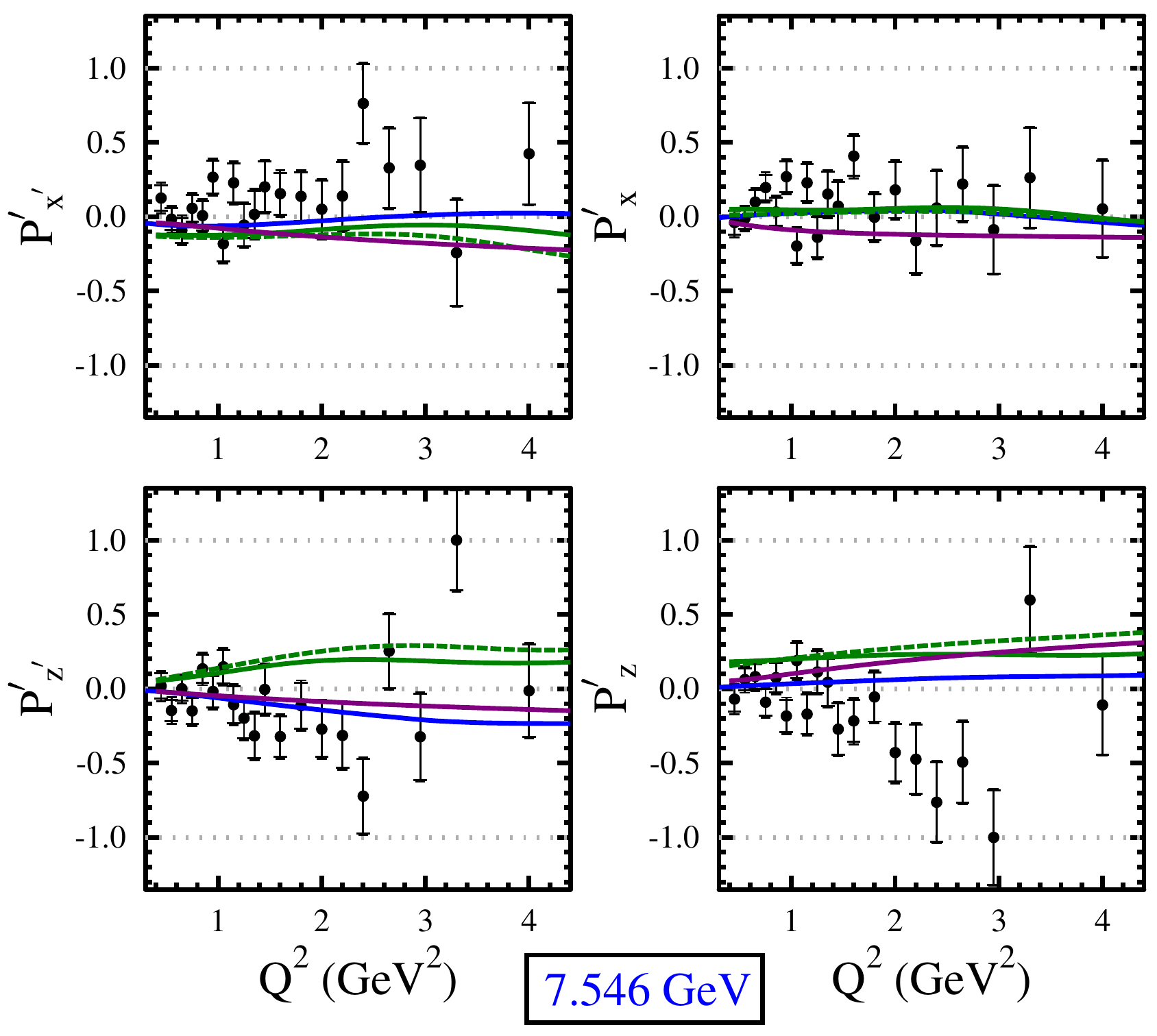}
\vspace{-4mm}
\caption{Transferred $\Sigma^0$ polarization components ${\cal P}'$ with respect to the $(x',z')$ and $(x,y)$ axes vs.~$Q^2$ for a
  beam energy of 6.535~GeV (left) and 7.546~GeV (right). The data are limited to $Q^2$ from 0.3 to 3.5~GeV$^2$ (6.535~GeV) and from
  0.4 to 4.5~GeV$^2$ (7.546~GeV), and $W$ from 1.625 to 2.4~GeV. In the text this is referred to as the 1D sort (see
  Section~\ref{binning} for details). The inner error bars on the data points represent the statistical uncertainties and the outer
  error bars represent the total uncertainties. The curves are calculations from RPR~\cite{rpr} (solid green - full RPR-2007 model, 
  dashed green - RPR-2007 model with resonance terms off), Kaon-MAID~\cite{kaon-maid1,kaon-maid2,kaon-maid3} (solid blue), and
  SL~\cite{saclay-lyon} (solid purple).}
\label{stpol1a}
\end{figure*}

\begin{figure*}[htbp]
\centering
\includegraphics[width=0.45\textwidth]{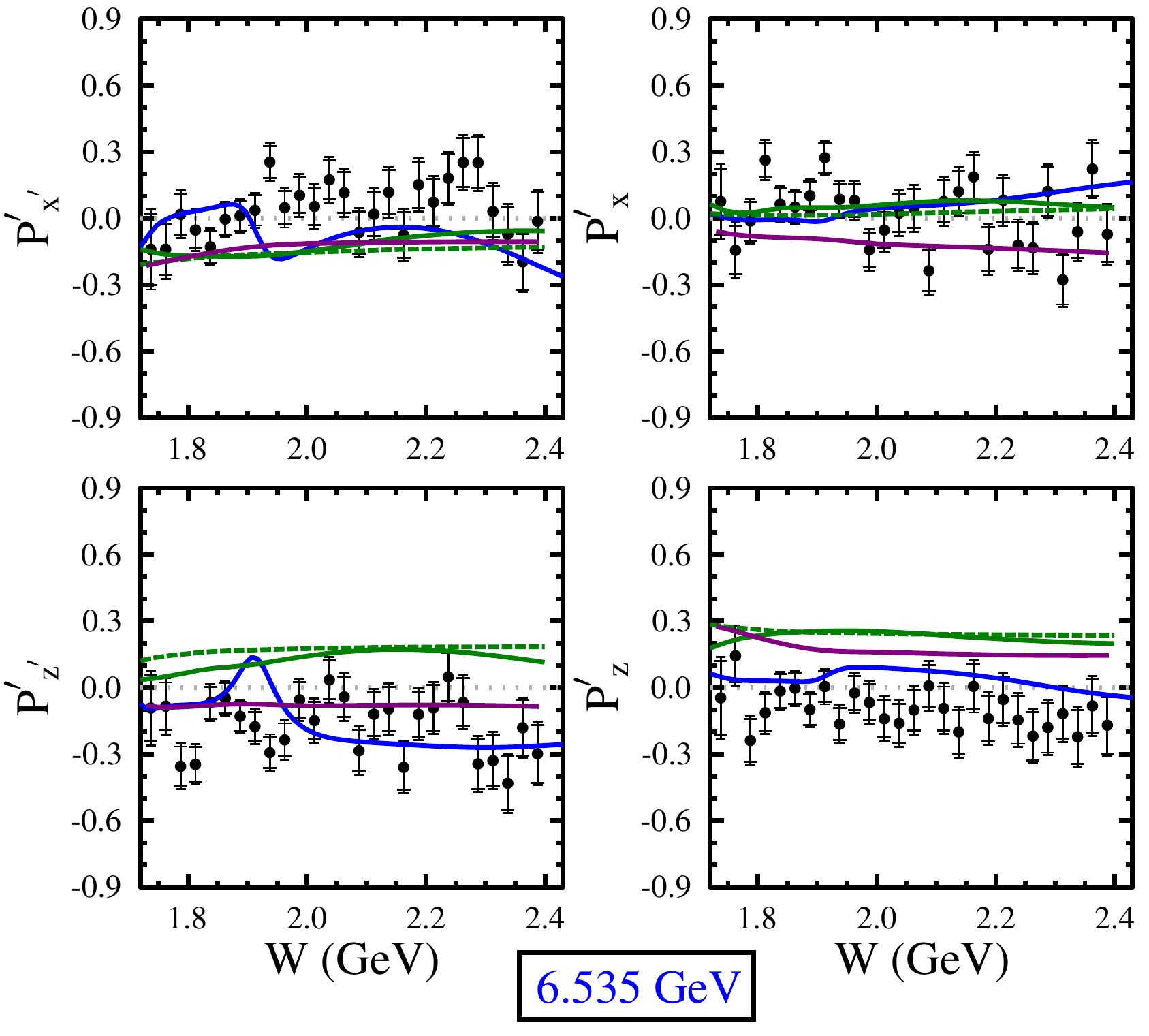}
\includegraphics[width=0.451\textwidth]{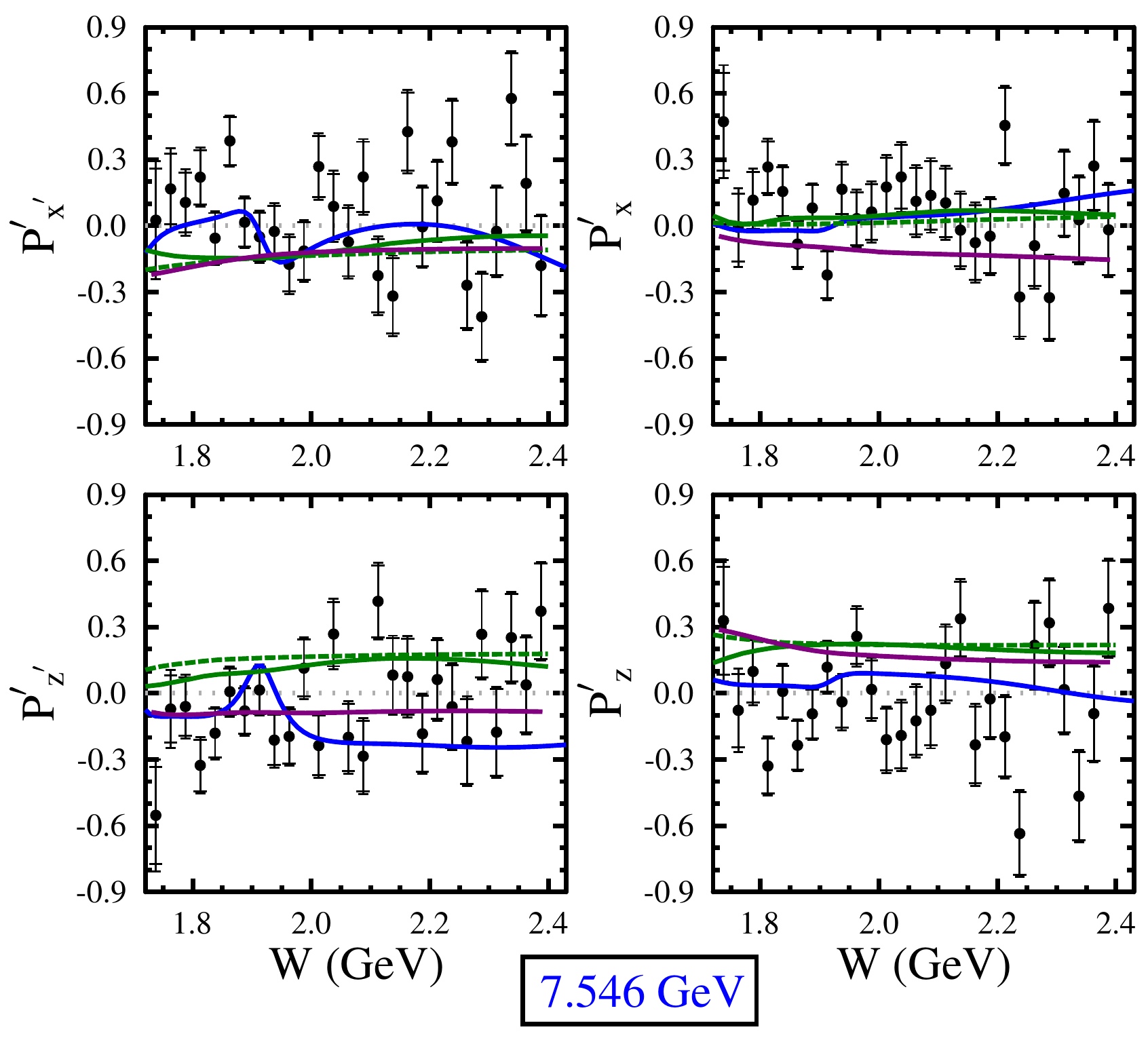}
\vspace{-4mm}
\caption{Transferred $\Sigma^0$ polarization components ${\cal P}'$ with respect to the $(x',z')$ and $(x,y)$ axes vs.~$W$ for a beam
  energy of 6.535~GeV (left) and 7.546~GeV (right). See the Fig.~\ref{stpol1a} caption for a description of the model curves.}
\label{stpol1b}
\end{figure*}

\begin{figure*}[htbp]
\centering
\includegraphics[width=0.45\textwidth]{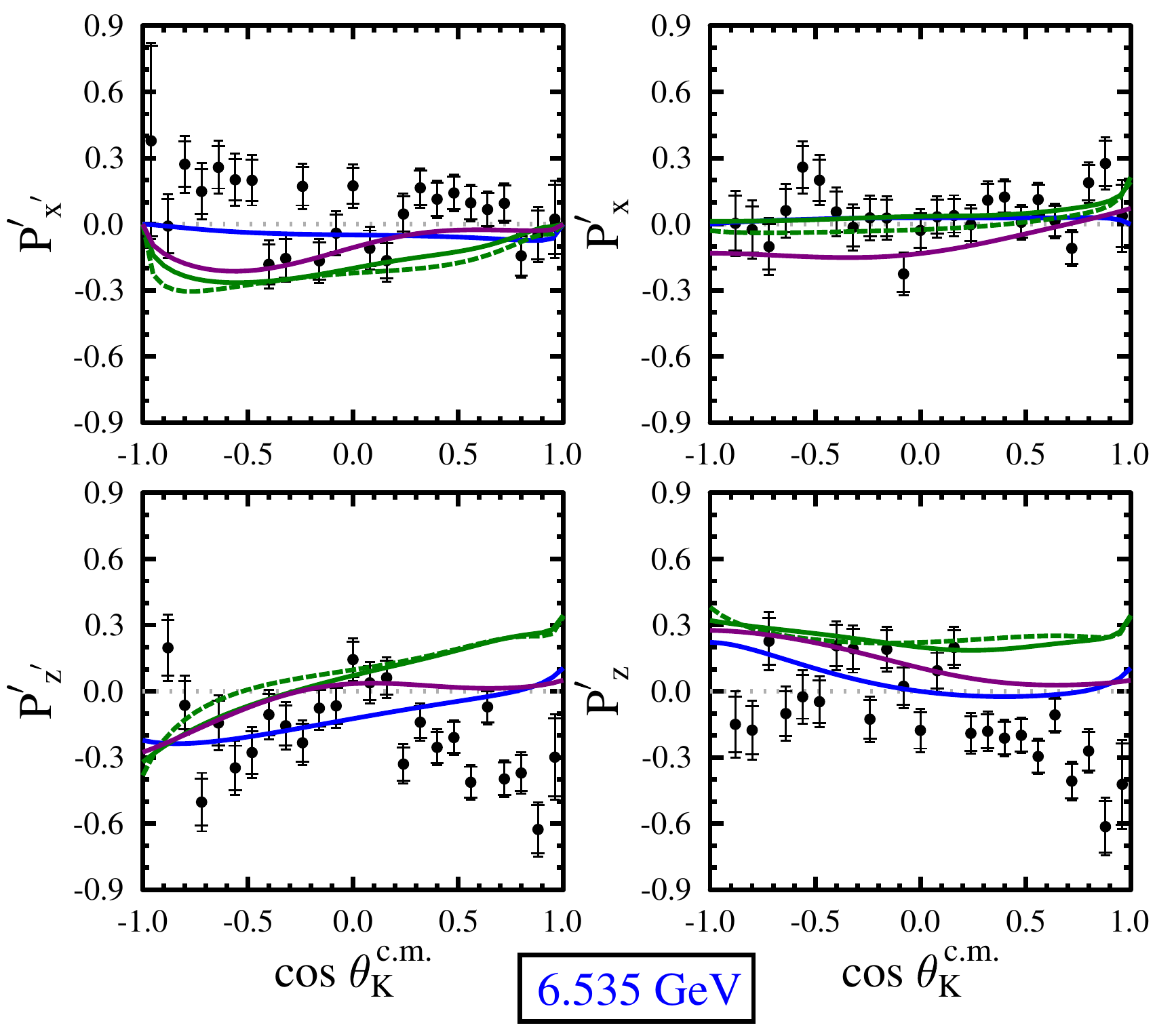}
\includegraphics[width=0.45\textwidth]{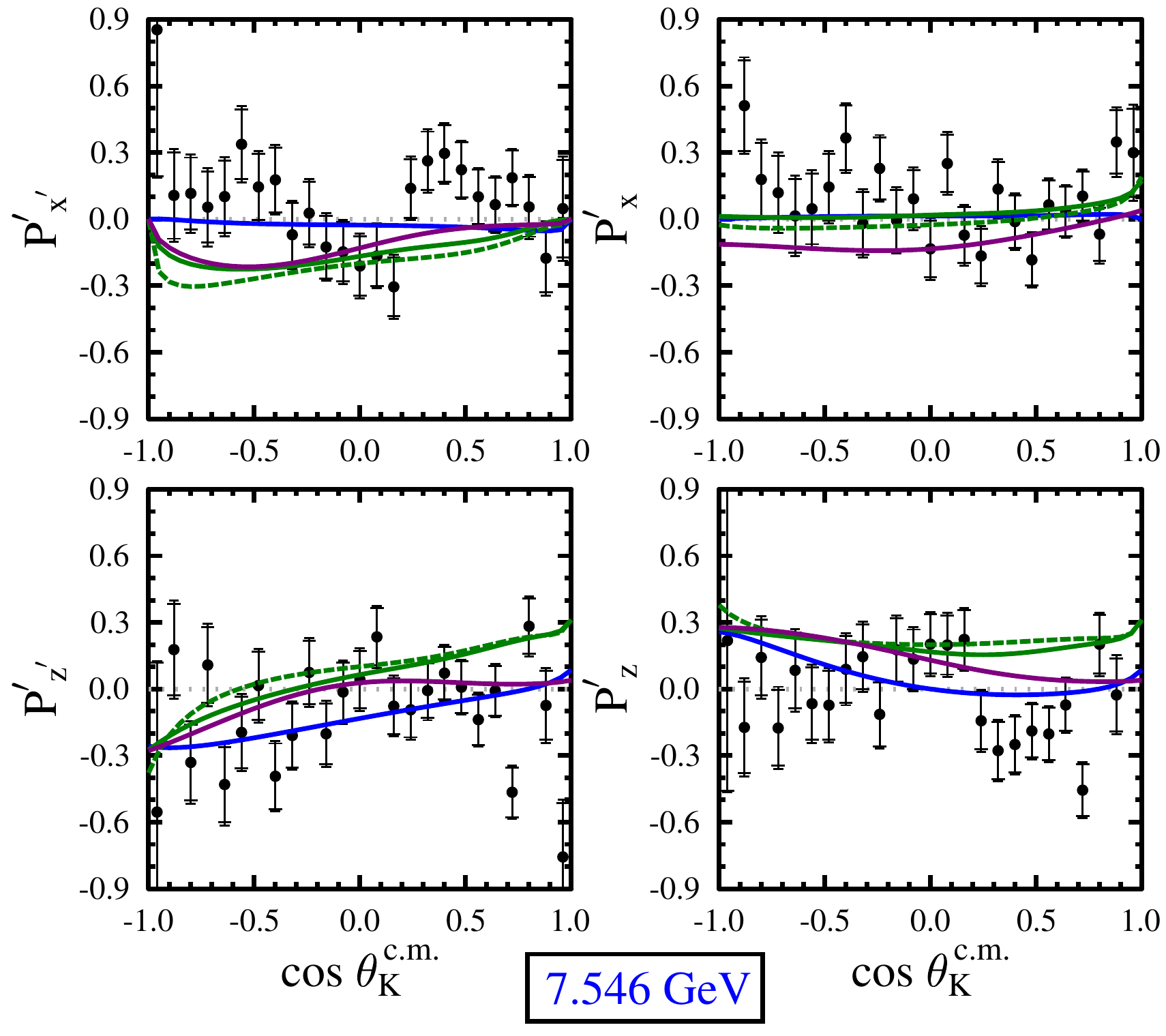}
\vspace{-4mm}
\caption{Transferred $\Sigma^0$ polarization components ${\cal P}'$ with respect to the $(x',z')$ and $(x,y)$ axes vs.~$\cos \theta_K^{c.m.}$
  for a beam energy of 6.535~GeV (left) and 7.546~GeV (right). See the Fig.~\ref{stpol1a} caption for a description of the model curves.}
\label{stpol1c}
\end{figure*}

\begin{figure*}[htbp]
\centering
\includegraphics[width=0.45\textwidth]{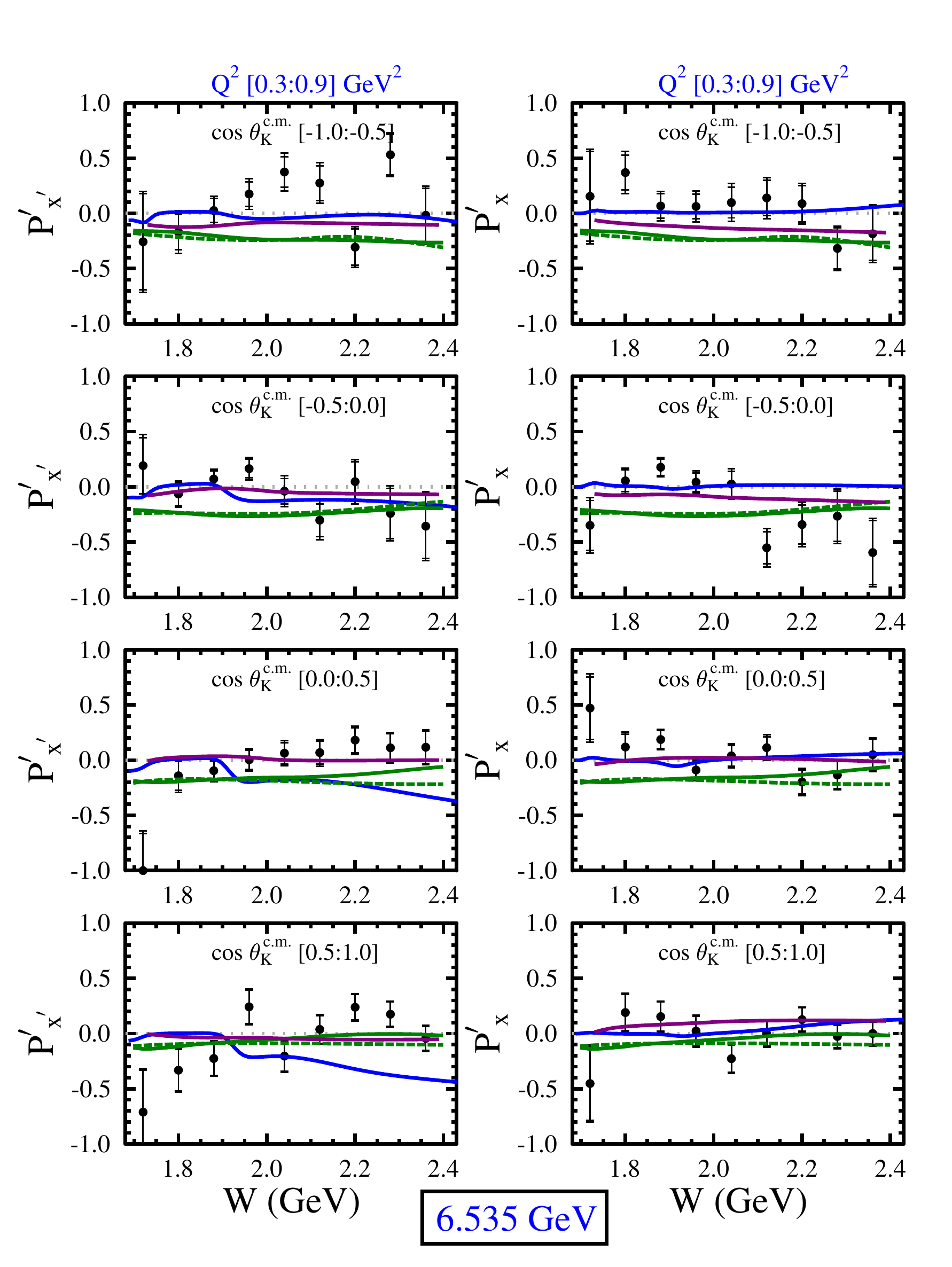}\hspace{5mm}
\includegraphics[width=0.45\textwidth]{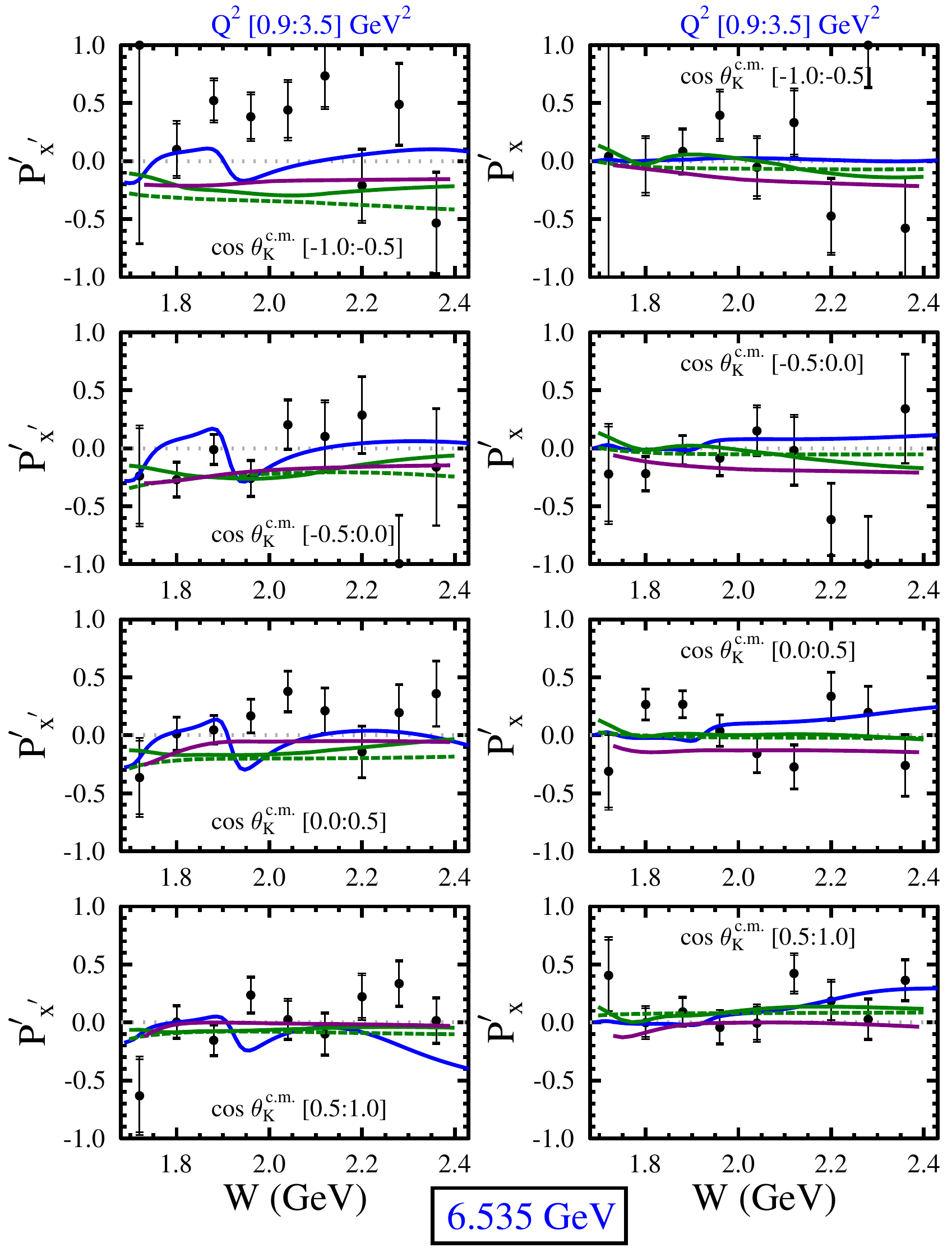}
\vspace{-4mm}
\caption{Transferred $\Sigma^0$ polarization components ${\cal P}'_{x'}$ and ${\cal P}'_x$ vs.~$W$ for a beam energy of 6.535~GeV. The
  data are binned in $Q^2$ from 0.3 to 0.9~GeV$^2$ (left 2$\times$4 plots) and $Q^2$ from 0.9 to 3.5~GeV$^2$ (right 2$\times$4 plots)
  for four different bins in $\cos \theta_K^{c.m.}$. In the text this is referred to as the 3D sort. The inner error bars on the data 
  points represent the statistical uncertainties and the outer error bars represent the total uncertainties. See the Fig.~\ref{stpol1a} 
  caption for a description of the model curves.}
\label{stpol3}
\end{figure*}

\begin{figure*}[htbp]
\centering
\includegraphics[width=0.45\textwidth]{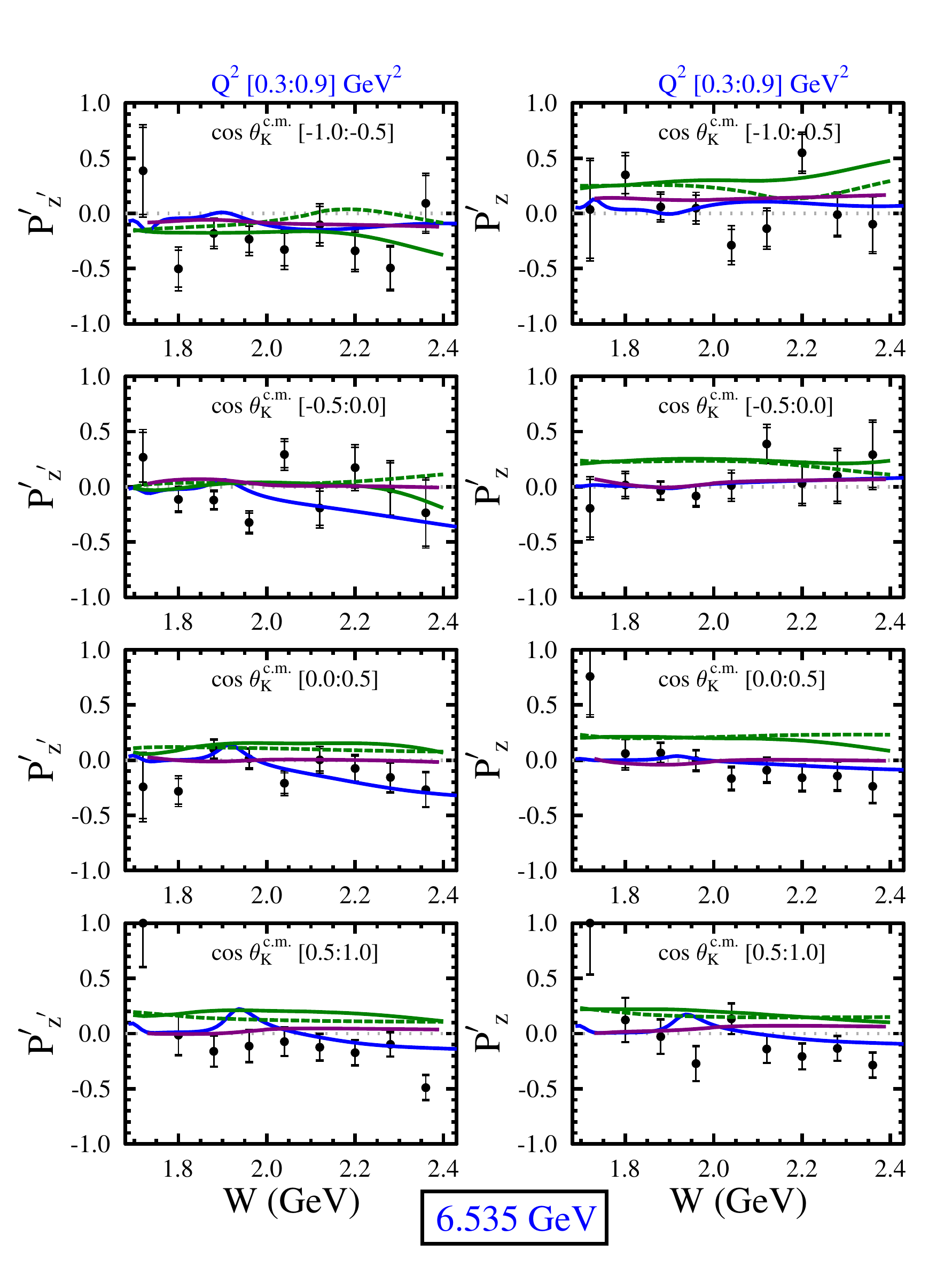}\hspace{5mm}
\includegraphics[width=0.455\textwidth]{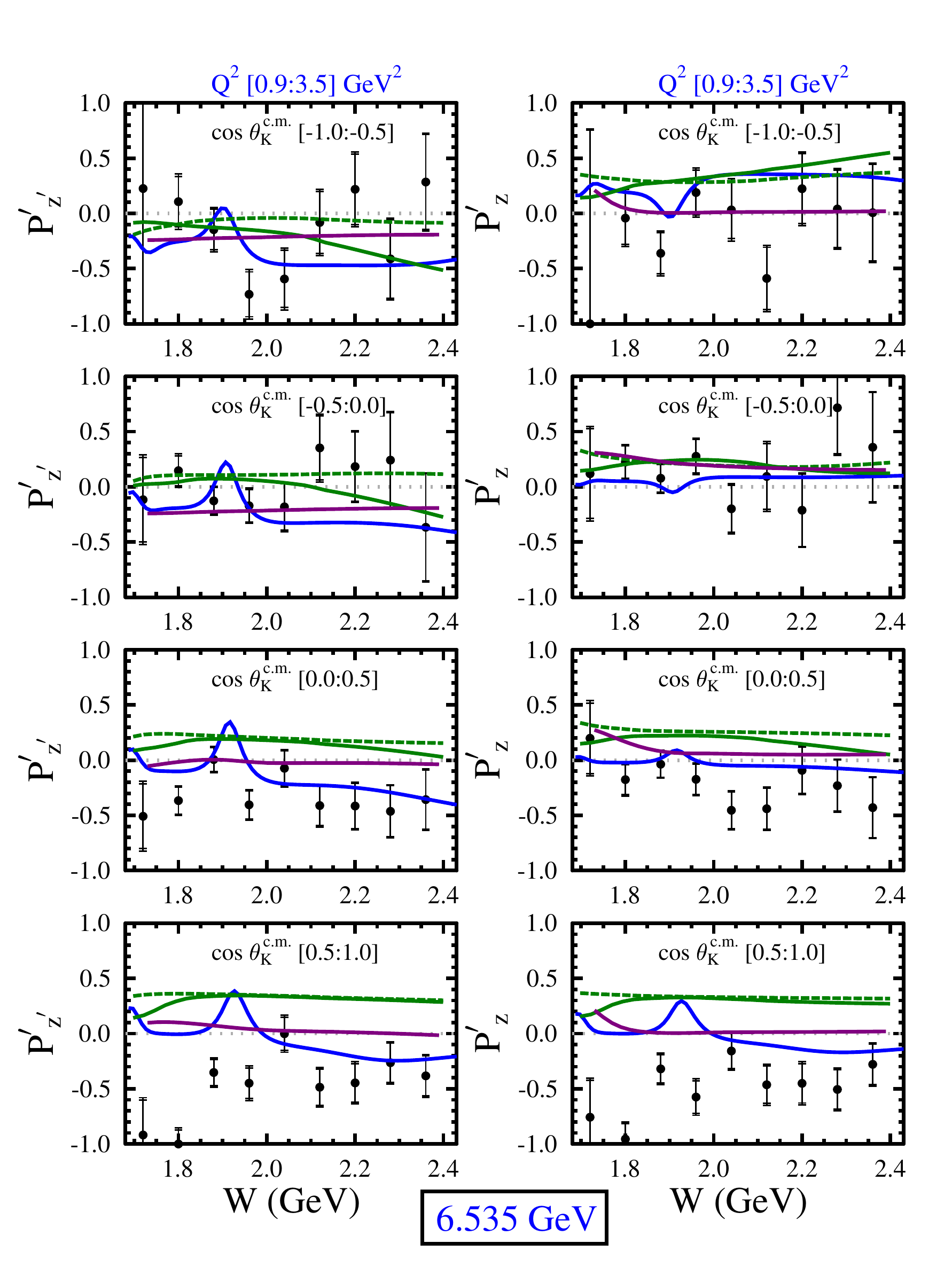}
\vspace{-4mm}
\caption{Transferred $\Sigma^0$ polarization components ${\cal P}'_{z'}$ and ${\cal P}'_z$ vs.~$W$ for a beam energy of 6.535~GeV. See
  the Fig.~\ref{stpol1a} for a description of the model curves and the Fig.~\ref{stpol3} caption for details on the data.}
\label{stpol4}
\end{figure*}

The results for the beam-recoil transferred $\Sigma^0$ polarization for the 6.535~GeV and 7.546~GeV datasets are shown in
Figs.~\ref{stpol1a} through \ref{stpol4} compared to several model calculations. The error bars in these figures include statistical
and total uncertainties (statistical + point-to-point systematic). The data uncertainties also include an overall scale uncertainty
of 0.04 on the polarization. The full set of results is contained in the CLAS physics database~\cite{physicsdb}.

These transferred $\Sigma^0$ polarization data are less sensitive to the detailed kinematic dependence of the observables compared to
the $\Lambda$ polarization components shown in Section~\ref{lambda} due to the larger statistical uncertainties. As seen in the 1D
analysis of Figs.~\ref{stpol1a} to \ref{stpol1c}, the components ${\cal P}'_{x'}$ and ${\cal P}'_x$ are largely consistent with zero vs.
$Q^2$, $W$, and $\cos \theta_K^{c.m.}$ within the uncertainties. The ${\cal P}'_{z'}$ and ${\cal P}'_z$ components are relatively flat vs.
$W$ and $\cos \theta_K^{c.m.}$ with ${\cal P}'_{z',z} \approx -0.2$. The $Q^2$ dependence of ${\cal P}'_{z'}$ and ${\cal P}'_z$ is consistent
with a shallow increase in magnitude with increasing $Q^2$. Despite the limitations of these $\Sigma^0$ polarization observables, they
should ultimately prove valuable as they effectively represent the first substantive measurement of this observable given the very low
statistics in the CLAS measurement included in Ref.~\cite{carman09} that was barely sufficient to determine the sign of the polarization.

Figures~\ref{stpol3} and \ref{stpol4} show the 3D analysis of the beam-recoil $\Sigma^0$ polarization from the 6.535~GeV dataset with
binning as detailed in Section~\ref{binning}. These data reveal trends very much in accord with the general observations noted for the
1D data sort. Both ${\cal P}'_{x'}$ and ${\cal P}'_x$ shown in Fig.~\ref{stpol3} and ${\cal P}'_{z'}$ and ${\cal P}'_z$ shown in
Fig.~\ref{stpol4} are relatively flat with $W$ and the components show a gradual, shallow increase in polarization going from forward
to backward angles.

\subsection{Model Comparisons}
\label{models}

There are several different single channels models shown in this work to compare against the polarization observables. In this section the 
main features of the different models are discussed to set the stage for the their comparisons to the data.

\vskip 0.2cm

\underline{Kaon-MAID}:
Kaon-MAID is a tree-level isobar model~\cite{kaon-maid1,kaon-maid2,kaon-maid3} that includes Born terms, $K^*(892)$ and $K_1(1290)$
exchanges in the $t$-channel, and a limited set of spin 1/2 and 3/2 $s$-channel resonances. These include the $N(1650)1/2^-$,
$N(1710)1/2^+$, and $N(1720)3/2^+$, along with the $N(1900)3/2^+$ for $K^+\Lambda$ and the $\Delta(1900)1/2^-$ and $\Delta(1910)1/2^+$ for
$K^+\Sigma^0$. The Born, vector meson, and resonance couplings are based on fits to the $\gamma p \to K^+Y$ and $\pi^-p \to K^0\Lambda$
data available in the late 1990s when the model was developed. Kaon-MAID is not constrained by any $K^+Y$ electroproduction data. This
isobar model, like most of the others described below, leaves the resonant term couplings as free parameters in fits to the data. The 
couplings are required to respect the limits imposed by SU(3), allowing for symmetry breaking at the level of about 20\%. The inclusion 
of hadronic form factors, with cut-off values fixed by the data, leads to a breaking of gauge invariance that is restored by the inclusion 
of non-resonant counter terms.

The Kaon-MAID model shows very sharp features in the $W$ dependence of ${\cal P}'_\Lambda$ and ${\cal P}'_\Sigma$ from the resonance terms 
that are not seen in the data. However, the dependence of ${\cal P}'$ vs.~$Q^2$ and $\cos \theta_K^{c.m.}$ varies smoothly. The model generally 
reproduces the polarization sign and qualitative features of the data. However, the model mainly fails to describe the kinematic dependence 
of ${\cal P}'_{z'}$ and ${\cal P}'_z$ for the $K^+\Lambda$ channel, defined by the $R_{TT'}^{x'0}$ and $R_{TT'}^{z'0}$ response functions. 
This model is archived online~\cite{kaon-maid} and the results included were integrated over the finite bins of this work by its
developer~\cite{mart-comm}. 

\vskip 0.2cm

\underline{Saclay-Lyon}
The Saclay-Lyon (SL) isobar model~\cite{saclay-lyon} is similar to the Kaon-MAID model with the same kaon resonances and SU(3) constraints 
on the main coupling constants. The model version used is limited to the inclusion of only spin 1/2 and 3/2 $s$-channel resonances
that match what is included in Kaon-MAID. It differs in that instead of hadronic form factors, this model includes a number of $u$-channel
terms to counterbalance the strength of the Born terms. As was the case for the Kaon-MAID model, the data used to constrain the parameters
of the SL model were very limited given that it was developed before the release of any of the data produced from CLAS. In this work the
SL model is shown only for the $K^+\Sigma^0$ data.

The SL model should not be expected to match the hyperon polarizations well given the lack of data available for constraints. The quality
of its match to the data is similar to that from the Kaon-MAID model and is no worse than later models developed based on fits to the 
photoproduction data from CLAS. Of course, without proper constraints from data at finite $Q^2$, there should be no expectation of good 
agreement from this archival model. The SL calculations were provided by Ref.~\cite{byd-comm} and were integrated over the finite bins of 
this work.

\vskip 0.2cm

\underline{RPR}:
The hybrid Regge plus Resonance (RPR) model was developed by the Ghent group~\cite{rpr}, and is based on a tree-level effective Lagrangian
model for $K^+\Lambda$ and $K^+\Sigma^0$ photoproduction from the proton. It differs from traditional isobar approaches in its description
of the non-resonant diagrams, which involve the exchange of $K$ and $K^*$ Regge trajectories. The RPR model includes all well-established
$s$-channel resonances below 2~GeV. The two variants of the RPR model included (RPR-2011 for the $K^+\Lambda$ channel and RPR-2007 for
the $K^+\Sigma^0$ channel) have been constrained by fits to the CLAS $\gamma p \to K^+Y$ photoproduction data with no constraints from the
CLAS $K^+Y$ electroproduction data. This model is archived online~\cite{rpr-online} and the calculations based on this model were integrated
over the finite bins of this work using the output from the webpage. The RPR model calculations shown here include the full calculations 
with all contributions turned on and a version with the $s$-channel resonances turned off (amounting effectively to a pure Regge model).

The RPR model varies smoothly vs.~kinematics for both $K^+Y$ final states. For $K^+\Lambda$ there is agreement of the RPR-2011 model with
the ${\cal P}'$ sign of the data but the magnitude of the polarization is not in accord with the data. As shown in Figs.~\ref{ltpol3} and
\ref{ltpol4}, accounting for the resonant contributions provides a reasonable description of our results on ${\cal P}'$ at $W < 2.0$~GeV 
and $Q^2$ from 0.3 to 0.9~GeV$^2$, although discrepancies are apparent in the range $W > 2$~GeV and for the $Q^2$ bin from 0.9 to 
3.5~GeV$^2$. For $K^+\Sigma^0$ the model agrees reasonably well with the small polarization magnitudes of the data. The model versions with 
the resonances turned on do not agree any better with the data than the versions with the resonances turned off.

\vskip 0.2cm

\underline{BS3}:
The Byd\v{z}ovsk\'{y}-Skoupil model (BS3)~\cite{skoupil} is another tree-level isobar model similar in design to the models detailed above. 
However, it represents a significant evolution beyond the 20 year old Kaon-MAID and SL models and the 10 year old RPR model in that it was 
based on fits to some of the available $\gamma p \to K^+\Lambda$ photoproduction data (differential cross sections, recoil polarization, 
beam spin asymmetry) and to some of the available $ep \to e'K^+\Lambda$ electroproduction data ($\sigma_U$, $\sigma_T$, $\sigma_L$, 
$\sigma_{LT'}$) from CLAS. The full set of 3- and 4-star PDG $N^*$ and $\Delta^*$ resonances of spins up to 5/2 and $W$ up to 2~GeV are 
included. Like the other isobar models, it includes Born terms and exchanges in the $t$- and $u$-channels to account for the non-resonant 
backgrounds. The BS3 model is presently only available for the $K^+\Lambda$ final state.

The BS3 model, like the other isobar models included in this work, qualitatively accounts for the sign and kinematic trends of the 
polarization observables. However, it does not provide any better description of the data compared to the existing models. Given that the 
response functions relevant for the beam-recoil transferred polarization in BS3 have not been constrained by any existing data, perhaps
this is not so surprising. The comparisons of the model predictions to the data show that the model parameters for the form factors and
coupling constants could be improved if it were to include these new data as part of its constraints. The BS3 calculations were provided
by Ref.~\cite{byd-comm} and were integrated over the finite bins of this work.

\vskip 0.2cm

None of the models included is able to reproduce the kinematic dependence seen in the data with their current parameters. Given that 
they were mainly determined by the CLAS $K^+Y$ photoproduction data, these new electroproduction data, in addition to the full set of 
existing $K^+Y$ electroproduction cross section and polarization observables from CLAS detailed in Section~\ref{intro}, should serve to 
provide improved constraints. When the remainder of the data from this experiment is collected in the near future, amounting to roughly a 
factor of ten increase from what is included here, much improved statistical precision with reduced bin sizes in $Q^2$, $W$, and 
$\cos \theta_K^{c.m.}$ will be possible, which can be expected to shed light on the presence of additional mechanisms that are relevant 
for electroproduction. These mechanisms may gradually emerge with increasing $Q^2$ or be related to the contribution from the amplitudes 
for longitudinally polarized photons that are absent in photoproduction. Further tests turning individual $N^*$ states on and off within 
these models could also provide insight into how individual states affect the polarization transfer observables. Finally, we note that as 
the models have not been fit to electroproduction data, the $Q^2$ dependence of the form factors is not well constrained, and for these 
$K^+Y$ models to advance, a realistic $Q^2$ dependence will have to be included. These form factors have been determined for $N^*$ states 
up to $W \approx 1.8$~GeV based on analysis of $\pi N$, $\eta p$, and $\pi \pi N$ data from CLAS~\cite{fbs-carman}. Ultimately, however, 
it will be important to move beyond the single-channel models to include the full dynamics from coupled-channel approaches that make 
possible a combined global analysis of all available data on exclusive meson photo-, electro-, and hadroproduction.


\section{Summary and Conclusions}
\label{summary}

In this paper the beam-recoil transferred polarization for the electroproduction of the $K^+\Lambda$ and $K^+\Sigma^0$ final states from
a proton target at beam energies of 6.535~GeV and 7.546~GeV are presented based on analysis of data from CLAS12 taken in Dec. 2018.
The observables were measured in the nucleon resonance region spanning the kinematic range of $Q^2$ from 0.3-4.5~GeV$^2$, $W$ from 1.6
to 2.4~GeV, and covering the full center-of-mass phase space of the final state $K^+$. The $\Lambda$ polarization measurements presented
in this work extend the available data from the CLAS program. However, the data for the $\Sigma^0$ hyperon represent the first statistically
meaningful dataset available to date.

These new CLAS12 data have been compared to predictions from several available single-channel models that have varying sensitivities to
the $s$-channel resonance contributions. The different models mainly account for the sign of the hyperon polarization and qualitatively
reproduce at least some of the kinematic trends vs.~$Q^2$, $W$, and $\cos \theta_K^{c.m.}$ for the two different coordinate systems connected
to the hadronic production plane and the electron scattering plane. However, a detailed comparison shows that these new data from CLAS12
will allow for improved constraints on any reaction model. It is also important to consider that reaction models whose development is
based only on the fits to the available $\gamma p \to K^+Y$ photoproduction data are not able to reproduce the electroproduction data. A
proper reaction model will necessarily require a simultaneous fit to both $K^+Y$ photo- and electroproduction data over the broad
kinematic range of the available data. Analyses of the CLAS12 data within a broad $Q^2$ range will allow us to establish the additional
mechanisms contributing to $KY$ electroproduction that cannot be seen in photoproduction. These new mechanisms can be related either
with the longitudinal electroproduction amplitudes or emerge gradually as $Q^2$ increases. Accounting for all mechanisms seen in the
experimental data is critical for the extraction of the $\gamma_vpN^*$ electrocouplings.

It is expected that these new polarization transfer data from CLAS12, along with the measurement of additional observables from CLAS12
in the $K^+Y$ channels that are in progress, will spur the development of reaction models that can be used to access the rich underlying
information to which these channels are expected to be sensitive. This includes determination of the contributing $N^*$ and $\Delta^*$
states in the $s$-channel at the upper end of the nucleon resonance region, as well as the electrocoupling amplitudes for the excited
nucleon states that provide access to the underlying structure of these states in terms of the interplay between the meson-baryon and
quark-gluon degrees of freedom.


\begin{acknowledgments}

We thank Petr Byd\v{z}ovsk\'{y}, Dalibor Skoupil, and Terry Mart for their efforts in preparing the model calculations for this paper
and the CLAS12 RG-K team for their feedback throughout this analysis work. We acknowledge the outstanding efforts of the staff of the
Accelerator and the Physics Divisions at Jefferson Lab in making this experiment possible. This work was supported in part by the U.S.
Department of Energy, the National Science Foundation (NSF), the Italian Istituto Nazionale di Fisica Nucleare (INFN), the French Centre
National de la Recherche Scientifique (CNRS), the French Commissariat pour l'Energie Atomique, the UK Science and Technology Facilities
Council, the National Research Foundation (NRF) of Korea, the HelmholtzForschungsakademie Hessen f{\"u}r FAIR (HFHF), the Chilean Agencia 
Nacional de Investigacion y Desarollo ANID PIA/APOYO AFB180002, and the Skobeltsyn Nuclear Physics Institute and Physics Department at the 
Lomonosov Moscow State University. The Southeastern Universities Research Association (SURA) operates the Thomas Jefferson National 
Accelerator Facility for the U.S. Department of Energy under Contract No. DE-AC05-06OR23177.

$^\dag$ Corresponding author: carman@jlab.org

\end{acknowledgments}


\end{document}